%% file: arXiv_V3.tex
\documentclass [superscriptaddress, reprint, amsmath, assume, showkeys, prm]{revtex4-2}
\usepackage{graphicx}
\usepackage{dcolumn}
\usepackage{bm}
\usepackage{color}
\usepackage{float}
\usepackage[utf8]{inputenc}
\usepackage[mathlines]{lineno}
\usepackage{comment}
\usepackage{amsmath}
\usepackage{hyperref}
\usepackage{xr}
\usepackage{tabularray}
\usepackage{xr-hyper}
\usepackage{titlesec}
\usepackage{soul,xcolor}
\soulregister\cite7
\soulregister\ref7
\setstcolor{blue}
\newcommand{\white}[1]{{\color[rgb]{1,1,1}{#1}}}

\definecolor{omer}{RGB}{240,226,182}
\definecolor{kev}{RGB}{192,225,215}
\definecolor{matan}{RGB}{240, 160, 80}
\definecolor{raf}{RGB}{90, 148, 166}
\definecolor{flo}{HTML}{FADF63}

\usepackage{todonotes}

\newcommand{\CPB}{CsPbBr$_3$}
\newcommand{\BZS}{BaZrS$_3$}

\newcommand{\wn}{cm$^{-1}$}
\newcommand{\app}{$\approx$}

\newcommand{\BE}{Bose-Einstein}

\newcommand{\RS}{R$^2$}
\renewcommand{\tt}[1]{\text{{#1}}} 

\newcommand   {\avg}  [1] {\ensuremath{\left\langle#1\right\rangle}}

\newcommand{\Ipl}{I_{\rm PL}}
\newcommand{\Iin}{I_{\rm in}}
\newcommand{\Inull}{I_{0}}
\newcommand{\Rs}{R_{\rm s}}
\newcommand{\Ea}{E_{\rm a}}
\newcommand{\Eb}{E_{\rm b}}
\newcommand{\Eg}{E_{g}}
\newcommand{\Eph}{E_{\rm ph}}
\newcommand{\kB}{k_{\rm B}}
\newcommand{\kr}{k_{\rm r}}
\newcommand{\krCPB}{k_{\rm r, CPB}}
\newcommand{\krBZS}{k_{\rm r, BZS}}
\newcommand{\knra}{k_{\rm nr, a}}
\newcommand{\knrb}{k_{\rm nr, b}}
\newcommand{\Gpl}{\Gamma_{\rm PL}}
\newcommand{\Gr}{\Gamma_{{\rm R} j}}
\newcommand{\Grp}{\Gamma_{{\rm R} j'}}
\newcommand{\Gnull}{\Gamma_{0}}
\newcommand{\Tdebye}{T_{\tt{Debye}}}

\makeatletter

\begin{document}

\title{Vibrational properties differ between halide and chalcogenide perovskite semiconductors, and it matters for optoelectronic performance}


\author{Kevin Ye}
\thanks{These authors contributed equally to this work}
\affiliation{Department of Materials Science and Engineering, Massachusetts Institute of Technology, Cambridge, MA 02139, USA}
\author{Matan Menahem}
\thanks{These authors contributed equally to this work}
\affiliation{Department of Chemical and Biological Physics, Weizmann Institute of Science, Rehovot 76100, Israel}
\author{Tommaso Salzillo}
\affiliation{Department of Industrial Chemistry, University of Bologna, Bologna, Italy}
\author{Florian Knoop}
\affiliation{Department of Physics, Chemistry and Biology (IFM), Link\"oping University, SE-581 83, Link\"oping, Sweden.}
\author{Boyang Zhao}
\affiliation{Mork Family Department of Chemical Engineering and Materials Science, University of Southern California, Los Angeles, CA 90089, USA}
\author{Shanyuan Niu}
\affiliation{Mork Family Department of Chemical Engineering and Materials Science, University of Southern California, Los Angeles, CA 90089, USA}
\author{Olle Hellman}
\affiliation{Department of Physics, Chemistry and Biology (IFM), Link\"oping University, SE-581 83, Link\"oping, Sweden.}
\affiliation{Department of Molecular Chemistry and Material Science, Weizmann Institute of Science, Rehovot 76100, Israel}
\author{Jayakanth Ravichandran}
\affiliation{Mork Family Department of Chemical Engineering and Materials Science, University of Southern California, Los Angeles, CA 90089, USA}
\affiliation{Ming Hsieh Department of Electrical and Computer Engineering, University of Southern California, Los Angeles, CA 90089, USA}
\affiliation{Core Center of Excellence in NanoImaging (CNI), University of Southern California, Los Angeles, CA 90089, USA}
\author{Rafael Jaramillo}
\email{rjaramil@mit.edu}
\affiliation{Department of Materials Science and Engineering, Massachusetts Institute of Technology, Cambridge, MA 02139, USA}
\author{Omer Yaffe}
\email{omer.yaffe@weizmann.ac.il}
\affiliation{Department of Chemical and Biological Physics, Weizmann Institute of Science, Rehovot 76100, Israel}

\date{\today}

\begin{abstract}

We report a comparative study of temperature-dependent photoluminescence and structural dynamics of two perovskite semiconductors, the chalcogenide \BZS\ and the halide \CPB. These materials have similar crystal structures and direct band gaps, but we find that they have quite distinct optoelectronic and vibrational properties. Both materials exhibit thermally-activated non-radiative recombination, but the non-radiative recombination rate in \BZS\ is four orders-of-magnitude faster than in \CPB, for the crystals studied here. Raman spectroscopy reveals that the effects of phonon anharmonicity are far more pronounced in \CPB\ than in \BZS. Further, although both materials feature a large dielectric response due to low-energy polar optical phonons, the phonons in \CPB\ are substantially lower in energy than in \BZS. Our results suggest that electron-phonon coupling in \BZS\ is more effective at non-radiative recombination than in \CPB, and that \BZS\ may also have a substantially higher concentration of non-radiative recombination centers than \CPB. The low defect concentration in \CPB\ may be related to the ease of lattice reconfiguration, typified by anharmonic bonding. It remains to be seen to what extent these differences are inherent to the chalcogenide and halide perovskites and to what extent they can be affected by materials processing.

\end{abstract}

\maketitle

\section{Introduction}

Halide perovskites exhibit outstanding optoelectronic properties and are under intensive development for photovoltaic (PV) applications~\cite{Brenner2016, Gratzel2014, Stranks2015}. They can be synthesized at near-room temperature and feature sharp band-edge optical absorption and long excited-state charge carrier lifetimes~\cite{He2016, Shi2015, Stranks2014, Yamada2015, Wehrenfennig2014}. They also suffer from the drawbacks of thermal and chemical instability and lead toxicity. For these reasons, research continues into alternative materials for thin-film photovoltaics. Chalcogenide perovskites are an intriguing alternative. They feature similar crystal structure and range of direct band gap as their halide counterparts~\cite{Niu2017, jaramillo_praise_2019, sun_chalcogenide_2015}. 
They also share a very large dielectric constant, which suggests similar lattice dynamics and screening of charged-defects~\cite{Filippone2020, ye_low-energy_2022}. However, chalcogenide perovskites are, in other ways, quite different. The materials are made of earth-abundant elements with little toxicity concerns. Synthesis is challenging and typically requires very high temperature and air-free processing. Once formed, they are very stable~\cite{Niu2017, Niu2018, sadeghi_making_2021, wei_realization_2020, surendran_epitaxial_2021}. The optoelectronic properties of chalcogenide perovskites have been little reported. Existing published results suggest that photoluminescence (PL) is not strong and that defect control is a challenge, but long PL transient decay times have been measured~\cite{Comparotto2020, Niu2017, wei_realization_2020, surendran_epitaxial_2021, ye_time-resolved_2022, pradhan_angew_2023, yang_low-temperature_2022}. No chalcogenide perovskite thin-film solar cells have been reported, likely due to the challenging conditions required for film synthesis.

Many researchers, including ourselves, have argued that the outstanding optoelectronic properties of halide perovskites are intertwined with their strongly anharmonic structural dynamics~\cite{SharmaMAPI1, YaffePRL2017, Guo2017b, Mayers2018, Whalley2016b, Egger2018, Panzer2017, Seidl2023, Gehrmann2019, Schilcher2021}. This poses questions of how the optoelectronic properties and lattice dynamics of chalcogenide and halide perovskites compare and what we can learn about photovoltaic performance from this comparison. 

Here, we employ temperature-dependent PL and Raman scattering spectroscopy, alongside first-principles self-consistent phonons simulations, to compare two exemplary single-crystal samples: the chalcogenide perovskite \BZS\ and the halide perovskite \CPB. 
\BZS\ and \CPB\ share the same distorted perovskite, corner-sharing, orthorhombic crystal structure (space group $Pnma$) within the studied temperature range~\cite{Niu2019, Lanigan-Atkins2021}. Between 80 and 300~K, the PL emission intensity of both materials decreases due to thermally-activated non-radiative recombination, with electron-phonon interactions broadening the PL linewidth in both cases. However, \CPB\ emission intensity surpasses that of single-crystal \BZS\ by over four orders of magnitude, suggesting a significantly higher concentration of very-low-barrier non-radiative recombination centers in \BZS.

We use Raman spectroscopy and first-principles simulations to show that the phonon spectrum in \BZS\ extends to nearly double the frequency in \CPB. Raman analysis demonstrates that \CPB\ exhibits pronounced anharmonic characteristics, even at low temperatures. In contrast, \BZS\ exhibits predominantly harmonic vibrational behavior akin to traditional tetrahedrally-bonded semiconductors. Our findings are consistent with the hypotheses that the lower-frequency and strongly anharmonic structural dynamics in \CPB\ may reduce carrier capture cross sections due to Franck-Condon effects~\cite{das_what_2020}. They may also decrease defect concentration through annealing, even at room temperature~\cite{cahen_are_2021}. Definitive tests of these hypotheses await detailed calculations of rate coefficients supported by first-principles quantum calculations and defect spectroscopy to experimentally identify and quantify recombination-active defects.

\section{\label{sec_methods} Methods}

We grew single crystals of \BZS\ with dimensions of approximately 100~$\mu$m using a flux method as reported previously~\cite{Niu2017}. 
We grew single crystals of \CPB\ with dimensions on the order of 2~mm using the vapor saturation of an antisolvent (VSA) method as reported previously~\cite{Rakita2016}.

We performed Raman scattering in a home-built system in back-scattering geometry. We used a 1.58~eV CW pump-diode laser (Toptica Inc., USA), which is below the band gap of \BZS\ (1.9~eV)~\cite{Niu2017, Nishigaki2020} and \CPB\ (2.32~eV)~\cite{Guo2018}.  
We performed PL spectroscopy using the same optical system and a 2.54~eV CW sapphire-pumped diode laser (Coherent Inc., USA). 
The signal was measured in a 1~m long spectrometer (FHR1000, Horiba Inc.) equipped with a holographic grating (1800~gr/mm for Raman and 150~gr/mm for PL) and a Synapse Plus CCD detector (Horiba Inc.).

The sample temperature was controlled using a liquid-N$_{\rm 2}$- or liquid He-cooled high-vacuum optical cryostat (Janis Inc., USA) for $T \leq 400$~K.
For higher temperature measurements, the temperature was controlled in an optical furnace (Linkam TS1000) under N$_{\rm 2}$ flow. 
We performed measurements for a series of increasing temperatures. We took precautions to isolate the effect of sample temperature on PL intensity by maintaining stable measurement conditions, refocusing the laser before every measurement, and taking repeated measurements across the whole temperature range. The measurement system and protocols are further described in Refs.~\cite{Menahem2021, SharmaMAPI2, Asher2020, Cohen2022, Asher2023}.

We calculate normalized PL yield ($Y$) by analyzing the integrated PL intensity, including both the band-to-band peaks and the shoulders. The PL data is measured in counts, and we compute intensity by dividing by the acquisition time. We define $Y$ as the ratio between the PL intensity emitted from the sample ($\Ipl$, measured in counts/sec) and the excitation laser intensity that reaches the interior of the sample ($\Iin$):
\begin{equation}
    Y = \frac{\Ipl}{\Iin} = \frac{\Ipl}{\Inull}(1-\Rs)~.
\label{eq:yield_raw}
\end{equation}
 We measure the laser intensity $\Inull$ (in photons/sec) at the microscope entrance. We compute $\Iin$ from $\Inull$ by accounting for losses due to reflection from the cryostat windows and the sample surface. The same cryostat window reflection losses affect the collected PL intensity and cancel in the expression for $Y$, leaving only a dependence on the sample surface power reflection coefficient ($\Rs$). Further details are presented in Sec.~\ref{secSM_PL} in the Supplemental Material (SM)~\cite{SM}. 


\section{Results}
\label{Sec:Crystals}

\subsection{Photoluminescence and Non-Radiative Recombination}

\begin{figure}
    \centering
    \includegraphics[width = 8.5 cm]{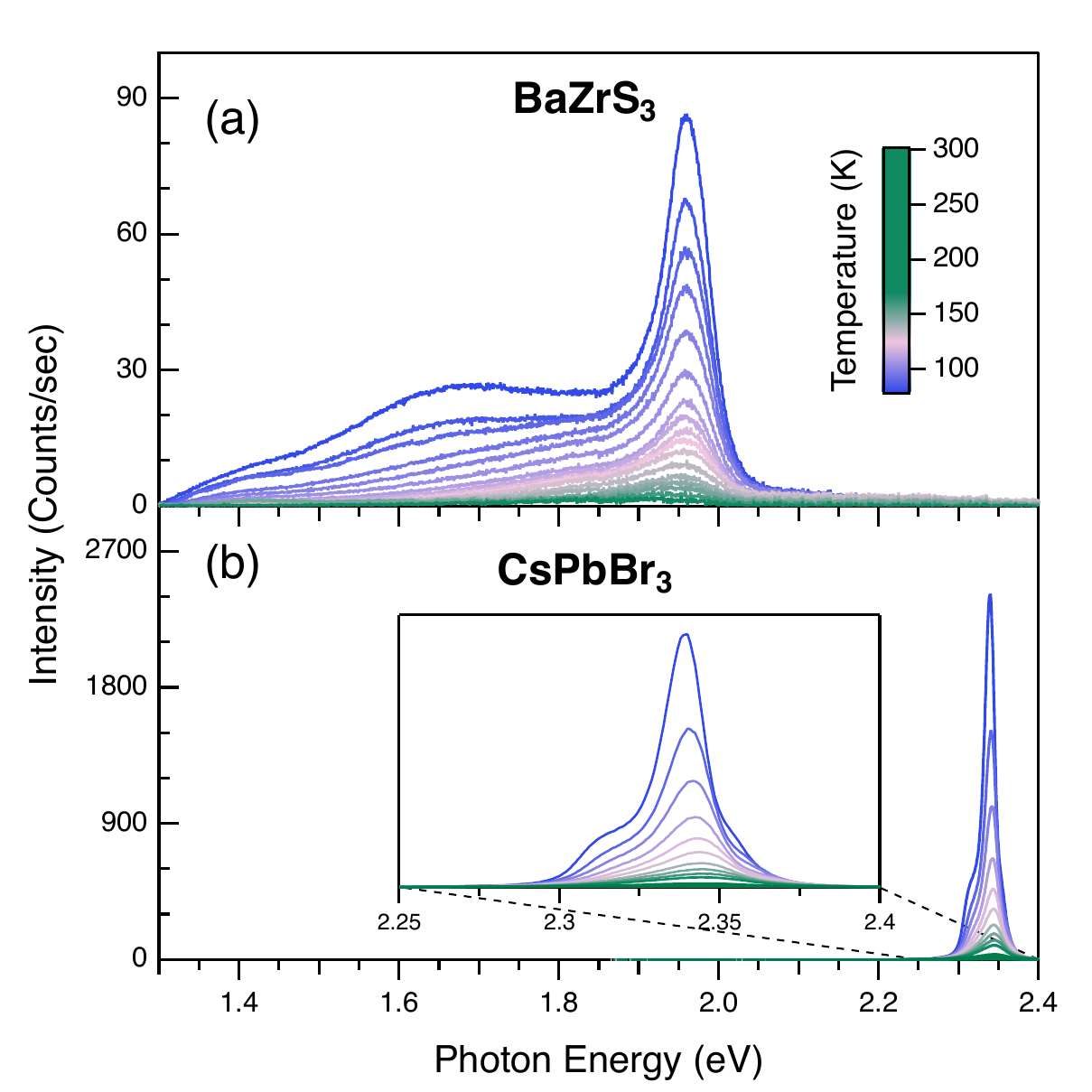}
    \caption{Temperature-dependent PL spectra. PL spectra measured between 80~K and 300~K for (a) \BZS\ and (b) \CPB, after subtraction of the background spectra at 170~K and 300~K, respectively. The line color indicates temperature.
    \BZS\ and \CPB\ were measured with different pump laser fluence: 4.3~MJ/cm$^2$ for \BZS, and 2.7~$\times$10$^{-4}$~MJ/cm$^2$ for \CPB.}
    \label{Fig:PL1}
\end{figure}

In Fig.~\ref{Fig:PL1}, we present temperature-dependent PL spectra of \BZS\ and \CPB\ measured between 80--300~K.
At low temperatures, spectra are dominated by band-edge recombination at $\Eg$\app 1.96~eV and 2.34~eV for \BZS\ and \CPB, respectively. The low-temperature spectra for both samples feature low-energy shoulders, which are quenched at elevated temperatures. The band-to-band peak and the shoulder are much broader in \BZS\ than in \CPB. In both materials, PL is quenched with increasing temperature, and for \BZS, it becomes too weak to measure above 170~K. To eliminate external features in the PL spectra, the high-temperature spectra were subtracted from those in Fig.~\ref{Fig:PL1}. 

\begin{figure}
    \centering
    \includegraphics[width = 8.5 cm]{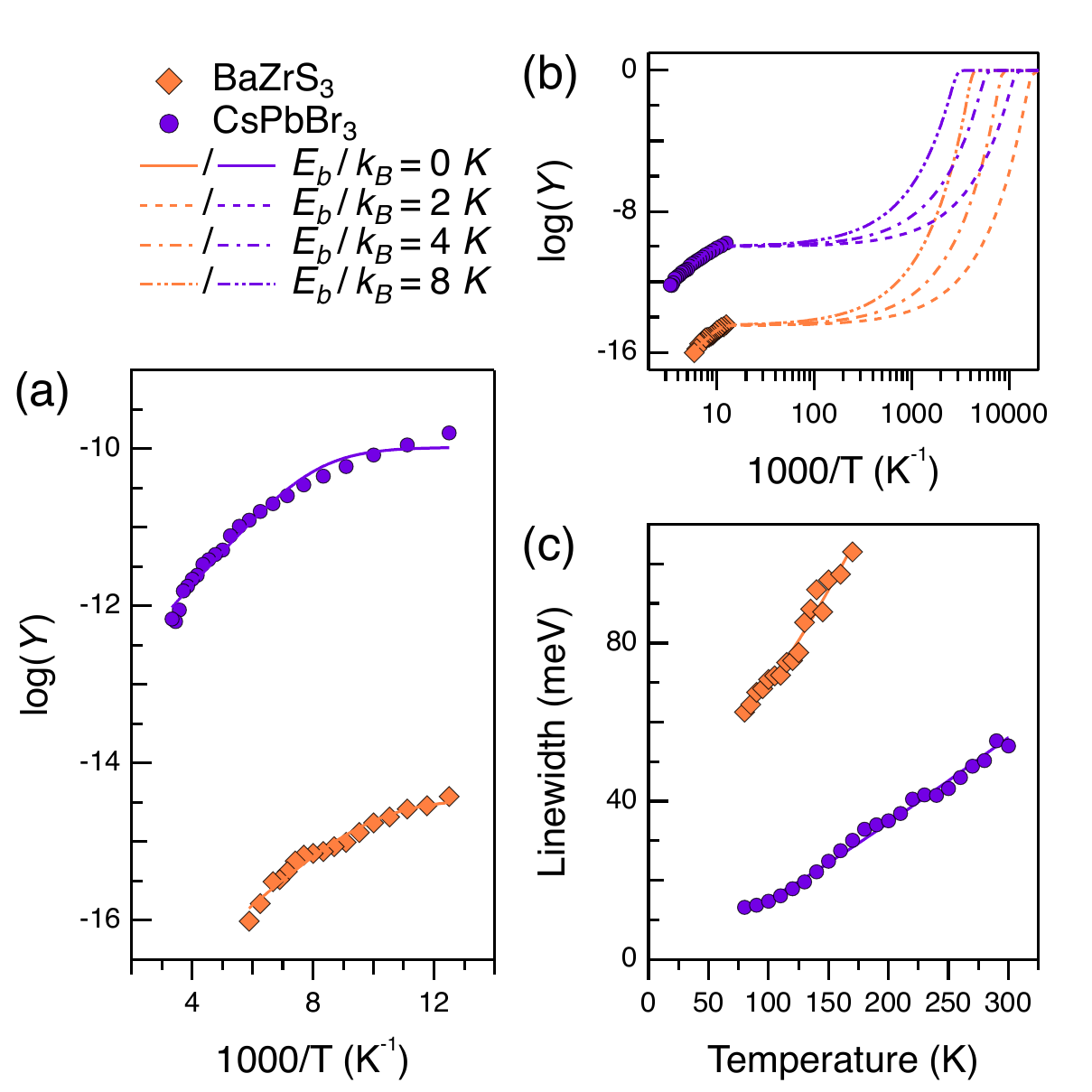}
    \caption{Model-based analysis of normalized PL yield ($Y$, Eq.~\eqref{eq:yield}) and linewidth ($\Gpl$).  (a) $Y$ as a function of the inverse temperature. (b) Fit to a model with a second non-radiative recombination pathway. $\Eb / \kB$ fall below our measurement temperature window. (c) $\Gpl$ of the band-to-band peak as a function of the temperature. All fitted values can be found in Sec.~\ref{secSM_PL} of the SM~\cite{SM}.
    Purple (circles) and orange (diamonds) traces (markers) are for \CPB\ and \BZS, respectively.}
    \label{Fig:PL2}
\end{figure}

\BZS\ and \CPB\ are direct band-gap semiconductors and strongly absorb the pump laser. At the pump laser energy of 2.54~eV, the absorption coefficient of \BZS\ is 1.68$\times \tt{10}^{5}$ $\tt{cm}^{-1}$, and that of \CPB\ is 4.04$\times \tt{10}^{5}$ $\tt{cm}^{-1}$~\cite{sadeghi_making_2021,ermolaev_giant_2023}. Therefore, the pump laser absorption depth is much smaller than the thickness for both single-crystal samples, and all the light entering the samples is absorbed.
However, the PL emission intensity of \CPB\ dwarfs that of \BZS. 

In Fig.~\ref{Fig:PL2}(a), we present an Arrhenius plot of the normalized PL yield ($Y$). We fit the data to a minimal model of thermally-activated non-radiative recombination~\cite{Bimberg1971, Lambkin1994}:
\begin{equation}
    Y(T) = \frac{1}{1 + A e^{-\Ea / \kB T}+Be^{-\Eb / \kB T}}~,
\label{eq:yield}
\end{equation}
$\Ea$ and $\Eb$ are the activation energies of two non-radiative recombination mechanisms, and $A = \knra^0 / \kr$ and $B = \knrb^0 / \kr$ are ratios of exponential prefactors that compare non-radiative to radiative recombination rates. As discussed below, two non-radiative recombination mechanisms are required for a minimal model of the data. This model assumes that all radiative recombination has a sufficiently small activation energy that appears barrierless within the measurement window. The effect of this assumption is that the PL internal quantum efficiency (IQE) approaches unity at low temperatures. The best-fit model parameters are presented in Table~\ref{tab:PL_Fits}.

\begin{table*}[!ht]
\centering
\caption{Best-fit parameters for PL yield, Eq.~\eqref{eq:yield}. Results are presented here for $\Eb / \kB=0$ K. In Table~\ref{tab:SM_PL_Fits} in the SM~\cite{SM}, we present results for other choices of $\Eb / \kB$.}
\begin{tblr}{c|c|c|c|c|c}

        Material &   $\Inull$ (photons/sec) &   $\Ea$ (meV)  &  ${\Eb} / \kB$ (K) &  $A$ &  $B$  \\
    \hline
         \BZS\ &  6.7$\times \tt{10}^{15}$ & 68 $\pm$ 7  & 0 &  (6.9 $\pm$ 3.8) $\times \tt{10}^{17}$ &  (2.8 $\pm$ 0.5) $\times \tt{10}^{14}$\\
    \hline
         \CPB\ &  4.3$\times \tt{10}^{11}$  &  86 $\pm$ 5 & 0 & (2.8 $\pm$ 0.8) $\times \tt{10}^{13}$ &  (9.6 $\pm$ 1.4) $\times \tt{10}^{9}$ \\
    \hline
\end{tblr}
\label{tab:PL_Fits}
\end{table*}

We find that \BZS\ and \CPB\ feature a non-radiative recombination process with similar activation energy $\Ea = 68 \pm 7$ and $86 \pm 5$~meV, respectively. This is apparent by eye in the Arrhenius plot (Fig.~\ref{Fig:PL2}(a)), as both data sets have similar curvature and a knee at approximately the same temperature. However, the PL yield of \CPB\ is over four orders of magnitude larger than that of \BZS, and the data do not extrapolate towards the same yield in the low-temperature limit. The presence of non-radiative recombination mechanisms with similar activation energy belies the difference in PL intensity. 

To explain the divergence in PL intensity, we invoke a second non-radiative recombination mechanism in \BZS, with activation energy $\Eb$ too low to be determined in our measurement window. We also assume a similar second mechanism in \CPB\ for consistency in the model function. By construction, the thermal scale $\Eb / \kB$ falls well below our experimental temperature window, and therefore, $\Eb$ cannot be determined by regression. This low-temperature process is saturated within the measured temperature range, and $e^{-\Eb / \kB T}\approx 1$. With this assumption, we fix the temperature scale $\Eb / \kB$ at a low value and fit the data to determine $B$. 
In Fig.~\ref{Fig:PL2}(b), we present a family of model functions with $\Eb$ varied manually; for $\Eb$ sufficiently small, its value has an insignificant impact on the fit results and our conclusions. The ratio $B_{\tt{BZS}}/B_{\tt{CPB}}$ is meaningful, as it determines the measured difference in low-temperature PL yield between \BZS\ (BZS) and \CPB\ (CPB). We find that:

\begin{equation} 
    \frac{B_{\text{BZS}}}{B_{\text{CPB}}} = \frac{\left(\knrb^0 / \kr \right)_{\text{BZS}}}{\left(\knrb^0 / \kr \right)_{\text{CPB}}} = 28,181~.
\end{equation}

\noindent Both materials have comparable optical absorption coefficients near their band edge; therefore, we can approximate $\left(\krCPB / \krBZS \right) \approx \mathcal{O}(1)$. We therefore conclude that the rate $\knrb^0$ of low- (or zero-) barrier non-radiative recombination is $\mathcal{O}(10^4)$ times faster in \BZS\ than in \CPB. 

We analyze the PL linewidth ($\Gpl$) for insight into how excited states couple to lattice vibrations. 
In Fig.~\ref{Fig:PL2}(b), we show the temperature dependence of $\Gpl$ of the main PL peak at the $\Eg$ for each material. We use multiple Gaussian peaks to fit the data, three for \CPB\ and four for \BZS. For both materials, the main PL peak is well modeled by two Gaussian peaks, and we determine $\Gpl$ numerically from the best-fit models. We fit $\Gpl$ to a relation describing thermal and non-thermal broadening, previously applied to halide perovskites~\cite{Guo2018}: 
\begin{equation}
    \Gpl (T) = \Gnull + \frac{\gamma}{{\rm e}^{\Eph / \kB T} - 1}~,
\label{eq:FWHM}
\end{equation}
\noindent $\Gnull$ represents temperature-independent contributions to $\Gpl$, such as heterogeneous broadening due to quenched structural disorder. The second term describes broadening due to coupling between excited states and lattice vibrations with characteristic energy $\Eph$ and electron-phonon coupling constant  $\gamma$. These lattice vibrations are likely longitudinal optical phonons (see below), and the electron-phonon interactions can be described through deformation potentials. 

We find that luminescent band-edge states couple to vibrations with characteristic energy that is similar in both materials: $\Eph=$~24~$\pm$~7~meV for \BZS, and 17~$\pm$~4~meV for \CPB\ (see below). The electron-phonon coupling constant $\gamma$ is larger for \BZS\ (200~$\pm$~100~meV) than for \CPB\ (50~$\pm$~10~meV).
$\gamma$ is related to deformation potentials and polar-optical coupling.
The higher value in \BZS\ may be related to the stronger covalent nature of the Zr-S bonds compared to Pb-Br bonds: 32\% covalent for Zr-S, compared to 26\% for Pb-Br, according to the Pauling scale. This implies enhanced sensitivity of electron energies to bond angles in the chalcogenide, as we have studied previously~\cite{li_band_2019,ye_low-energy_2022}. We also see that \BZS\ has larger heterogeneous broadening, $\Gnull$, than \CPB, reflecting a higher concentration of structural defects and other forms of non-thermal disorder. In addition to affecting the PL spectra, these factors affect how steeply optical absorption rises at the band gap, the width of the Urbach tail, and the phonon-limited charge transport mobility, and are therefore relevant to photovoltaic performance~\cite{wolter_progress_2021,wang_determination_2020,falsini_analysis_2022}.

\subsection{Raman Scattering and Phonon Dynamics}

\begin{figure}
    \centering
    \includegraphics[width = 8.5 cm]{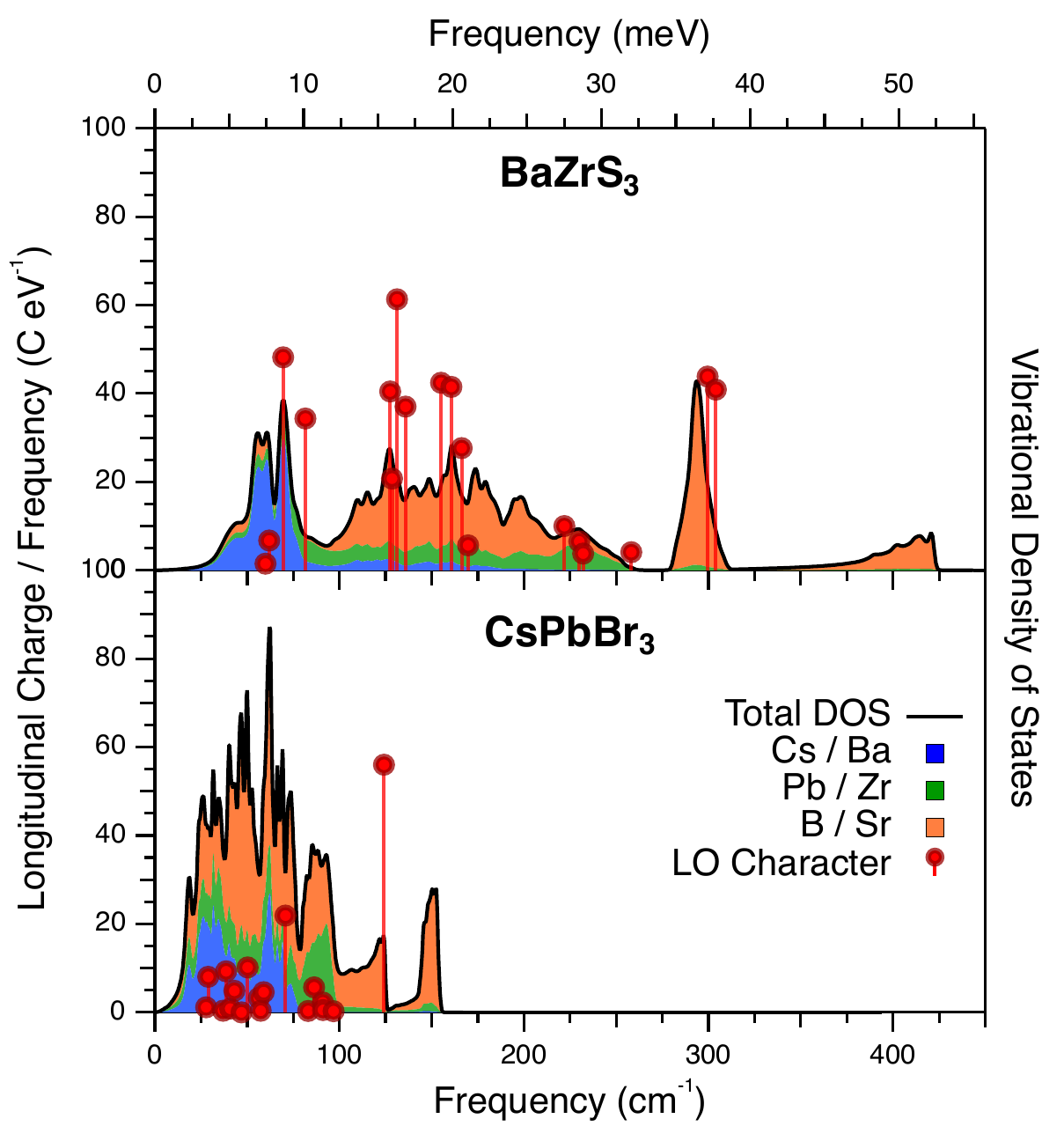}
    \caption{Calculated room-temperature phonon density of states (DOS, right axis) of \BZS\ and \CPB. The colors indicate the partial contribution of the A (blue), M (green), and X (orange) ions in the AMX$_3$ formula. The red lines are the measure (left axis) of the LO character.}
    \label{Fig:TDEP}
\end{figure}

The lower luminescence yield and larger PL broadening in \BZS, compared to \CPB, are related to differences in lattice vibrations. To better understand these relationships, we turn to theoretical and experimental studies of the phonon spectra. In Fig.~\ref{Fig:TDEP}, we present the total (black) and partial (orange, green, and blue) vibrational density of states of \BZS\ (top) and \CPB\ (bottom), calculated \emph{ab initio} using the temperature-dependent effective potential (TDEP) method at 300\,K (details in Sec.~\ref{SecSM_TDEP} the SM~\cite{SM, Knoop.2024, Benshalom2022, Roekeghem.2021, Giannozzi.2020, Bianco.2017, Shulumba.2017, Giannozzi.2017, Schlipf.2015ake, Hellman.2013oi5, Hamann.2013, Jain.2013, Giannozzi.2009, Blum.2009, Mattsson.2008, Armiento.2005, Perdew.1996}). The optical vibrations in \BZS\ span frequencies up to 430~$\text{cm}^{-1}$ (53~meV), while those in \CPB\ peak at 160~$\text{cm}^{-1}$ (20~meV). This is due to the higher mass of the atoms in \CPB\ and the more covalent character of the bonds in \BZS~\cite{FoxOPOS}.

Longitudinal optical (LO) phonons couple to electronic excitations and therefore likely contribute most to the temperature-dependent PL linewidth broadening. In Fig.~\ref{Fig:TDEP}, the vertical red lines provide a measure of the phonon LO character: As detailed in Sec.~\ref{SecSM_TDEP} in the SM~\cite{SM}, the LO character is estimated by computing the longitudinal charges for optical vibrations from the mode eigenvectors and atomic Born charge tensors projected on the [101] excitation direction and divided by the vibration frequency to account for the average mode amplitudes. We see that \BZS\ has many phonons with prominent LO character along the [101] excitation direction, whereas \CPB\ only shows one mode of appreciable LO character at $\approx 125\,{\rm cm}^{-1}$. This factor likely contributes to the stronger temperature-dependent PL broadening in \BZS\ in addition to the higher deformation potentials.

The characteristic energy of lattice vibrations responsible for PL broadening in \CPB, $\Eph \simeq \tt{17}$~meV ($\approx 140\: \text{cm}^{-1}$), is comparable to the most prominent LO phonon in the calculated spectrum. For \BZS, $\Eph \simeq \tt{24}$~meV ($\approx 190\: \text{cm}^{-1}$) falls roughly in the middle of the calculated spectrum. Likely, band-edge states couple to a diversity of lattice vibrations in both materials through polar-optical coupling and deformation potentials, and the characteristic energy determined by PL broadening represents a distribution of processes. Similar conclusions have been reached in studies of ultra-fast lattice rearrangement following optical excitation of halide perovskites, including \CPB~\cite{yazdani_coupling_2024}.

\begin{figure}
    \centering
    \includegraphics[width = 8.5 cm]{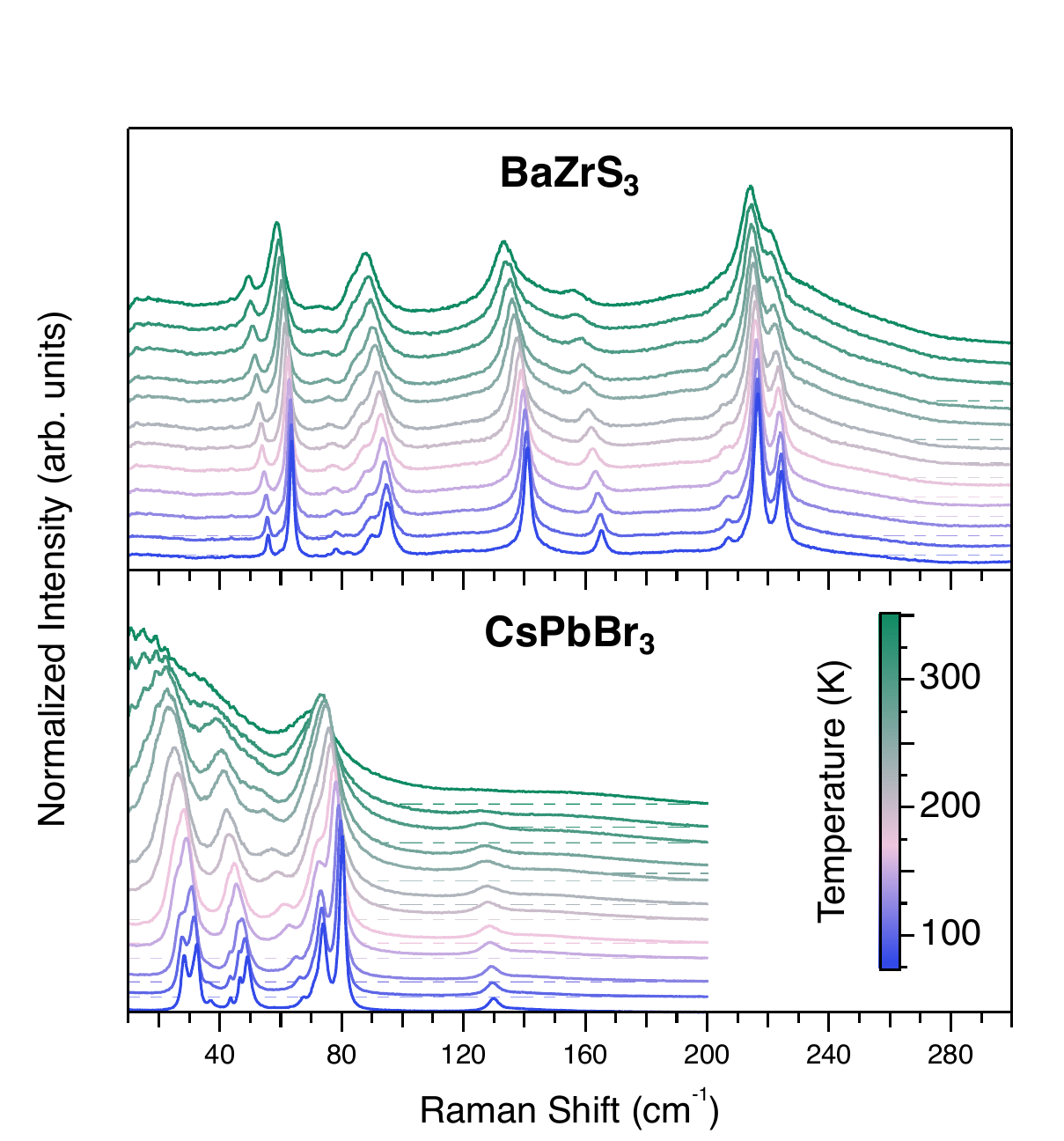}
    \caption{Temperature-dependent Raman scattering spectra of \BZS\ and \CPB. Temperature is indicated by color. The individual data sets are shifted vertically for clarity; the dashed horizontal lines indicate the zero level for each data set.}
    \label{Fig:Raman_Norm}
\end{figure}

In Fig.~\ref{Fig:Raman_Norm}, we present the low-frequency Raman spectra of \BZS\ and \CPB\ measured in the temperature range 80--350~K. The spectra of \BZS\ and \CPB\ are consistent with previous reports~\cite{Cohen2022, Gross2017}. In Sec.~\ref{SecSM_RamanData} in the SM~\cite{SM}, we present the Raman spectra of \BZS\ measured over a wider temperature range, from 10--875~K, showing no indication of a phase transition~\cite{Niu2018, Xu2022, Yu2021}. 

The Raman spectra of \CPB\ exhibit sharp first-order peaks (one photon scattered by one vibration)~\cite{Cardona1982} at low temperatures, which shift and broaden significantly with increasing temperature. These structural dynamics are characteristic of halide and certain oxide perovskites~\cite{Menahem2023, Perry1967, McMillan1988, ROuillon2002}.  
The broad, featureless spectra at high temperatures, combined with a rapid decrease in vibrational lifetime (evidenced by peak broadening), indicate the impact of anharmonic vibrational dynamics ~\cite{YaffePRL2017, Guo2018}. 

At low temperatures, the Raman spectra of \BZS\ resemble those of \CPB\ and other halide perovskites~\cite{Menahem2021, Reuveni2023}, with a similar number of prominent peaks divided into doublets and triplets. The vibrations in \BZS\ are higher in frequency than those in \CPB. Unlike \CPB, the Raman peaks in \BZS\ remain sharp and distinct with increasing temperature, and even at 350~K, there is scant signature of anharmonic dynamics. The harmonic dynamics in \BZS\ recall those in tetrahedrally-bonded semiconductors (\textit{e.g.} GaAs), in which vibrations along the stiff covalent bonds do not interact and change little with temperature~\cite{Born1947, Y.Yu2010}. 

\begin{figure*}
    \centering
    \includegraphics[width = 17 cm]{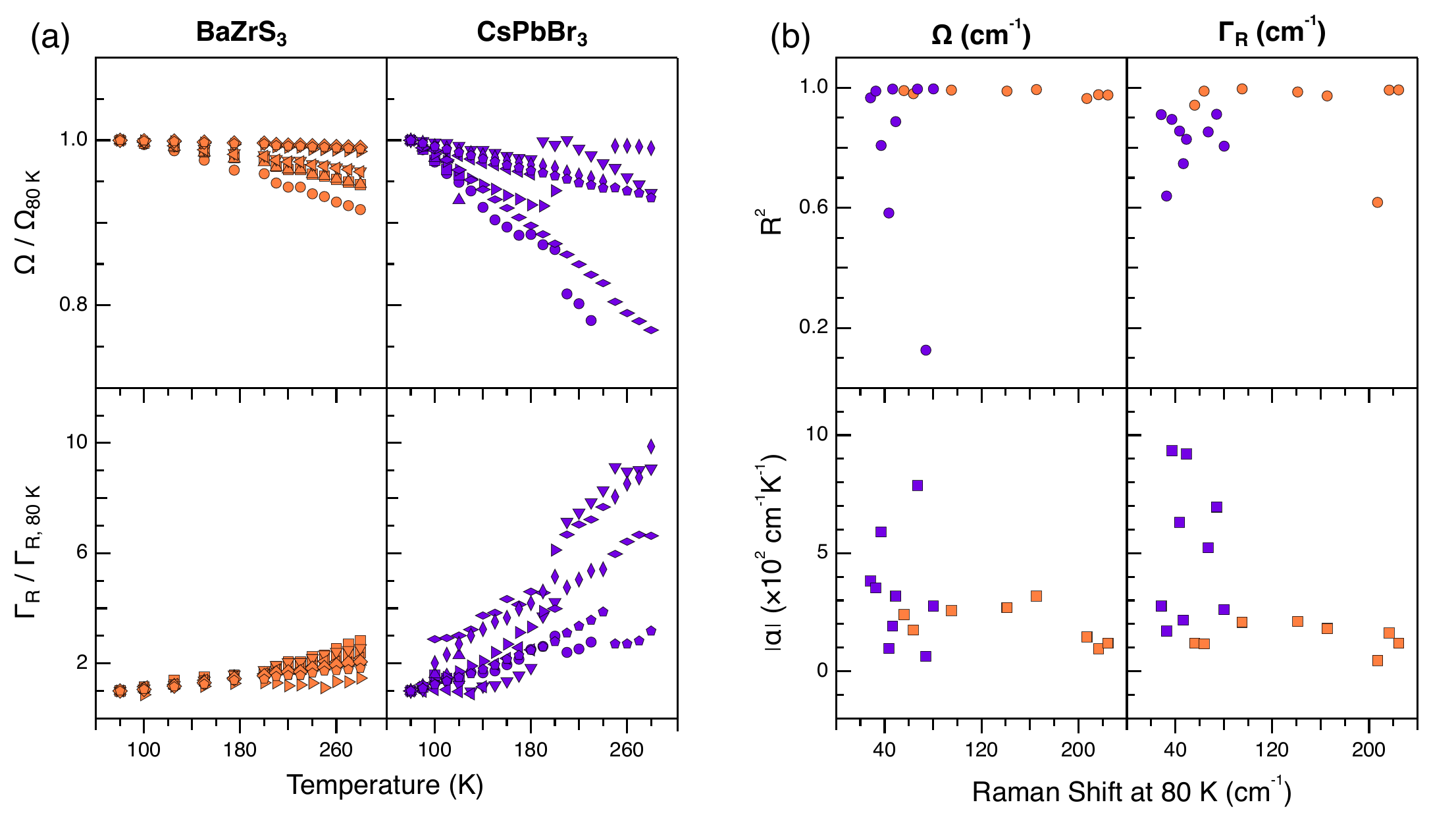}
    \caption{Analysis of Raman Spectra: (a) Frequencies ($\Omega$ - top) and linewidths ($\Gr$ - bottom) of the first-order peaks of \BZS\ (left) and \CPB\ (right) as a function of temperature, relative to the value at 80~K. Different markers mark different peaks. (b) Linear regression results (\RS\ - top, and slopes $\alpha$ - bottom) of the temperature-dependent frequencies (left) and linewidths (right) of \BZS\ and \CPB, between 80-280~K. The results are plotted as a function of peak frequency at 80~K. Orange and purple markers are for \BZS\ and \CPB, respectively.}
    \label{Fig:Raman_Analysis}
\end{figure*}

We can more quantitatively compare the structural dynamics of \BZS\ and \CPB\ by considering the temperature dependence of the phonon frequencies and linewidths. We model the Raman spectra as a convolution of pseudo-Voigt peaks multiplied by the Bose-Einstein occupation factor~\cite{Kwok1968, Maradudin1971, Safran1977} (see Sec.~\ref{SecSM_RamanData} in the SM~\cite{SM}). We focus on the prominent peaks because they yield the most confident fits. In Fig.~\ref{Fig:Raman_Analysis}(a) we present the vibrational frequency ($\Omega$) and linewidth ($\Gr$) of the prominent first-order peaks of \BZS\ (left panel, orange) and \CPB\ (right panel, purple). The frequency and linewidth data are divided by their values at 80~K for ease of comparison (for original values, see Sec.~\ref{SecSM_RamanData} in the SM~\cite{SM}). The Raman peaks of \BZS\ red shift and broaden linearly with increasing temperature, as is expected from perturbation theory for weak anharmonic effects~\cite{Haro1986, Menendez1984}. A slight change in slope near 200~K may relate to the observed increase in the low-frequency dielectric susceptibility in this same temperature range~\cite{Filippone2020}.

\CPB\ exhibits a similar monotonic and linear trend in frequency and linewidth with increasing temperature up to 280~K, above which the data deviate substantially, including sharp blue shifts and rapidly increasing linewidth (see Fig.~\ref{fig:SM_RamanTrend} is the SM~\cite{SM}). We attribute these changes to two causes: (1) the thermal population of all the optical phonons above $\Tdebye \approx 250$~K, and (2) the incipient orthorhombic-to-tetragonal phase transition at $T \approx 360$~K. Peak broadening with increasing temperature is attributed to the shortening of vibrational lifetime due to phonon-phonon scattering~\cite{Haro1986, Menendez1984, Maradudin1971}. It is apparent that the structural dynamics of \BZS\ exhibit minor thermal effects and are less impacted by anharmonic vibrational interactions than the dynamics of \CPB~\cite{Haro1986, Menendez1984}.

The extent of anharmonicity in the phonon dynamics can be further quantified by analyzing the deviation from linearity of the temperature dependence of the phonon frequencies and linewidths. A linear temperature dependence is expected within perturbation theory for weak phonon-phonon interactions. Therefore, strong and anharmonic interactions are indicated by deviations from linearity. In Fig.~\ref{Fig:Raman_Analysis}(b), we present the \RS\ parameter and the absolute value of the slope ($\lvert\alpha\rvert$) of linear regression to the $\Omega(T)$ and $\Gr(T)$ (for mean values, see Sec.~\ref{SecSM_RamanData} in the SM~\cite{SM}). For \CPB, the \RS\ values are far from unity, and the slopes are much larger than for \BZS. This analysis further supports the conclusion that anharmonic effects in \BZS\ can be modeled as a weak perturbation, whereas in \CPB, they substantially change the lattice vibrational properties.

\section{Discussion}

Lead halide perovskites excel as PV absorbers largely due to their slow rates of defect-assisted Shockley-Read-Hall (SRH) non-radiative free energy loss. Electron-phonon interactions mediate SRH recombination processes, converting excited-state energy into phonons and heat~\cite{Kirchartz2018, Schilcher2021, Egger2018, Mayers2018}. Therefore, by focusing on vibrational properties and electron-phonon interactions, we can better understand the origin of the excellent PV performance of the halides and identify promising new semiconductors for thin-film PV. 

Chalcogenide perovskites have an untested potential for PV mainly due to challenges in thin film synthesis. There are few published measurements of the optoelectronic properties of chalcogenide perovskites, and the results are ambiguous. We reported time-resolved PL data that suggest long radiative lifetimes in \BZS\ and the Ruddlesden-Popper compound Ba$_3$Zr$_2$S$_7$, comparable to lead halides and other high-performing PV materials such as CIGS~\cite{ye_time-resolved_2022}. Several research groups have reported band-edge PL measured from \BZS\ samples of various form factors, including solution-processed nanoparticles and thin films made by sulfurizing oxide precursors~\cite{yang_low-temperature_2022,pradhan_angew_2023, Comparotto2020,wei_realization_2020}. However, quantitative PL measurements (albeit on early-stage powder samples) suggest low quantum efficiency and low quasi-Fermi level splitting (\textit{i.e.} open-circuit voltage potential)~\cite{Niu2017}. Our experience of measuring PL on \BZS\ single crystals is that many samples have weak or no measurable band-edge PL and that the band-edge PL can be strongly position-dependent. We have observed similar variability in measuring PL on epitaxial thin films (unpublished). Substantial sample-to-sample variation, combined with challenges in synthesis and defect control, makes it difficult to assess the technological potential of chalcogenide perovskites. 

SRH recombination is much faster in \BZS\ than in \CPB\ due to an unidentified recombination pathway that is active even at low temperature, implying that it has a very low thermal activation energy (Fig.~\ref{Fig:PL2}(b)). The theory of multi-phonon non-radiative recombination highlights the importance of the overlap between the vibrational wavefunctions of the initial and final electronic states, consistent with the Franck-Condon principle~\cite{das_what_2020}. Model calculations for a harmonic lattice indicate a steep dependence of SRH recombination rate on phonon energy~\cite{Kirchartz2018}. Lower phonon energy means more phonons are needed to participate in non-radiative recombination processes. Within this model, the overlap of vibrational wavefunctions is suppressed with increasing phonon number. Therefore, the lower phonon frequency in \CPB\ compared to \BZS\ may explain the difference in SRH recombination rates, even if the concentration and properties of defects are similar.

The same harmonic model also predicts that SRH recombination rates increase with the extent of lattice distortion accompanying defect charge capture; this distortion can be parameterized by the Huang-Rhys factor~\cite{Kirchartz2018,das_what_2020}. These lattice distortions often result in long-lived charge trapping and are responsible for persistent photoconductivity in many semiconductors; it is related to DX-center phenomena in III-V compounds, and we termed it defect-level switching in the context of resistive switches~\cite{lang_large-lattice-relaxation_1977,mooney_deep_1990,yin_large_2018,yin_defect-level_2021}. There have been many studies of persistent photoconductivity in halide perovskites. Still, there is no clear model for defects in \CPB~\cite{Liu2023, Cohen2019, Kang2017, Sebastian2015}. The timescale of photoconductivity decay in \BZS\ crystals depends on sample polishing, suggesting that defects introduced during plastic deformation (\textit{e.g.} dislocations) may be effective charge traps with large Huang-Rhys factors~\cite{zhao_photoconductive_2023}. However, these observations may not be directly relevant to the Huang-Rhys factors of recombination-active defects, which may differ from the traps that influence photoconductivity decay. Further, it has been suggested that polaronic effects resulting from strong charge-lattice coupling - including Fröhlich (\textit{i.e.} large, effective mass states) polarons and defect-localized polarons - may suppress SRH recombination rates by screening dissimilar charges~\cite{emin_barrier_2018,franchini_polarons_2021}.
It is likely that a full quantum calculation of electron-phonon interactions, combined with first-principles modeling of the known crystal structures and the most likely recombination-active defects, will be required to resolve these questions over the role of strong electron-phonon coupling on SRH recombination rates.

The effect of anharmonic phonon-phonon interactions on SRH recombination remains an open question. It would be enlightening to compare a model calculation of vibrational wavefunction overlap during multi-phonon recombination with and without anharmonic lattice vibrations and for varying Huang-Rhys factor for otherwise similar phonon energies. Model system calculations could help to understand the conflated effects of low phonon energy, lattice distortion at defects, and anharmonic vibrations on SRH recombination rates. However, full quantum calculations are likely needed to accurately model the differences in carrier capture rates between halide and chalcogenide perovskites.

The discussion to this point has focused on the effects of lattice vibrations on SRH recombination rates via the carrier capture processes. However, another significant unknown is the concentration of recombination-active defects. Anharmonic vibrations indicate softer bonds, a more pliable crystal lattice, and lower activation energy for ionic diffusion~\cite{Brenner2020, LimmerJCP2020, Zhao2016}. 
Therefore, anharmonic vibrations in lead halide perovskites may improve optoelectronic performance by accelerating the elimination of point defects by annealing, even at room temperature (termed ``self-healing'')~\cite{cahen_are_2021, Aharon2022, Finkenauer2022, Ceratti2018, Parida2023}.
\BZS, in contrast, requires high temperature to form and, therefore, is likely to retain quenched-in defects (\textit{e.g.} chemical impurities). This suggests that the PL yield and PV performance of chalcogenide perovskites will depend more strongly on defect control during sample processing than the performance of halide perovskites. 

Defect characterization by methods such as deep-level transient spectroscopy (DLTS) and related forms of impedance spectroscopy could directly measure defect concentrations and quantify the extent of self-healing~\cite{Hutter2015, Bi2016, Xing2014, Lei2021, LeCorre2021, Herz2016, Rosenberg2017}. Interpretation of DLTS data is challenging for halide perovskites due to mixed ionic-electronic conduction, but progress has been reported~\cite{reichert_probing_2020}.
Similar studies have not yet been undertaken for chalcogenide perovskites. 

\section{Conclusions}
We have presented a detailed spectroscopic comparison between  two perovskite semiconductors, the chalcogenide \BZS\ and the well-studied halide \CPB. We find that the PL emission yield of \CPB\ is over four orders of magnitude higher than that of \BZS\ for the single-crystal samples considered here, indicative of a far higher concentration of very-low-barrier, non-radiative recombination centers in \BZS. \emph{Ab initio} simulations show that \BZS\ exhibits higher LO vibrational frequencies than \CPB: this may increase carrier capture cross sections due to Franck-Condon effects~\cite{das_what_2020}. 
Raman scattering results suggest that the strongly anharmonic structural dynamics in \CPB, relative to \BZS, may reduce defect concentration through accelerated solid-state diffusion and annealing~\cite{cahen_are_2021}. 

We are optimistic that the fast SRH recombination rates observed here for \BZS\ can be reduced through sample processing. Large sample-to-sample variability, and preliminary results on \BZS\ epitaxial thin films (not shown), suggest that low PL yield may not be intrinsic to \BZS\. The usefulness of \BZS\ (and likely other chalcogenide perovskites) for optoelectronic applications will hinge on continued progress in defect control during synthesis and processing - as has been historically required for other semiconductors.


\section*{Acknowledgements}

We acknowledge the support of Dr. I. Pinkas (WIS) for help designing the experimental setup, Dr. L. Segev (WIS) for developing the Raman software, and Dr. S. Aharon (Princeton) for help in the synthesis of \CPB\ crystals.
OY acknowledges funding from the European Research Council starting grant (850041 - ANHARMONIC). We acknowledge support from the MIT-Israel Zuckerman STEM Fund and the Sagol Weizmann-MIT Bridge Program. We acknowledge support from the United States-Israel Binational Science Foundation, grant no. 2020270. We acknowledge support from the National Science Foundation (NSF) under grant no. 1751736. We acknowledge support from the Skolkovo Institute of Science and Technology and the MIT-Skoltech Next Generation Program. KY acknowledges support from the NSF Graduate Research Fellowship, grant no. 1745302. B.Z. and J.R. acknowledge support from an ARO MURI with award number W911NF-21-1-0327, an NSF grant with award number DMR-2122071, and an ONR grant with award number N00014-23-1-2818.
F.K. acknowledges support from the Swedish Research Council (VR) program 2020-04630, and the Swedish e-Science Research Centre (SeRC).
The computations were enabled by resources provided by the National Academic Infrastructure for Supercomputing in Sweden (NAISS) at NSC and PDC partially funded by the Swedish Research Council through grant agreement no. 2022-06725.

\white{\section*{References Main}}

\input{arXiv_V3_MainBBL.bbl}

\clearpage
\onecolumngrid

\renewcommand{\thepage}{S\arabic{page}}  
\renewcommand{\thesection}{S\arabic{section}}   
\renewcommand{\thesubsection}{S\arabic{section}.\alph{subsection}} 
\renewcommand{\thetable}{S\arabic{table}}   
\renewcommand{\thefigure}{S\arabic{figure}}
\renewcommand{\theequation}{S\arabic{equation}}
\setcounter{page}{1}
\setcounter{figure}{0}
\setcounter{equation}{0}
\setcounter{section}{0}

\titleformat{\section}
  {\normalfont\bfseries}{\thesection.}{1em}{}
\titleformat{\subsection}
  {\normalfont\bfseries}{\thesubsection.}{1em}{}

\linespread{1.6} 
\fontsize{12pt}{14pt}
\setlength{\parindent}{20pt} 
\setlength{\parskip}{10pt} 
\setlength{\baselineskip}{24pt}

\centering{{\Huge Supplemental Material}} \\ 
\vspace{20pt}
\centering{\textbf{Vibrational properties differ between halide and chalcogenide perovskite semiconductors, and it matters for optoelectronic performance}}

\begin{flushleft}

\section{\label{secSM_PL}Photoluminescence Data and Analysis}

\subsection{PL Data Processing}
Figure~\ref{fig:SM_PLRaw} shows the as-measured photoluminescence (PL) spectra of \BZS\ (left) and \CPB\ (right) in the temperature range of 80--300~K.
The measurement parameters are listed in Table~\ref{tab:PL_Measure}.

\begin{figure}[!hbt]
    \centering
    \includegraphics[width = 17 cm]{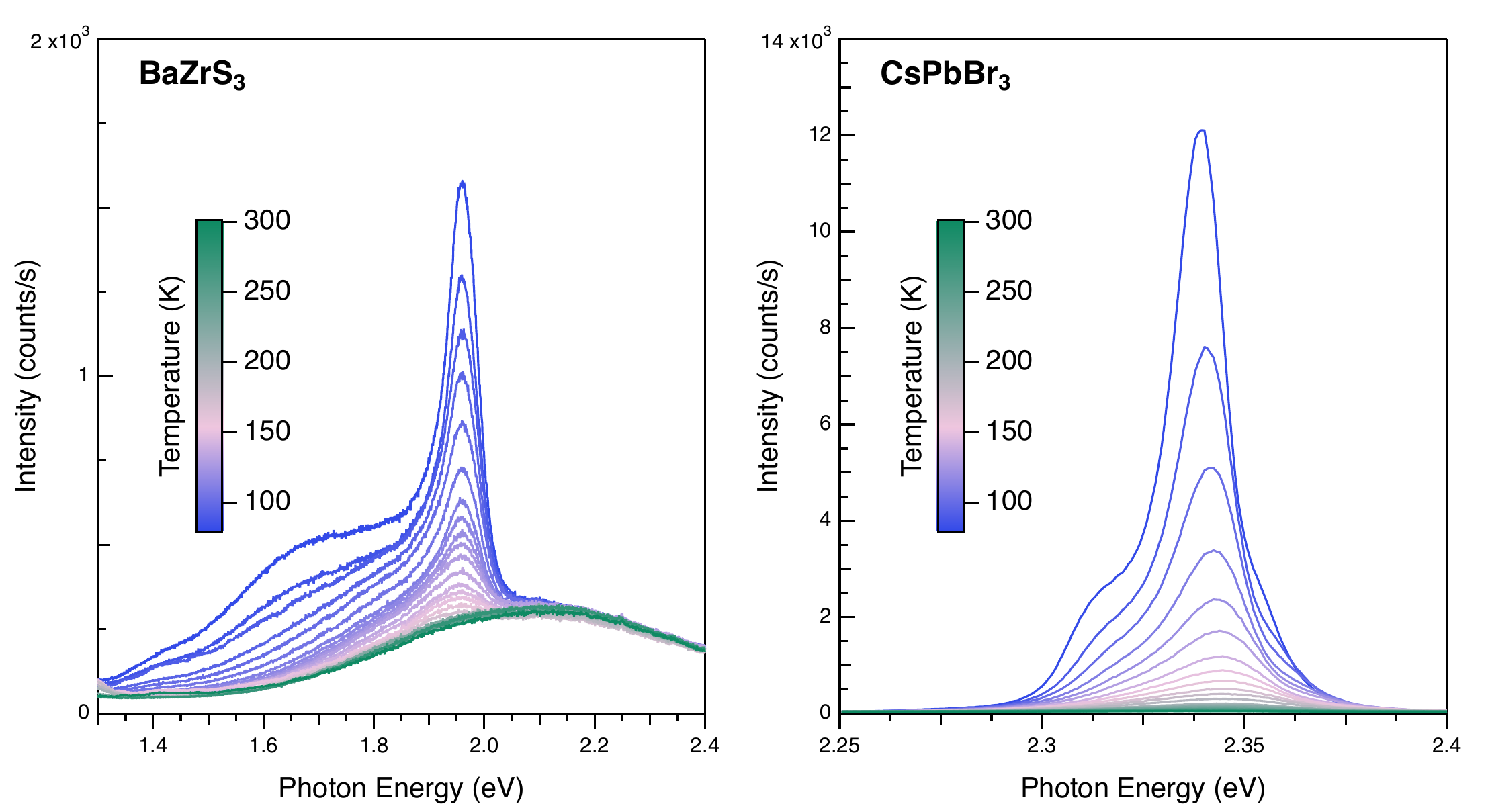}
    \caption{\textbf{Photoluminescence Raw Spectra} of \BZS\ (left) and \CPB\ (right) as measured, from 80--300~K.}
    \label{fig:SM_PLRaw}
\end{figure}

\begin{table}[!hbt]
\caption{Photoluminescence Measurement Parameters}
\centering
\begin{tblr}{c|c|c}
        Parameter    &  \BZS  &  \CPB  \\
        \hline
        Excitation Wavelength [nm] &  488  &  488  \\\hline
        Excitation Energy [eV] &  2.54  &  2.54  \\
        \hline
        Excitation Power [mW] &  2.74  &  1.74$\times \tt{10}^{-4}$  \\
        \hline
        Acquisition Time [s] &  15  &  5  \\
        \hline
        Spot Size [$\mu$m] & 1.1 & 1.1 \\
        \hline
        Excitation Fluence [MJ/cm$^2$] &  4.3  &  2.7~$\times$10$^{-4}$  \\
        \hline
\end{tblr}
\label{tab:PL_Measure}
\end{table}

We first subtract a wavelength-independent constant background from each spectrum from electronic noise to ensure that the tails of the spectra approach zero. In the case of \CPB, this was the only pre-processing step performed before integrating the entire spectra. In the case of \BZS, due to its low PL intensity, we found a small but non-negligible contribution from the excitation laser, reaching a maximum near 2.2~eV. The PL signal from the sample is quenched at $\approx$170~K, leaving only the temperature-independent laser contribution. Therefore, we remove it by subtracting the spectrum measured at high temperature from all lower-temperature data. 

We model the band-to-band PL intensity by fitting four Gaussian peaks to the spectra for \BZS\ and three in the case of \CPB. 
Figure~\ref{fig:SM_PLFit} shows an example of a fit at 80~K.

\begin{figure}[ht!]
    \centering
    \includegraphics[width = 17 cm]{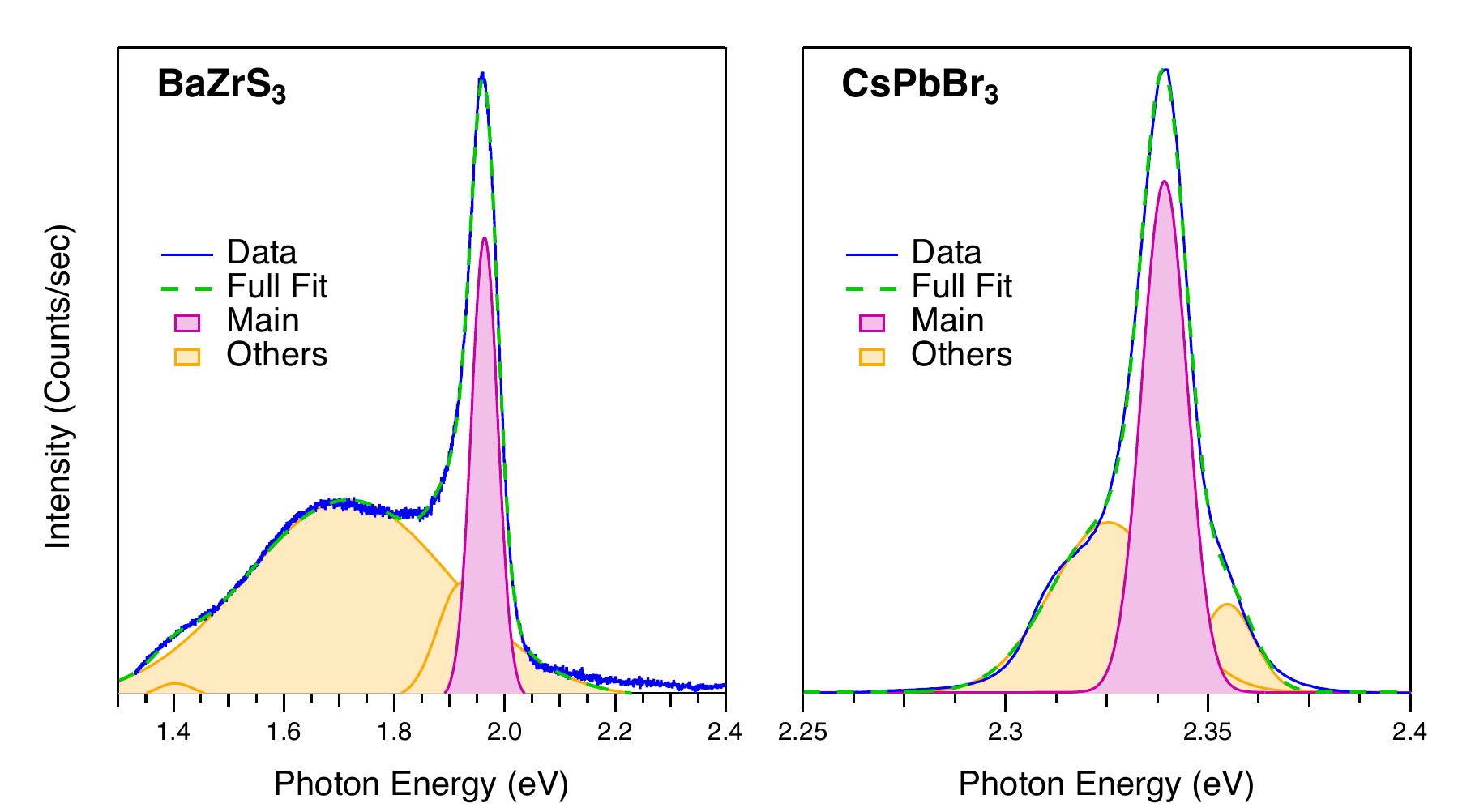}
    \caption{\textbf{Photoluminescence intensity modeling.} Gaussian deconvolution results for the photoluminescence spectra of \BZS\ (left) and \CPB\ (right) at 80~K. The blue and green traces are the data and fit, respectively. The shaded traces are individual peaks -- purple and orange for the main transition and other transitions, respectively.}
    \label{fig:SM_PLFit}
\end{figure}

\subsection{Determination of Normalized PL Yield}

To convert the measured spectra in counts per second to normalized PL yield ($Y$, Eq.~\eqref{eq:yield_raw}~\cite{Main}), we start by considering the incident excitation laser power measured upstream of the microscope ($\Inull$). The cryostat windows attenuate the incident laser light and the sample PL emission, and these effects cancel in the expression for $Y$. We calculate sample reflection loss using the Fresnel equation; with $\sqrt\epsilon = n$, we obtain $n=3.1$ for \BZS\ and $n=2.3$ for \CPB~\cite{bennett_effect_2009,ermolaev_giant_2023}.

We fit the normalized PL yield data to Eq.~\eqref{eq:yield} for fixed values of $\Eb$, as described in the main text~\cite{Main}. In Table~\ref{tab:SM_PL_Fits}, we report the excitation power $\Inull$ and the best-fit values.

\begin{table*}[!hbt]
\caption{Normalized PL Yield Fitted Parameters}
\centering
\small
\begin{tblr}{c|c|c|c|c|c}

         Material &  $\Inull$ (photons/sec) &  $\Ea$ (meV)  &  $\Eb/\kB$ (K) & A & B  \\
    \hline
        \SetCell[r=4]{m} \BZS\ &  6.7$\times \tt{10}^{15}$ & 68 $\pm$ 7  & 0 &  (6.9 $\pm$ 3.8) $\times \tt{10}^{17}$ &  (2.8 $\pm$ 0.5) $\times \tt{10}^{14}$\\
    \cline{2-6}
         &  6.7$\times \tt{10}^{15}$  &  68 $\pm$ 7  & 2 & (6.9 $\pm$ 3.8) $\times \tt{10}^{17}$ &  (2.9 $\pm$ 0.5) $\times \tt{10}^{14}$\\
    \cline{2-6}
        &  6.7$\times \tt{10}^{15}$  &  68 $\pm$ 7  & 4 &  (7.0 $\pm$ 3.9) $\times \tt{10}^{17}$ &  (3.0 $\pm$ 0.5) $\times \tt{10}^{14}$\\
    \cline{2-6}
        &  6.7$\times \tt{10}^{15}$  &  68 $\pm$ 7  & 8 &  (7.2 $\pm$ 4.0) $\times \tt{10}^{17}$ &  (3.1 $\pm$ 0.5) $\times \tt{10}^{14}$\\
    \hline
        \SetCell[r=4]{m} \CPB\ &  4.3$\times \tt{10}^{11}$  &  86 $\pm$ 5 & 0 & (2.8 $\pm$ 0.8) $\times \tt{10}^{13}$ &  (9.6 $\pm$ 1.4) $\times \tt{10}^{9}$ \\
    \cline{2-6}
         &  4.3$\times \tt{10}^{11}$  &  86 $\pm$ 5 & 2 & (2.8 $\pm$ 0.8) $\times \tt{10}^{13}$ &  (9.9 $\pm$ 1.5) $\times \tt{10}^{9}$ \\
    \cline{2-6}
         &  4.3$\times \tt{10}^{11}$  &  86 $\pm$ 5 & 4 & (2.9 $\pm$ 0.8) $\times \tt{10}^{13}$ &  (1.0 $\pm$ 0.2)$\times \tt{10}^{10}$ \\
    \cline{2-6}
        &  4.3$\times \tt{10}^{11}$  &  86 $\pm$ 5 & 8 & (2.9 $\pm$ 0.8) $\times \tt{10}^{13}$ &  (1.1 $\pm$ 0.2) $\times \tt{10}^{10}$ \\
    \hline
\end{tblr}
\label{tab:SM_PL_Fits}
\end{table*}

\subsection{Determination of PL Linewidth}

We require two considerably overlapping Gaussian peaks to model the band-to-band PL lineshape of both materials.
Therefore, we numerically calculate the linewidth (full-width-half-maximum, $\Gpl$) of these combined peaks. We fit the results using Eq.~\eqref{eq:FWHM}~\cite{Main}, and the best-fit parameters are presented in Table~\ref{tab:PL_FWHM_Fits}. 

\begin{table}[!hbt]
\caption{PL Linewidth Fitted Parameters}
\centering
\begin{tblr}{c|c|c|c}

       Material  &  $E_{ph}$ (meV) &  $\gamma$ (meV)  &   $\Gnull$ (meV)  \\
    \hline
        \BZS\ &  24 $\pm$ 7 & 205 $\pm$ 116  &  57 $\pm$ 4 \\
    \hline
       \CPB\ &  18 $\pm$ 4  &  48 $\pm$ 12 &  9 $\pm$ 2 \\
    \hline
\end{tblr}
\label{tab:PL_FWHM_Fits}
\end{table}

\vspace{2 cm}
\clearpage

\section{\label{SecSM_TDEP}Phonon density of states and LO character}

\subsection{Self-consistent phonon scheme \label{sm.sec:stdep}}

We use the \emph{stochastic temperature-dependent effective potential} (sTDEP) scheme to compute temperature-dependent phonons described by the force constants $\{ \Phi_{ij} (T) \}$ which depend on temperature $T$. This scheme is formally equivalent to the QSCAILD method described in Ref.~\cite{Roekeghem.2021} and relies on minimizing the mean-square error of the harmonic model forces,
\begin{align}
  S [\Phi, T]
    &\equiv 
      \sum_i \left\langle ( f_i - f_{\Phi; i} )^2 \right\rangle_{\Phi, T}~,
  \label{sm.eq:S1}
\end{align}\
where $\langle \cdot \rangle_{\Phi, T}$ denotes a thermal average with respect to the force constants $\Phi$ at temperature $T$, $i,j $ denote atomic and Cartesian indices, and the harmonic force $f_{\Phi; i}$ is given by $f_{\Phi; i} = - \sum_j \Phi_{ij} u_j$, where $u_j$ is a vector with atomic displacements from the thermal average positions.

As shown and discussed in Ref.~\cite{Roekeghem.2021, Bianco.2017}, the least-squares solution to Eq.~\eqref{sm.eq:S1} is given by
\begin{align}
  \Phi_{ij} (T)
    = - \left\langle u_i u_j \right\rangle_{ \Phi, T}^{-1} \left\langle f_i u_j \right\rangle_{ \Phi, T}^{\phantom{-1}}
    \equiv
      \left\langle \frac{\partial^2 V}{\partial u_i \partial u_j} \right\rangle_{\Phi, T}~,
\end{align}
\textit{i.e.}, the configuration average of the second derivative of the potential energy $V$ with respect to the effective harmonic distribution of atomic positions at target temperature $T$. This solution is equivalent to a Gibbs-Bogoliubov free energy minimization and yields the best trial free energy of harmonic form. Further details are given in \cite{Benshalom2022, Knoop.2024, Shulumba.2017}.

After obtaining the temperature-dependent force constants at room temperature, we can use them to compute the dynamical matrix
\begin{align}
  D_{ij} ({\bf q})
    = \frac{1}{\sqrt{m_i m_j}} \sum_L {\rm e}^{{\rm i} {\bf q} \cdot {\bf R}_{L}}
      \Phi_{i0, jL} ~,
  \label{sm.eq:dynamical.matrix}
\end{align}
via mass weighting with the atomic masses $m_i$ and Fourier-transforming to q-space by summing over lattice vectors ${\bf R}_L$, from which temperature-renormalized dynamical properties can be inferred. In particular, we define the eigenvalue equation diagonalizing the atomic basis $i,j$ to mode indices $s$,
\begin{align}
  \sum_j D_{ij} ({\bf q}) \epsilon_{j {\bf q} s}
    = \omega^2_{{\bf q} s} \epsilon_{i {\bf q} s}^{\phantom{I}}~,
\end{align}
with eigenfrequencies $\omega_{{\bf q} s}$ and eigenvectors $\epsilon_{i {\bf q} s}$.

Without going into details, we note that these frequencies do \emph{not} correspond to the quasiparticle poles of phonon excitations. For physical spectral properties, phonon-phonon interactions need to be taken into account, which can be done in the TDEP framework by obtaining third-order force constants \cite{Hellman.2013oi5}. We denote the fully thermally shifted quasiparticle frequencies as $\tilde \omega_{{\bf q} s}$ in the following.

\subsection{Density of states}

From the phonon quasiparticle spectrum $\{ \tilde \omega_{{\bf q} s} \}$ at room temperature, we can compute the \emph{density of states} (DOS)
\begin{align}
  {\rm DOS} (\omega) 
    = \sum_{{\bf q} s} \delta (\omega - \tilde \omega_{{\bf q} s})~.
\end{align}

\subsection{LO character}

Longitudinal optical (LO) modes are characterized by a longitudinal charge,
\begin{align}
  Z^{\rm LO}_{{\bf k} s} 
    = \frac{{\bf k}}{k} \cdot {\bf Z}_{{\bf k} s}  ~,
\end{align}
with
\begin{align}
  {\bf Z}_{{\bf k} s} 
    = \sum_{i}  Z_i \epsilon_{i {\bf k} s} / \sqrt{m_i}~,
\end{align}
where $Z_i$ is the Born effective charge tensor for atom $i$, and ${\bf k} \to 0$ denotes a small but finite wave vector characteristic for the probing laser in an optical experiment.

The longitudinal charge creates a macroscopic polarization along $\bf k$ when mode $s$ is excited

\begin{align}
  {\bf p}^{\rm LO}_{{\bf k} s} = Z^{\rm LO}_{{\bf k} s} A_{{\bf k} s}^{\phantom{I}} \frac{{\bf k}}{k}~,
\end{align}
where $A_{{\bf k} s}$ is the mode amplitude. Through this polarization, a mode $s$ can couple to electronic excitations via Fröhlich coupling. Since the mode amplitude goes with
\begin{align}
  \langle A_{{\bf k} s}^2 \rangle \propto 1/\omega_{{\bf k} s}^2~,
\end{align}
we take
\begin{align}
  Z^{\rm LO}_{{\bf k} s} / \omega_{{\bf k} s}
\end{align}
as an estimate for the ``longitudinal mode coupling strength'' or simply ``LO character''. Since this coupling strength is direction-dependent via the incident wave vector $\bf k$, we take $\bf k$ to point in $[101]$ direction in line with the experimental setup. However, we have verified that the qualitative picture does not change if $\bf k$ is chosen along another direction, \textit{e.g.}, one of the Cartesian axes.

\subsection{Computational details}

For both systems, we use a unit cell with 20 atoms, space group 62 (\emph{Pnma}) (Materials Project codes mp-540771 for \BZS\ and mp-567681 for \CPB~\cite{Jain.2013}). We use the temperature-dependent effective potential method (TDEP) with the stochastic sampling scheme described above (sTDEP). Details can be found in \cite{Knoop.2024} and references therein.
For force sampling, 3x2x3 supercells (360 atoms) and the all-electron DFT code FHI-aims were used, with \emph{light default} basis sets~\cite{Blum.2009}, 2x2x2 k-points, and am05 \emph{xc} functional~\cite{Armiento.2005, Mattsson.2008}.
To compute the Born effective charge tensor, Quantum Espresso v7.1 was used~\cite{Giannozzi.2009, Giannozzi.2017, Giannozzi.2020}, with 75~Ry energy cutoff, 4x3x4 k-points, PBE functional~\cite{Perdew.1996}, and Schlipf-Gygi norm-conserving pseudopotentials created with the ONCVPSP code~\cite{Hamann.2013, Schlipf.2015ake}.

\clearpage

\section{\label{SecSM_RamanData}Raman Data and Analysis}

\subsection{\BZS\
Raman Data Over a Wide Temperature Range}

In Fig.~\ref{fig:SM_BZSRaman}, we present the Raman spectra of \BZS\ from 10--875~K, where we see signs of sample degradation in the optical microscope.
Each spectrum was normalized between zero and one by reducing a constant background and dividing by the maximum intensity.
\begin{figure}[b!]
    \centering
    \includegraphics[width = 8.5 cm]{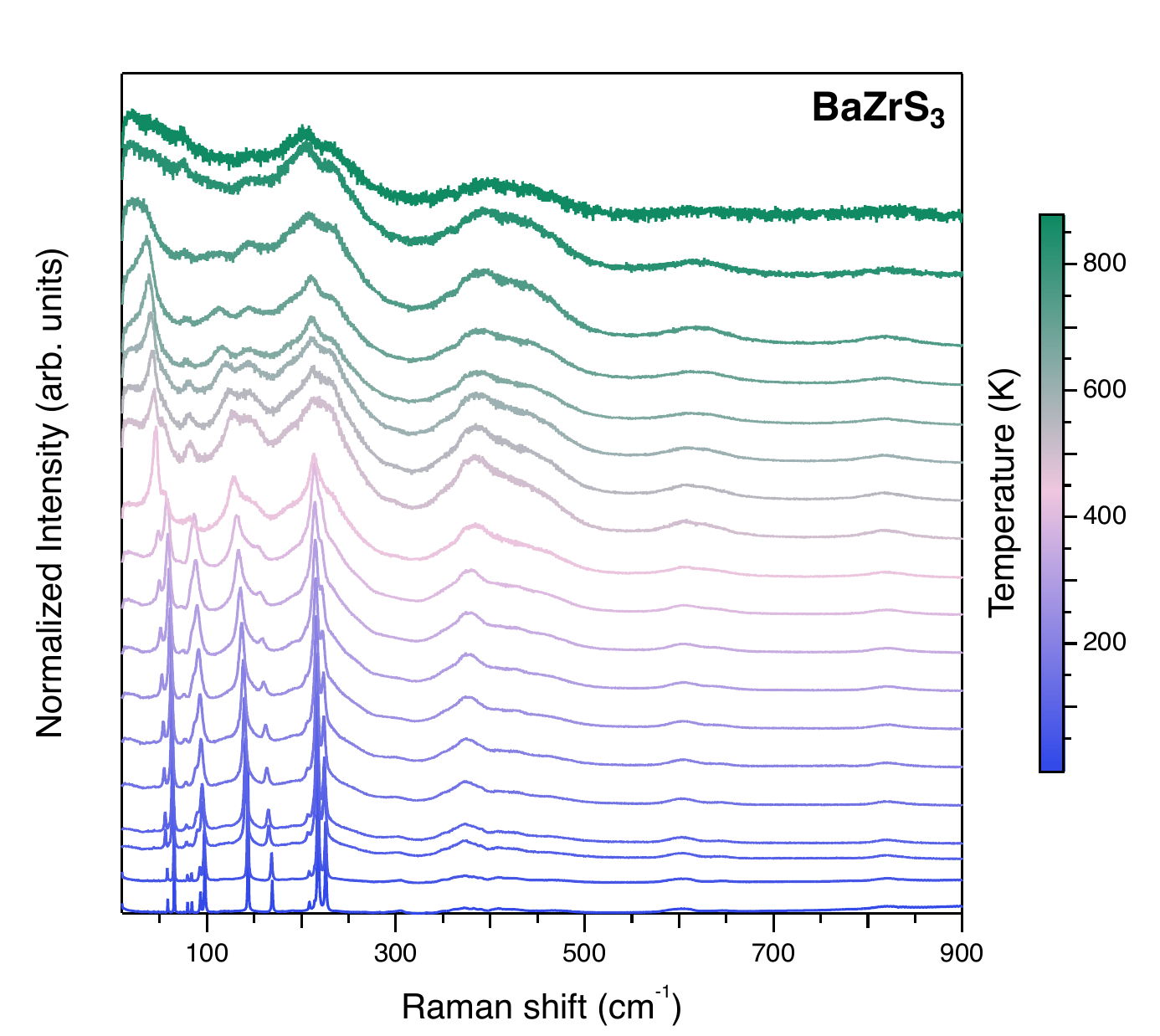}
    \caption{\textbf{Raman Spectra of \BZS\ crystal} in the range of 10--875~K. Spectra were normalized and shifted vertically for clarity.}
    \label{fig:SM_BZSRaman}
\end{figure}
The Raman spectra below 80~K were measured in the same optical system, using liquid-He as the cryogenic coolant.

At low temperatures, the Raman spectra are characterized by sharp isolated peaks assigned to first-order scattering from single normal-mode vibrations (one photon scattering by one vibration).
As often happens, the first-order peaks slightly red shift and broaden with increasing temperature~\cite{Menahem2023, Asher2023}.
The broad, higher-frequency ($\omega \geq 300$~\wn) features that emerge with increasing temperature are assigned to second-order scattering due to their absence at low temperatures, their rapid increase with temperature~\cite{Cardona1982}, and from the calculated phonon density of states. A broad background forms below the low-frequency peaks; this we assign to disorder-enhanced high-order scattering of unknown origin~\cite{Menahem2023}. We can distinguish first-order scattering peaks from the broad features at higher temperatures only by following the peaks at increasing temperatures from cryogenic conditions. Further assignment using integrated intensity temperature dependence is presented below.

The temperature-dependent Raman data show no evidence of a phase transformation; \BZS\ remains in the orthorhombic phase throughout~\cite{Gross2017}. 
This is consistent with reported calorimetric measurements~\cite{Niu2018} and x-ray studies~\cite{Comparotto2020, Xu2022, Yu2021} showing no phase transformation up to \app\ 1300~K, at which point degradation begins. 

\subsection{Deconvolution}

To assign and analyze the temperature-dependent trends of the Raman peaks, each spectrum was deconvolved into a product of the \BE\ distribution ($n_{\omega, T}$) and multiple pseudo-Voigt oscillators:
\begin{equation}
\begin{split}
    \mathcal{S}(\omega) = & \left[ n_{\omega,T} + 1 \right] \times \\
    & \sum_{j} c_{j} \left[ (1-\zeta_j) \frac{ \omega | \Omega _{j} | \Gr^{2} }{\omega ^{2} \Gr^{2} +(\omega^{2} -\Omega _{j}^{2})^{2}} + \zeta_j \exp{\left( \frac{ -(\omega - \Omega_{j})^2 \ln{16}}{\Gr^2} \right) } \right]~,
\end{split}
\label{eq:pvoigt}
\end{equation}
$\mathcal{S}(\omega)$ and $\omega$ are the Raman intensity and Raman shift, and $\Omega_j$, $\Gr$ and $c_j$ are the $j$'th mode's vibrational frequency, FWHM, and intensity, respectively.
The \BE\ distribution is given by $n_{\omega, T} = (e^{\hbar \omega / \kB T} -1)^{-1}$, where $\hbar$, $\kB$ and $T$ are the reduced Planck's constant, the Boltzmann constant and temperature. The parameter $0 \leq \zeta_j \leq 1$ is the Gaussian fraction of the pseudo-Voigt line shape, used to capture broadening unrelated to vibrational lifetime, such as higher-order scattering. Overall, $\zeta_j \approx 0$ was used for the sharp first-order peaks.

Figure~\ref{fig:SM_RamanFit} shows the deconvolution results for the 80~K spectra of \BZS\ (left) and \CPB\ (right). The blue and green traces are the measured and fitted spectra, while the red traces are individual peaks. The intensity of a Raman peak depends on temperature through the \BE\ distribution according to the scattering order~\cite{Cardona1982}.
As the temperature is lowered, first-order peaks are more prominent than higher orders. Therefore, we identify the first-order scattering peaks by their presence in the spectra at 10~K (black traces), which contain primarily first-order peaks. The inset shows the deconvolution of the high-frequency modes of \BZS, which we assign as second-order scattering. This assignment is corroborated by the calculated vibrational density of states, which shows a high density of optical modes up to 250~\wn\ and a few optical modes at 300~\wn\ and 400~\wn, with a large gap between them. Assuming low Raman activity of the 300~\wn\ and 400~\wn\ branches, our assignment nicely matches the calculated vibrational density of states. 

In the case of \CPB, the sharp peaks at 80~K broaden significantly at higher temperatures, making it possible to determine their scattering order only at low temperatures (see Fig.~\ref{Fig:Raman_Norm}~\cite{Main}).
\begin{figure}[t!]
    \centering
    \includegraphics[width = 17 cm]{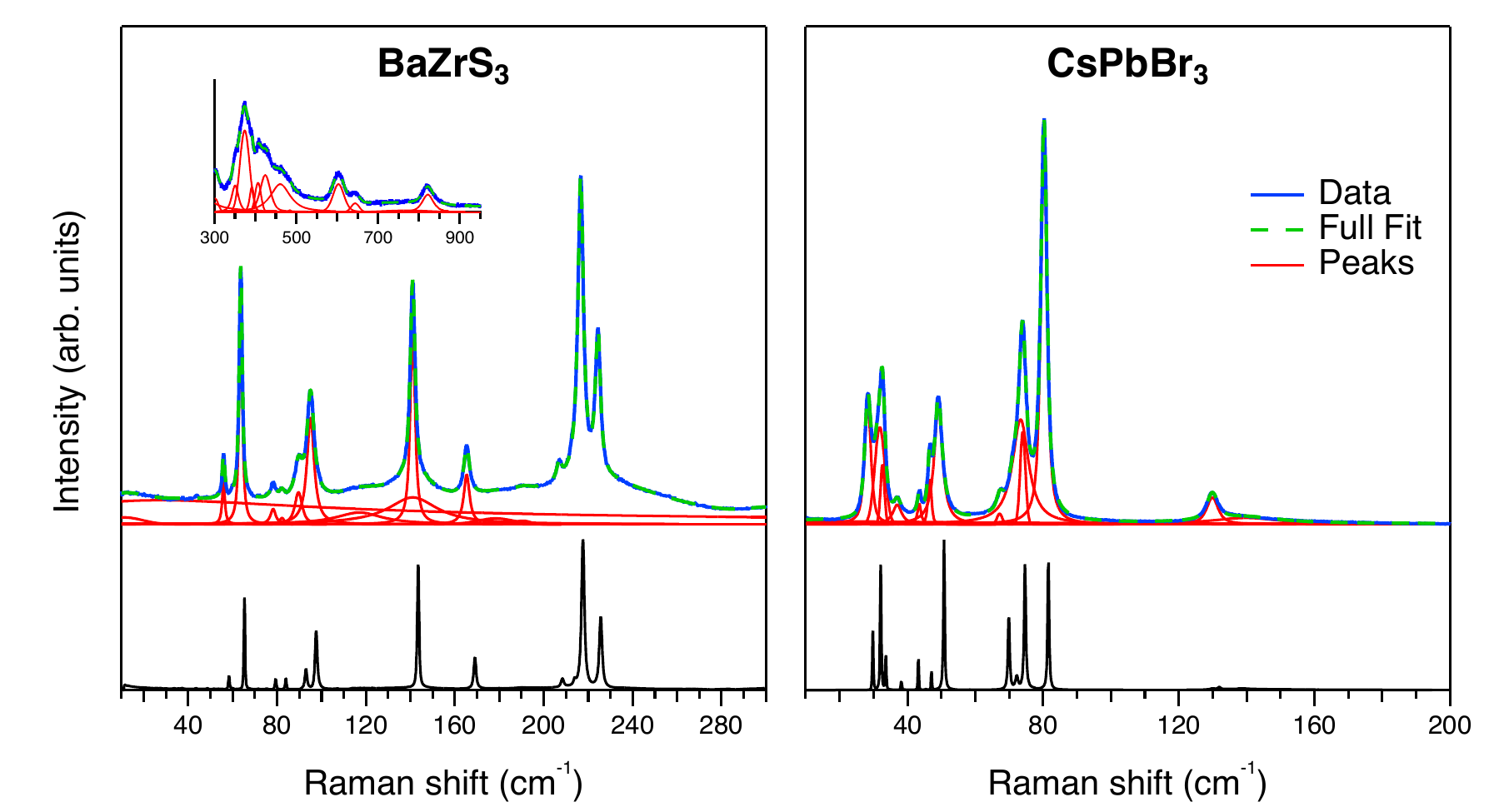}
    \caption{\textbf{Raman deconvolution.} pseudo-Voigt deconvolution results for the Raman spectra of \BZS\ (left) and \CPB\ (right) at 80~K.
    The blue and green traces are the data and fit, respectively. The red traces are individual peaks.
    Black traces are the spectra at 10~K that were used to identify the first-order peaks at higher temperatures.
    Inset - deconvolution of the high-frequency modes of \BZS.}
    \label{fig:SM_RamanFit}
\end{figure}

\subsection{Assignment of Scattering Order}

To substantiate our assignment of the modes of \BZS, we compare the temperature dependence of the integrated intensity to the expected dependence of first- and second-order scattering~\cite{Menahem2023}.
The intensity of first-order Raman scattering is proportional to the product of the \BE\ distribution and the mode spectral function ($\mathcal{J}_j$)~\cite{Kwok1968}.
Therefore, for Stokes scattering, it follows that~\cite{Maradudin1971, Safran1977} 
\begin{equation}
    \mathcal{S}_j(\omega) = \Xi \int_0^\infty e^{-i\omega t} \avg{U^*_j , U_j}~\tt{d}t = \Xi ~ [n_{\omega,T}+1] ~ \mathcal{J}_j(\Omega_j,\Gr)~,
\label{eq:S1}
\end{equation}
where $\Xi$ is the proportionality factor that contains the Raman tensor and properties of the material affecting the scattering, and $\mathcal{J}_j$ is defined as the temperature-independent Fourier transform of the pair displacement correlation function ($\avg{U^*_j, U_j}$) such that
\begin{equation}
    \int_0^\infty \mathcal{J}_j~\tt{d}\omega = 1~.
\label{eq:J1}
\end{equation}
Therefore, we expect the integrated intensity of any pseudo-Voigt peak originating from first-order scattering to remain constant at all temperatures.
The intensity of second-order Raman scattering (one photon scattered by two vibrations of opposite momentum) is more complicated to handle because, at each frequency, there are contributions from many combinations of modes.
Therefore, we can not describe the expected dependence of a specific mode and must calculate the dependence of an extensive spectral range.
To do that, we model second-order scattering as the convolution of mode spectra functions~\cite{Menahem2023},
\begin{equation}
\begin{split}
    \mathcal{S}(\omega) &= ~\Xi \int_0^\infty e^{-i\omega t} \sum_{j,j'} \avg{U^*_j, U^*_{j'}, U_j, U_{j'}}~\tt{d}t \\
    &= ~ \Xi ~ \times \sum_{j,j'} [n_{\omega,T}+1]\mathcal{J}_j(\Omega_j,\Gr) * [n_{\omega,T}+1]\mathcal{J}_{j'}(\Omega_{j'},\Grp)~,
\end{split}
\label{eq:S2}
\end{equation}
following the same notation and conditions from Eq.~\eqref{eq:S1} and Eq.~\eqref{eq:J1}.
The set of {$\mathcal{J}_{j}$} that reproduces the experimental line shape is not unique; However, if we find one solution that can reproduce the line shape and integrated intensity, we prove that second-order scattering dominates in the spectral range in question. 
To that end, we estimate that combination modes ($j \neq j'$) have negligible contribution compared to overtones ($j=j'$), and simplify Eq.~\eqref{eq:S2} to a single sum,
\begin{equation}
    \mathcal{S}(\omega) = \Xi ~ \times \sum_{j} [n_{\omega,T}+1]\mathcal{J}_j(\Omega_j,\Gr) * [n_{\omega,T}+1]\mathcal{J}_{j}(\Omega_{j},\Gr)~.
\end{equation}
\begin{figure}[b!]
    \centering
    \includegraphics[width = 17 cm]{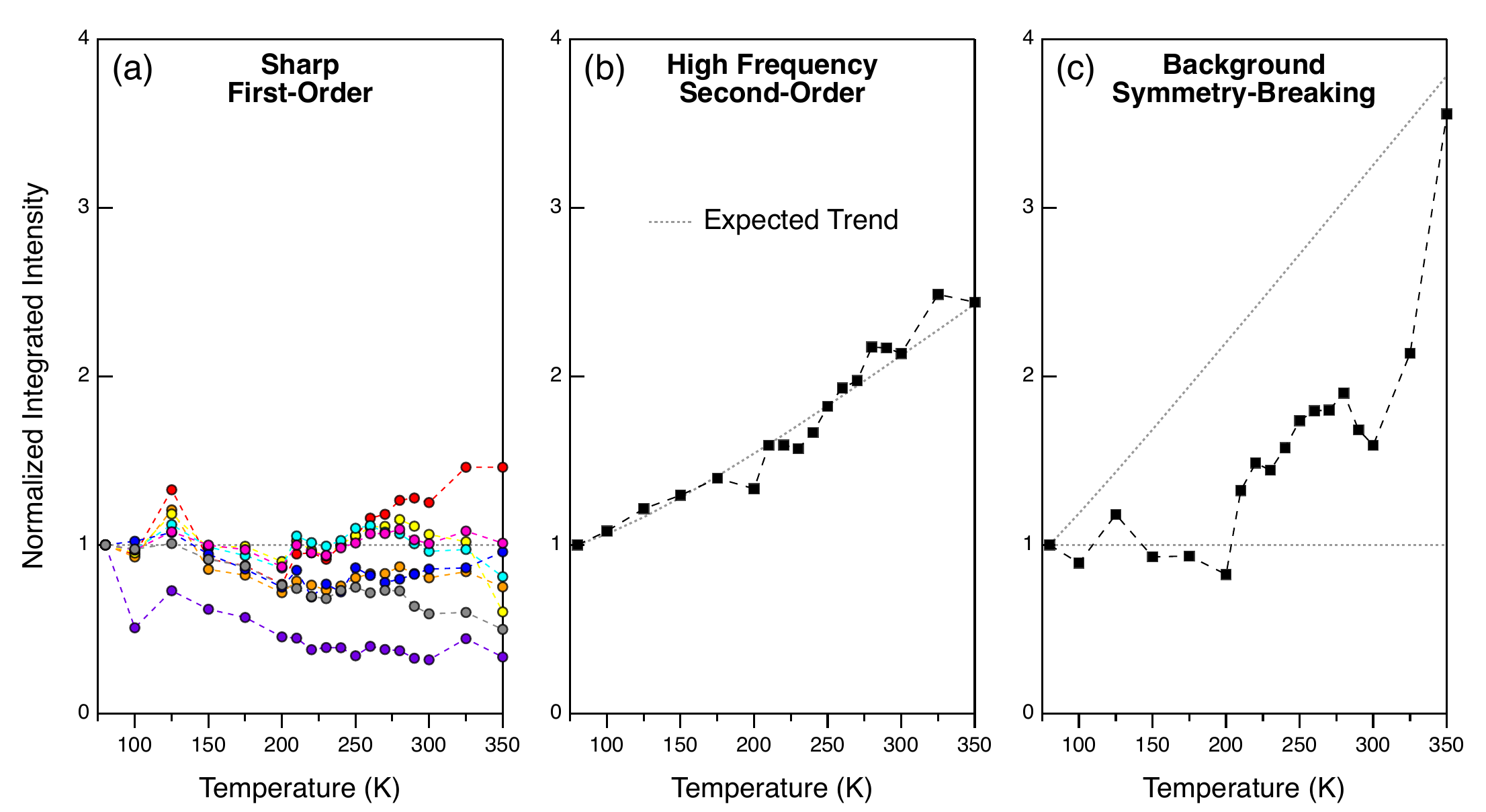}
    \caption{\textbf{pseuo-Voigt relative integrated intensities compared to expected trends.} 
    (a) low frequency ($\omega < 300$~\wn) sharp peak compared to expectation for first-order scattering.
    (b) - high-frequency ($\omega \geq 300$~\wn) peaks compared to expectation for second-order scattering. 
    (c) - low-frequency broad features ($\Gr > 10$~\wn) compared to expectation for first- and second-order scattering.
    Colored markers represent the different fitted peaks with a dashed line to guide the eye, and the dotted gray lines are expectation curves.}
    \label{fig:SM_RamanII}
\end{figure}

The simplified version allows us to solve each $\mathcal{J}_{j}$ separately and then sum the contributions. As mentioned, we assign all the high-frequency peaks ($\omega \geq 300$~\wn) to second-order Raman scattering. Therefore, we tried to find a set of {$\mathcal{J}_{j}$} to reproduce the line shape of the high-frequency spectral range and compare the temperature dependence of the integrated intensities, as follows:
\begin{enumerate}
    \item Draw the high-frequency pseudo-Voigt peak using the fitted parameters at $T_0$.
    
    \item Find a $\mathcal{J}_{j}$ such that $\mathcal{S}_{pV}(\omega) = \cfrac{[n_{\omega,T_0}+1]\mathcal{J}_j(\Omega_j,\Gr) * [n_{\omega,T_0}+1]\mathcal{J}_{j}(\Omega_{j},\Gr)}{[n_{\omega,T_0}+1]}$ \\
    and $\int_0^\infty \mathcal{J}_{j}~\tt{d}\omega = 1$.
    
    \item Integrate the fitted pseudo-Voigt peaks and the sum of reproduced line shapes separately.
    
    \item Assuming the set of {$\mathcal{J}_{j}$} is temperature-independent, recalculate the expected line shape at $T \neq T_0$ for the given set through the \BE\ distribution.
    
    \item Compare the temperature-dependent integrated intensity of the fitted peaks and the recalculated line shapes. 
\end{enumerate}

Figure~\ref{fig:SM_RamanII} shows the relative integrated intensities of the low-frequency sharp peaks (a) assigned as first-order scattering, the high-frequency peaks (b) assigned as second-order scattering, and the low-frequency broad background peaks ($\Gr > 10$~\wn) compared to the expected trends for first- and second-order scattering (c). Dashed lines are to guide the eyes, and grey-dotted lines are expectation curves.

\begin{figure}[b!]
    \centering
    \includegraphics[width = 17 cm]{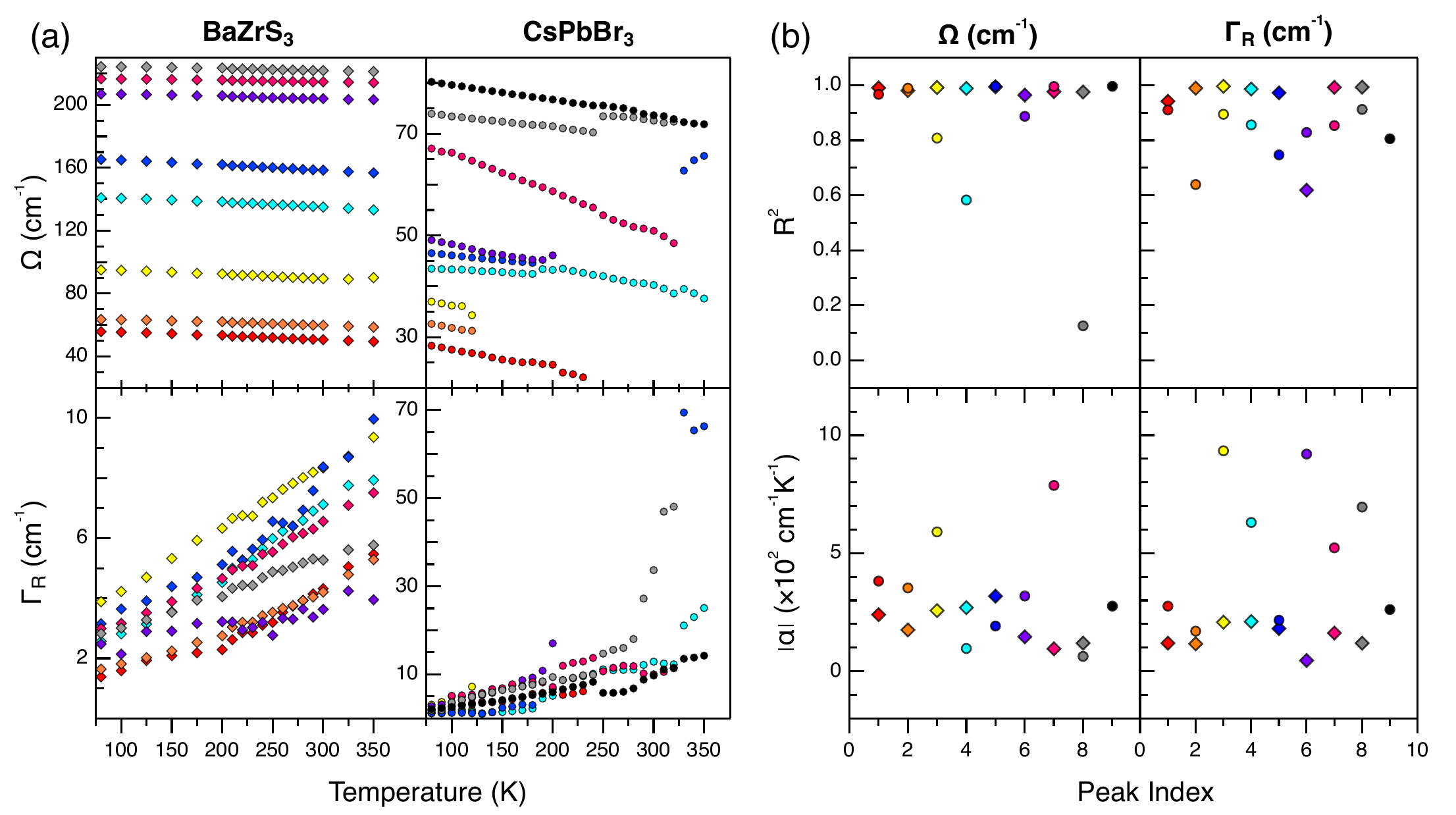}
    \caption{\textbf{Analysis of Raman Spectra:} (a) $\Omega_j$ (top) and $\Gr$ (bottom) of the first-order peaks of \BZS\ (left) and \CPB\ (right) as a function of temperature. 
    (b) Linear regression results (\RS\ - top, and slopes, $alpha$ - bottom) of $\Omega_j(T)$ (left) and $\Gr(T)$ (right) of \BZS\ (diamonds) and \CPB\ (circles), between 80--280~K. A different color represents each peak.}
    \label{fig:SM_RamanTrend}
\end{figure}

The integrated intensity of the low-frequency sharp modes varies by $\approx 50\%$ within the temperature range, which is within our allowed error range, considering the complexity of a real crystal and the experiment. Furthermore, according to our analysis of second-order Raman scattering, there is an excellent correlation between the integrated intensity of the high-frequency peaks and the expectation. The expected curve was drawn using fitting parameters from the 80~K fit. Nevertheless, using the fitting parameters of other temperatures produced similar results (data not shown). The case of the broad low-frequency features presented in the right panel is more complicated - their integrated intensity remains constant for $T \leq 200~K$, then increases dramatically with temperature, almost following a second-order scattering trend. The dramatic change correlates well with a reported increase in dielectric response at kHz frequencies~\cite{Filippone2020}.
Therefore, we faithfully assign the sharp $\omega < 300$~\wn\ modes and the $\omega \geq 300$~\wn\ peaks to first- and second-order scattering, respectively.
Following previous work~\cite{Menahem2023}, we assign these features to high-order scattering due to symmetry-breaking.

\subsection{Linear Trends}

Figure~\ref{fig:SM_RamanTrend}(a) shows the $\Omega_j$ (top) and $\Gr$ (bottom) of the prominent first-order peaks of \BZS\ (left) and \CPB\ (right) as a function of temperature, resulting from the pseudo-Voigt fit. Figure~\ref{fig:SM_RamanTrend}(b) shows the results of the linear regression of $\Omega_j(T)$ (left) and $\Gr(T)$ (right) of \BZS\ (diamonds) and \CPB\ (circles) as a function of the peak index $j$.
The \RS\ value and slope are presented in the top and bottom panels, respectively.

Table~\ref{tab:meanlin} shows the average \RS\ values and slopes of the linear regression of $\Omega_j(T)$ and $\Gr(T)$ of the peaks of \BZS\ and \CPB\ in the range of 80--280~K. The average \RS\ values and their standard deviation (STD) indicate that the frequencies and linewidths of \BZS\ are more linear than those of \CPB, showing average values closer to 1 and lower STD.
Moreover, the average slope values indicate that the peaks of \CPB\ shift and broaden more than the peaks of \BZS.
The drastic change in the structural dynamics and the anharmonicity of \CPB\ compared to \BZS\ are indicated by the larger STDs.

\vspace{2.5cm}

\begin{table}[h!]
    \centering
    \caption{Mean values of the linear regression presented in Fig.~\ref{fig:SM_RamanTrend}}
     \label{tab:meanlin}

   \centering
    \begin{tblr}{c|c|c|c}
        Material  &  Property  &  Parameter  &  Mean $\pm$ STD \\
        \hline
        \SetCell[r=4]{m} \BZS\  &  \SetCell[r=2]{m}$\Omega_j (T)$  &  \RS  &  $0.98 \pm 0.01$ \\
        \cline{3-4}
                                &  &  Slope [cm K]$^{-1}$  &  $0.020 \pm 0.008$  \\
        \cline{2-4}
                    &  \SetCell[r=2]{m}$\Gr (T)$  &  \RS  &  $0.9 \pm 0.1$ \\
        \cline{3-4}
                                &  &  Slope [cm K]$^{-1}$  &  $0.015 \pm 0.006$  \\
        \hline
        \SetCell[r=4]{m} \CPB\  &  \SetCell[r=2]{m}$\Omega_j (T)$  &  \RS  &  $0.8 \pm 0.3$ \\
        \cline{3-4}
                                &   &  Slope [cm K]$^{-1}$  &  $0.03 \pm 0.02$  \\
        \cline{2-4}
                    &  \SetCell[r=2]{m}$\Gr (T)$  &  \RS  &  $0.83 \pm 0.09$ \\
        \cline{3-4}
                                &  &  Slope [cm K]$^{-1}$  &  $0.05 \pm 0.03$  \\
        \hline    
    \end{tblr}
\end{table}

\newpage

\end{flushleft}

\twocolumngrid
\clearpage

\white{\section*{References SM}}

\linespread{1}

\input{arXiv_V3_SMBBL.bbl}
\end{document}

%% file: arXiv_V3_MainBBL.bbl
%

%% file: arXiv_V3.bbl
\begin{thebibliography}{112}%
\makeatletter
\providecommand \@ifxundefined [1]{%
 \@ifx{#1\undefined}
}%
\providecommand \@ifnum [1]{%
 \ifnum #1\expandafter \@firstoftwo
 \else \expandafter \@secondoftwo
 \fi
}%
\providecommand \@ifx [1]{%
 \ifx #1\expandafter \@firstoftwo
 \else \expandafter \@secondoftwo
 \fi
}%
\providecommand \natexlab [1]{#1}%
\providecommand \enquote  [1]{``#1''}%
\providecommand \bibnamefont  [1]{#1}%
\providecommand \bibfnamefont [1]{#1}%
\providecommand \citenamefont [1]{#1}%
\providecommand \href@noop [0]{\@secondoftwo}%
\providecommand \href [0]{\begingroup \@sanitize@url \@href}%
\providecommand \@href[1]{\@@startlink{#1}\@@href}%
\providecommand \@@href[1]{\endgroup#1\@@endlink}%
\providecommand \@sanitize@url [0]{\catcode `\\12\catcode `\$12\catcode
  `\&12\catcode `\#12\catcode `\^12\catcode `\_12\catcode `\%12\relax}%
\providecommand \@@startlink[1]{}%
\providecommand \@@endlink[0]{}%
\providecommand \url  [0]{\begingroup\@sanitize@url \@url }%
\providecommand \@url [1]{\endgroup\@href {#1}{\urlprefix }}%
\providecommand \urlprefix  [0]{URL }%
\providecommand \Eprint [0]{\href }%
\providecommand \doibase [0]{https://doi.org/}%
\providecommand \selectlanguage [0]{\@gobble}%
\providecommand \bibinfo  [0]{\@secondoftwo}%
\providecommand \bibfield  [0]{\@secondoftwo}%
\providecommand \translation [1]{[#1]}%
\providecommand \BibitemOpen [0]{}%
\providecommand \bibitemStop [0]{}%
\providecommand \bibitemNoStop [0]{.\EOS\space}%
\providecommand \EOS [0]{\spacefactor3000\relax}%
\providecommand \BibitemShut  [1]{\csname bibitem#1\endcsname}%
\let\auto@bib@innerbib\@empty
\bibitem [{\citenamefont {Brenner}\ \emph {et~al.}(2016)\citenamefont
  {Brenner}, \citenamefont {Egger}, \citenamefont {Kronik}, \citenamefont
  {Hodes},\ and\ \citenamefont {Cahen}}]{Brenner2016}%
  \BibitemOpen
  \bibfield  {author} {\bibinfo {author} {\bibfnamefont {T.~M.}\ \bibnamefont
  {Brenner}}, \bibinfo {author} {\bibfnamefont {D.~A.}\ \bibnamefont {Egger}},
  \bibinfo {author} {\bibfnamefont {L.}~\bibnamefont {Kronik}}, \bibinfo
  {author} {\bibfnamefont {G.}~\bibnamefont {Hodes}},\ and\ \bibinfo {author}
  {\bibfnamefont {D.}~\bibnamefont {Cahen}},\ }\bibfield  {title} {\bibinfo
  {title} {{Hybrid organic—inorganic perovskites: low-cost semiconductors
  with intriguing charge-transport properties}},\ }\href
  {https://doi.org/10.1038/natrevmats.2016.11} {\bibfield  {journal} {\bibinfo
  {journal} {Nature Reviews Materials}\ ,\ \bibinfo {pages} {16011}} (\bibinfo
  {year} {2016})}\BibitemShut {NoStop}%
\bibitem [{\citenamefont {Gr{\"{a}}tzel}(2014)}]{Gratzel2014}%
  \BibitemOpen
  \bibfield  {author} {\bibinfo {author} {\bibfnamefont {M.}~\bibnamefont
  {Gr{\"{a}}tzel}},\ }\bibfield  {title} {\bibinfo {title} {{The light and
  shade of perovskite solar cells}},\ }\href {https://doi.org/10.1038/nmat4065}
  {\bibfield  {journal} {\bibinfo  {journal} {Nature Materials}\ }\textbf
  {\bibinfo {volume} {13}},\ \bibinfo {pages} {838} (\bibinfo {year}
  {2014})}\BibitemShut {NoStop}%
\bibitem [{\citenamefont {Stranks}\ and\ \citenamefont
  {Snaith}(2015)}]{Stranks2015}%
  \BibitemOpen
  \bibfield  {author} {\bibinfo {author} {\bibfnamefont {S.~D.}\ \bibnamefont
  {Stranks}}\ and\ \bibinfo {author} {\bibfnamefont {H.~J.}\ \bibnamefont
  {Snaith}},\ }\bibfield  {title} {\bibinfo {title} {{Metal-halide perovskites
  for photovoltaic and light-emitting devices}},\ }\href
  {https://doi.org/10.1038/nnano.2015.90} {\bibfield  {journal} {\bibinfo
  {journal} {Nature Nanotechnology}\ }\textbf {\bibinfo {volume} {10}},\
  \bibinfo {pages} {391} (\bibinfo {year} {2015})}\BibitemShut {NoStop}%
\bibitem [{\citenamefont {He}\ \emph {et~al.}(2016)\citenamefont {He},
  \citenamefont {Yu}, \citenamefont {Li}, \citenamefont {Li}, \citenamefont
  {Si}, \citenamefont {Jin}, \citenamefont {Wang}, \citenamefont {Wang},
  \citenamefont {He}, \citenamefont {Wang}, \citenamefont {Zhang},\ and\
  \citenamefont {Ye}}]{He2016}%
  \BibitemOpen
  \bibfield  {author} {\bibinfo {author} {\bibfnamefont {H.}~\bibnamefont
  {He}}, \bibinfo {author} {\bibfnamefont {Q.}~\bibnamefont {Yu}}, \bibinfo
  {author} {\bibfnamefont {H.}~\bibnamefont {Li}}, \bibinfo {author}
  {\bibfnamefont {J.}~\bibnamefont {Li}}, \bibinfo {author} {\bibfnamefont
  {J.}~\bibnamefont {Si}}, \bibinfo {author} {\bibfnamefont {Y.}~\bibnamefont
  {Jin}}, \bibinfo {author} {\bibfnamefont {N.}~\bibnamefont {Wang}}, \bibinfo
  {author} {\bibfnamefont {J.}~\bibnamefont {Wang}}, \bibinfo {author}
  {\bibfnamefont {J.}~\bibnamefont {He}}, \bibinfo {author} {\bibfnamefont
  {X.}~\bibnamefont {Wang}}, \bibinfo {author} {\bibfnamefont {Y.}~\bibnamefont
  {Zhang}},\ and\ \bibinfo {author} {\bibfnamefont {Z.}~\bibnamefont {Ye}},\
  }\bibfield  {title} {\bibinfo {title} {{Exciton localization in
  solution-processed organolead trihalide perovskites}},\ }\href
  {https://doi.org/10.1038/ncomms10896} {\bibfield  {journal} {\bibinfo
  {journal} {Nature Communications}\ }\textbf {\bibinfo {volume} {7}},\
  \bibinfo {pages} {1} (\bibinfo {year} {2016})}\BibitemShut {NoStop}%
\bibitem [{\citenamefont {Shi}\ \emph {et~al.}(2015)\citenamefont {Shi},
  \citenamefont {Adinolfi}, \citenamefont {Comin}, \citenamefont {Yuan},
  \citenamefont {Alarousu}, \citenamefont {Buin}, \citenamefont {Chen},
  \citenamefont {Hoogland}, \citenamefont {Rothenberger}, \citenamefont
  {Katsiev}, \citenamefont {Losovyj}, \citenamefont {Zhang}, \citenamefont
  {Dowben}, \citenamefont {Mohammed}, \citenamefont {Sargent},\ and\
  \citenamefont {Bakr}}]{Shi2015}%
  \BibitemOpen
  \bibfield  {author} {\bibinfo {author} {\bibfnamefont {D.}~\bibnamefont
  {Shi}}, \bibinfo {author} {\bibfnamefont {V.}~\bibnamefont {Adinolfi}},
  \bibinfo {author} {\bibfnamefont {R.}~\bibnamefont {Comin}}, \bibinfo
  {author} {\bibfnamefont {M.}~\bibnamefont {Yuan}}, \bibinfo {author}
  {\bibfnamefont {E.}~\bibnamefont {Alarousu}}, \bibinfo {author}
  {\bibfnamefont {A.}~\bibnamefont {Buin}}, \bibinfo {author} {\bibfnamefont
  {Y.}~\bibnamefont {Chen}}, \bibinfo {author} {\bibfnamefont {S.}~\bibnamefont
  {Hoogland}}, \bibinfo {author} {\bibfnamefont {A.}~\bibnamefont
  {Rothenberger}}, \bibinfo {author} {\bibfnamefont {K.}~\bibnamefont
  {Katsiev}}, \bibinfo {author} {\bibfnamefont {Y.}~\bibnamefont {Losovyj}},
  \bibinfo {author} {\bibfnamefont {X.}~\bibnamefont {Zhang}}, \bibinfo
  {author} {\bibfnamefont {P.~A.}\ \bibnamefont {Dowben}}, \bibinfo {author}
  {\bibfnamefont {O.~F.}\ \bibnamefont {Mohammed}}, \bibinfo {author}
  {\bibfnamefont {E.~H.}\ \bibnamefont {Sargent}},\ and\ \bibinfo {author}
  {\bibfnamefont {O.~M.}\ \bibnamefont {Bakr}},\ }\bibfield  {title} {\bibinfo
  {title} {{Low trap-state density and long carrier diffusion in organolead
  trihalide perovskite single crystals}},\ }\href
  {https://doi.org/10.1126/science.aaa2725} {\bibfield  {journal} {\bibinfo
  {journal} {Science}\ }\textbf {\bibinfo {volume} {347}},\ \bibinfo {pages}
  {519} (\bibinfo {year} {2015})}\BibitemShut {NoStop}%
\bibitem [{\citenamefont {Stranks}\ \emph {et~al.}(2014)\citenamefont
  {Stranks}, \citenamefont {Burlakov}, \citenamefont {Leijtens}, \citenamefont
  {Ball}, \citenamefont {Goriely},\ and\ \citenamefont {Snaith}}]{Stranks2014}%
  \BibitemOpen
  \bibfield  {author} {\bibinfo {author} {\bibfnamefont {S.~D.}\ \bibnamefont
  {Stranks}}, \bibinfo {author} {\bibfnamefont {V.~M.}\ \bibnamefont
  {Burlakov}}, \bibinfo {author} {\bibfnamefont {T.}~\bibnamefont {Leijtens}},
  \bibinfo {author} {\bibfnamefont {J.~M.}\ \bibnamefont {Ball}}, \bibinfo
  {author} {\bibfnamefont {A.}~\bibnamefont {Goriely}},\ and\ \bibinfo {author}
  {\bibfnamefont {H.~J.}\ \bibnamefont {Snaith}},\ }\bibfield  {title}
  {\bibinfo {title} {{Recombination Kinetics in Organic-Inorganic Perovskites:
  Excitons, Free Charge, and Subgap States}},\ }\href
  {https://doi.org/10.1103/PhysRevApplied.2.034007} {\bibfield  {journal}
  {\bibinfo  {journal} {Physical Review Applied}\ }\textbf {\bibinfo {volume}
  {2}},\ \bibinfo {pages} {1} (\bibinfo {year} {2014})}\BibitemShut {NoStop}%
\bibitem [{\citenamefont {Yamada}\ \emph {et~al.}(2015)\citenamefont {Yamada},
  \citenamefont {Yamada}, \citenamefont {Phuong}, \citenamefont {Maruyama},
  \citenamefont {Nishimura}, \citenamefont {Wakamiya}, \citenamefont {Murata},\
  and\ \citenamefont {Kanemitsu}}]{Yamada2015}%
  \BibitemOpen
  \bibfield  {author} {\bibinfo {author} {\bibfnamefont {Y.}~\bibnamefont
  {Yamada}}, \bibinfo {author} {\bibfnamefont {T.}~\bibnamefont {Yamada}},
  \bibinfo {author} {\bibfnamefont {L.~Q.}\ \bibnamefont {Phuong}}, \bibinfo
  {author} {\bibfnamefont {N.}~\bibnamefont {Maruyama}}, \bibinfo {author}
  {\bibfnamefont {H.}~\bibnamefont {Nishimura}}, \bibinfo {author}
  {\bibfnamefont {A.}~\bibnamefont {Wakamiya}}, \bibinfo {author}
  {\bibfnamefont {Y.}~\bibnamefont {Murata}},\ and\ \bibinfo {author}
  {\bibfnamefont {Y.}~\bibnamefont {Kanemitsu}},\ }\bibfield  {title} {\bibinfo
  {title} {{Dynamic Optical Properties of CH 3 NH 3 PbI 3 Single Crystals As
  Revealed by One- and Two-Photon Excited Photoluminescence Measurements}},\
  }\href {https://doi.org/10.1021/jacs.5b04503} {\bibfield  {journal} {\bibinfo
   {journal} {Journal of the American Chemical Society}\ }\textbf {\bibinfo
  {volume} {137}},\ \bibinfo {pages} {10456} (\bibinfo {year}
  {2015})}\BibitemShut {NoStop}%
\bibitem [{\citenamefont {Wehrenfennig}\ \emph {et~al.}(2014)\citenamefont
  {Wehrenfennig}, \citenamefont {Eperon}, \citenamefont {Johnston},
  \citenamefont {Snaith},\ and\ \citenamefont {Herz}}]{Wehrenfennig2014}%
  \BibitemOpen
  \bibfield  {author} {\bibinfo {author} {\bibfnamefont {C.}~\bibnamefont
  {Wehrenfennig}}, \bibinfo {author} {\bibfnamefont {G.~E.}\ \bibnamefont
  {Eperon}}, \bibinfo {author} {\bibfnamefont {M.~B.}\ \bibnamefont
  {Johnston}}, \bibinfo {author} {\bibfnamefont {H.~J.}\ \bibnamefont
  {Snaith}},\ and\ \bibinfo {author} {\bibfnamefont {L.~M.}\ \bibnamefont
  {Herz}},\ }\bibfield  {title} {\bibinfo {title} {{High Charge Carrier
  Mobilities and Lifetimes in Organolead Trihalide Perovskites}},\ }\href
  {https://doi.org/10.1002/adma.201305172} {\bibfield  {journal} {\bibinfo
  {journal} {Adv. Mater.}\ }\textbf {\bibinfo {volume} {26}},\ \bibinfo {pages}
  {1584} (\bibinfo {year} {2014})}\BibitemShut {NoStop}%
\bibitem [{\citenamefont {Niu}\ \emph {et~al.}(2017)\citenamefont {Niu},
  \citenamefont {Huyan}, \citenamefont {Liu}, \citenamefont {Yeung},
  \citenamefont {Ye}, \citenamefont {Blankemeier}, \citenamefont {Orvis},
  \citenamefont {Sarkar}, \citenamefont {Singh}, \citenamefont {Kapadia},\ and\
  \citenamefont {Ravichandran}}]{Niu2017}%
  \BibitemOpen
  \bibfield  {author} {\bibinfo {author} {\bibfnamefont {S.}~\bibnamefont
  {Niu}}, \bibinfo {author} {\bibfnamefont {H.}~\bibnamefont {Huyan}}, \bibinfo
  {author} {\bibfnamefont {Y.}~\bibnamefont {Liu}}, \bibinfo {author}
  {\bibfnamefont {M.}~\bibnamefont {Yeung}}, \bibinfo {author} {\bibfnamefont
  {K.}~\bibnamefont {Ye}}, \bibinfo {author} {\bibfnamefont {L.}~\bibnamefont
  {Blankemeier}}, \bibinfo {author} {\bibfnamefont {T.}~\bibnamefont {Orvis}},
  \bibinfo {author} {\bibfnamefont {D.}~\bibnamefont {Sarkar}}, \bibinfo
  {author} {\bibfnamefont {D.~J.}\ \bibnamefont {Singh}}, \bibinfo {author}
  {\bibfnamefont {R.}~\bibnamefont {Kapadia}},\ and\ \bibinfo {author}
  {\bibfnamefont {J.}~\bibnamefont {Ravichandran}},\ }\bibfield  {title}
  {\bibinfo {title} {{Bandgap Control via Structural and Chemical Tuning of
  Transition Metal Perovskite Chalcogenides}},\ }\href
  {https://doi.org/10.1002/adma.201604733} {\bibfield  {journal} {\bibinfo
  {journal} {Advanced Materials}\ }\textbf {\bibinfo {volume} {29}},\ \bibinfo
  {pages} {16} (\bibinfo {year} {2017})}\BibitemShut {NoStop}%
\bibitem [{\citenamefont {Jaramillo}\ and\ \citenamefont
  {Ravichandran}(2019)}]{jaramillo_praise_2019}%
  \BibitemOpen
  \bibfield  {author} {\bibinfo {author} {\bibfnamefont {R.}~\bibnamefont
  {Jaramillo}}\ and\ \bibinfo {author} {\bibfnamefont {J.}~\bibnamefont
  {Ravichandran}},\ }\bibfield  {title} {\bibinfo {title} {In praise and in
  search of highly-polarizable semiconductors: {Technological} promise and
  discovery strategies},\ }\href {https://doi.org/10.1063/1.5124795} {\bibfield
   {journal} {\bibinfo  {journal} {APL Materials}\ }\textbf {\bibinfo {volume}
  {7}},\ \bibinfo {pages} {100902} (\bibinfo {year} {2019})}\BibitemShut
  {NoStop}%
\bibitem [{\citenamefont {Sun}\ \emph {et~al.}(2015)\citenamefont {Sun},
  \citenamefont {Agiorgousis}, \citenamefont {Zhang},\ and\ \citenamefont
  {Zhang}}]{sun_chalcogenide_2015}%
  \BibitemOpen
  \bibfield  {author} {\bibinfo {author} {\bibfnamefont {Y.-Y.}\ \bibnamefont
  {Sun}}, \bibinfo {author} {\bibfnamefont {M.~L.}\ \bibnamefont
  {Agiorgousis}}, \bibinfo {author} {\bibfnamefont {P.}~\bibnamefont {Zhang}},\
  and\ \bibinfo {author} {\bibfnamefont {S.}~\bibnamefont {Zhang}},\ }\bibfield
   {title} {\bibinfo {title} {Chalcogenide {Perovskites} for {Photovoltaics}},\
  }\href {https://doi.org/10.1021/nl504046x} {\bibfield  {journal} {\bibinfo
  {journal} {Nano Letters}\ }\textbf {\bibinfo {volume} {15}},\ \bibinfo
  {pages} {581} (\bibinfo {year} {2015})}\BibitemShut {NoStop}%
\bibitem [{\citenamefont {Filippone}\ \emph {et~al.}(2020)\citenamefont
  {Filippone}, \citenamefont {Zhao}, \citenamefont {Niu}, \citenamefont
  {Koocher}, \citenamefont {Silevitch}, \citenamefont {Fina}, \citenamefont
  {Rondinelli}, \citenamefont {Ravichandran},\ and\ \citenamefont
  {Jaramillo}}]{Filippone2020}%
  \BibitemOpen
  \bibfield  {author} {\bibinfo {author} {\bibfnamefont {S.}~\bibnamefont
  {Filippone}}, \bibinfo {author} {\bibfnamefont {B.}~\bibnamefont {Zhao}},
  \bibinfo {author} {\bibfnamefont {S.}~\bibnamefont {Niu}}, \bibinfo {author}
  {\bibfnamefont {N.~Z.}\ \bibnamefont {Koocher}}, \bibinfo {author}
  {\bibfnamefont {D.}~\bibnamefont {Silevitch}}, \bibinfo {author}
  {\bibfnamefont {I.}~\bibnamefont {Fina}}, \bibinfo {author} {\bibfnamefont
  {J.~M.}\ \bibnamefont {Rondinelli}}, \bibinfo {author} {\bibfnamefont
  {J.}~\bibnamefont {Ravichandran}},\ and\ \bibinfo {author} {\bibfnamefont
  {R.}~\bibnamefont {Jaramillo}},\ }\bibfield  {title} {\bibinfo {title}
  {{Discovery of highly polarizable semiconductors BaZrS$_3$ and
  Ba$_3$Zr$_2$S$_7$}},\ }\bibfield  {journal} {\bibinfo  {journal} {Physical
  Review Materials}\ }\textbf {\bibinfo {volume} {4}},\ \href
  {https://doi.org/10.1103/PhysRevMaterials.4.091601}
  {10.1103/PhysRevMaterials.4.091601} (\bibinfo {year} {2020})\BibitemShut
  {NoStop}%
\bibitem [{\citenamefont {Ye}\ \emph {et~al.}(2022{\natexlab{a}})\citenamefont
  {Ye}, \citenamefont {Koocher}, \citenamefont {Filippone}, \citenamefont
  {Niu}, \citenamefont {Zhao}, \citenamefont {Yeung}, \citenamefont {Bone},
  \citenamefont {Robinson}, \citenamefont {Vora}, \citenamefont {Schleife},
  \citenamefont {Ju}, \citenamefont {Boubnov}, \citenamefont {Rondinelli},
  \citenamefont {Ravichandran},\ and\ \citenamefont
  {Jaramillo}}]{ye_low-energy_2022}%
  \BibitemOpen
  \bibfield  {author} {\bibinfo {author} {\bibfnamefont {K.}~\bibnamefont
  {Ye}}, \bibinfo {author} {\bibfnamefont {N.~Z.}\ \bibnamefont {Koocher}},
  \bibinfo {author} {\bibfnamefont {S.}~\bibnamefont {Filippone}}, \bibinfo
  {author} {\bibfnamefont {S.}~\bibnamefont {Niu}}, \bibinfo {author}
  {\bibfnamefont {B.}~\bibnamefont {Zhao}}, \bibinfo {author} {\bibfnamefont
  {M.}~\bibnamefont {Yeung}}, \bibinfo {author} {\bibfnamefont
  {S.}~\bibnamefont {Bone}}, \bibinfo {author} {\bibfnamefont {A.~J.}\
  \bibnamefont {Robinson}}, \bibinfo {author} {\bibfnamefont {P.}~\bibnamefont
  {Vora}}, \bibinfo {author} {\bibfnamefont {A.}~\bibnamefont {Schleife}},
  \bibinfo {author} {\bibfnamefont {L.}~\bibnamefont {Ju}}, \bibinfo {author}
  {\bibfnamefont {A.}~\bibnamefont {Boubnov}}, \bibinfo {author} {\bibfnamefont
  {J.~M.}\ \bibnamefont {Rondinelli}}, \bibinfo {author} {\bibfnamefont
  {J.}~\bibnamefont {Ravichandran}},\ and\ \bibinfo {author} {\bibfnamefont
  {R.}~\bibnamefont {Jaramillo}},\ }\bibfield  {title} {\bibinfo {title}
  {Low-energy electronic structure of perovskite and {Ruddlesden}-{Popper}
  semiconductors in the {Ba}-{Zr}-{S} system probed by bond-selective polarized
  x-ray absorption spectroscopy, infrared reflectivity, and {Raman}
  scattering},\ }\href {https://doi.org/10.1103/PhysRevB.105.195203} {\bibfield
   {journal} {\bibinfo  {journal} {Physical Review B}\ }\textbf {\bibinfo
  {volume} {105}},\ \bibinfo {pages} {195203} (\bibinfo {year}
  {2022}{\natexlab{a}})}\BibitemShut {NoStop}%
\bibitem [{\citenamefont {Niu}\ \emph {et~al.}(2018)\citenamefont {Niu},
  \citenamefont {Milam-Guerrero}, \citenamefont {Zhou}, \citenamefont {Ye},
  \citenamefont {Zhao}, \citenamefont {Melot},\ and\ \citenamefont
  {Ravichandran}}]{Niu2018}%
  \BibitemOpen
  \bibfield  {author} {\bibinfo {author} {\bibfnamefont {S.}~\bibnamefont
  {Niu}}, \bibinfo {author} {\bibfnamefont {J.}~\bibnamefont {Milam-Guerrero}},
  \bibinfo {author} {\bibfnamefont {Y.}~\bibnamefont {Zhou}}, \bibinfo {author}
  {\bibfnamefont {K.}~\bibnamefont {Ye}}, \bibinfo {author} {\bibfnamefont
  {B.}~\bibnamefont {Zhao}}, \bibinfo {author} {\bibfnamefont {B.~C.}\
  \bibnamefont {Melot}},\ and\ \bibinfo {author} {\bibfnamefont
  {J.}~\bibnamefont {Ravichandran}},\ }\bibfield  {title} {\bibinfo {title}
  {{Thermal stability study of transition metal perovskite sulfides}},\ }\href
  {https://doi.org/10.1557/jmr.2018.419} {\bibfield  {journal} {\bibinfo
  {journal} {Journal of Materials Research}\ }\textbf {\bibinfo {volume}
  {33}},\ \bibinfo {pages} {4135} (\bibinfo {year} {2018})}\BibitemShut
  {NoStop}%
\bibitem [{\citenamefont {Sadeghi}\ \emph {et~al.}(2021)\citenamefont
  {Sadeghi}, \citenamefont {Ye}, \citenamefont {Xu}, \citenamefont {Li},
  \citenamefont {LeBeau},\ and\ \citenamefont
  {Jaramillo}}]{sadeghi_making_2021}%
  \BibitemOpen
  \bibfield  {author} {\bibinfo {author} {\bibfnamefont {I.}~\bibnamefont
  {Sadeghi}}, \bibinfo {author} {\bibfnamefont {K.}~\bibnamefont {Ye}},
  \bibinfo {author} {\bibfnamefont {M.}~\bibnamefont {Xu}}, \bibinfo {author}
  {\bibfnamefont {Y.}~\bibnamefont {Li}}, \bibinfo {author} {\bibfnamefont
  {J.~M.}\ \bibnamefont {LeBeau}},\ and\ \bibinfo {author} {\bibfnamefont
  {R.}~\bibnamefont {Jaramillo}},\ }\bibfield  {title} {\bibinfo {title}
  {Making {BaZrS} $_{\textrm{3}}$ {Chalcogenide} {Perovskite} {Thin} {Films} by
  {Molecular} {Beam} {Epitaxy}},\ }\href
  {https://doi.org/10.1002/adfm.202105563} {\bibfield  {journal} {\bibinfo
  {journal} {Advanced Functional Materials}\ }\textbf {\bibinfo {volume}
  {31}},\ \bibinfo {pages} {2105563} (\bibinfo {year} {2021})}\BibitemShut
  {NoStop}%
\bibitem [{\citenamefont {Wei}\ \emph {et~al.}(2020)\citenamefont {Wei},
  \citenamefont {Hui}, \citenamefont {Zhao}, \citenamefont {Deng},
  \citenamefont {Han}, \citenamefont {Yu}, \citenamefont {Sheng}, \citenamefont
  {Roy}, \citenamefont {Chen}, \citenamefont {Lin}, \citenamefont {Watson},
  \citenamefont {Sun}, \citenamefont {Thomay}, \citenamefont {Yang},
  \citenamefont {Jia}, \citenamefont {Zhang},\ and\ \citenamefont
  {Zeng}}]{wei_realization_2020}%
  \BibitemOpen
  \bibfield  {author} {\bibinfo {author} {\bibfnamefont {X.}~\bibnamefont
  {Wei}}, \bibinfo {author} {\bibfnamefont {H.}~\bibnamefont {Hui}}, \bibinfo
  {author} {\bibfnamefont {C.}~\bibnamefont {Zhao}}, \bibinfo {author}
  {\bibfnamefont {C.}~\bibnamefont {Deng}}, \bibinfo {author} {\bibfnamefont
  {M.}~\bibnamefont {Han}}, \bibinfo {author} {\bibfnamefont {Z.}~\bibnamefont
  {Yu}}, \bibinfo {author} {\bibfnamefont {A.}~\bibnamefont {Sheng}}, \bibinfo
  {author} {\bibfnamefont {P.}~\bibnamefont {Roy}}, \bibinfo {author}
  {\bibfnamefont {A.}~\bibnamefont {Chen}}, \bibinfo {author} {\bibfnamefont
  {J.}~\bibnamefont {Lin}}, \bibinfo {author} {\bibfnamefont {D.~F.}\
  \bibnamefont {Watson}}, \bibinfo {author} {\bibfnamefont {Y.-Y.}\
  \bibnamefont {Sun}}, \bibinfo {author} {\bibfnamefont {T.}~\bibnamefont
  {Thomay}}, \bibinfo {author} {\bibfnamefont {S.}~\bibnamefont {Yang}},
  \bibinfo {author} {\bibfnamefont {Q.}~\bibnamefont {Jia}}, \bibinfo {author}
  {\bibfnamefont {S.}~\bibnamefont {Zhang}},\ and\ \bibinfo {author}
  {\bibfnamefont {H.}~\bibnamefont {Zeng}},\ }\bibfield  {title} {\bibinfo
  {title} {Realization of {Bazrs3} chalcogenide perovskite thin films for
  optoelectronics},\ }\href {https://doi.org/10.1016/j.nanoen.2019.104317}
  {\bibfield  {journal} {\bibinfo  {journal} {Nano Energy}\ }\textbf {\bibinfo
  {volume} {68}},\ \bibinfo {pages} {104317} (\bibinfo {year}
  {2020})}\BibitemShut {NoStop}%
\bibitem [{\citenamefont {Surendran}\ \emph {et~al.}(2021)\citenamefont
  {Surendran}, \citenamefont {Chen}, \citenamefont {Zhao}, \citenamefont
  {Thind}, \citenamefont {Singh}, \citenamefont {Orvis}, \citenamefont {Zhao},
  \citenamefont {Han}, \citenamefont {Htoon}, \citenamefont {Kawasaki},
  \citenamefont {Mishra},\ and\ \citenamefont
  {Ravichandran}}]{surendran_epitaxial_2021}%
  \BibitemOpen
  \bibfield  {author} {\bibinfo {author} {\bibfnamefont {M.}~\bibnamefont
  {Surendran}}, \bibinfo {author} {\bibfnamefont {H.}~\bibnamefont {Chen}},
  \bibinfo {author} {\bibfnamefont {B.}~\bibnamefont {Zhao}}, \bibinfo {author}
  {\bibfnamefont {A.~S.}\ \bibnamefont {Thind}}, \bibinfo {author}
  {\bibfnamefont {S.}~\bibnamefont {Singh}}, \bibinfo {author} {\bibfnamefont
  {T.}~\bibnamefont {Orvis}}, \bibinfo {author} {\bibfnamefont
  {H.}~\bibnamefont {Zhao}}, \bibinfo {author} {\bibfnamefont {J.-K.}\
  \bibnamefont {Han}}, \bibinfo {author} {\bibfnamefont {H.}~\bibnamefont
  {Htoon}}, \bibinfo {author} {\bibfnamefont {M.}~\bibnamefont {Kawasaki}},
  \bibinfo {author} {\bibfnamefont {R.}~\bibnamefont {Mishra}},\ and\ \bibinfo
  {author} {\bibfnamefont {J.}~\bibnamefont {Ravichandran}},\ }\bibfield
  {title} {\bibinfo {title} {Epitaxial {Thin} {Films} of a {Chalcogenide}
  {Perovskite}},\ }\href {https://doi.org/10.1021/acs.chemmater.1c02202}
  {\bibfield  {journal} {\bibinfo  {journal} {Chemistry of Materials}\ }\textbf
  {\bibinfo {volume} {33}},\ \bibinfo {pages} {7457} (\bibinfo {year}
  {2021})}\BibitemShut {NoStop}%
\bibitem [{\citenamefont {Comparotto}\ \emph {et~al.}(2020)\citenamefont
  {Comparotto}, \citenamefont {Davydova}, \citenamefont {Ericson},
  \citenamefont {Riekehr}, \citenamefont {Moro}, \citenamefont {Kubart},\ and\
  \citenamefont {Scragg}}]{Comparotto2020}%
  \BibitemOpen
  \bibfield  {author} {\bibinfo {author} {\bibfnamefont {C.}~\bibnamefont
  {Comparotto}}, \bibinfo {author} {\bibfnamefont {A.}~\bibnamefont
  {Davydova}}, \bibinfo {author} {\bibfnamefont {T.}~\bibnamefont {Ericson}},
  \bibinfo {author} {\bibfnamefont {L.}~\bibnamefont {Riekehr}}, \bibinfo
  {author} {\bibfnamefont {M.~V.}\ \bibnamefont {Moro}}, \bibinfo {author}
  {\bibfnamefont {T.}~\bibnamefont {Kubart}},\ and\ \bibinfo {author}
  {\bibfnamefont {J.}~\bibnamefont {Scragg}},\ }\bibfield  {title} {\bibinfo
  {title} {{Chalcogenide Perovskite BaZrS$_3$ : Thin Film Growth by Sputtering
  and Rapid Thermal Processing}},\ }\href
  {https://doi.org/10.1021/acsaem.9b02428} {\bibfield  {journal} {\bibinfo
  {journal} {ACS Applied Energy Materials}\ }\textbf {\bibinfo {volume} {3}},\
  \bibinfo {pages} {2762} (\bibinfo {year} {2020})}\BibitemShut {NoStop}%
\bibitem [{\citenamefont {Ye}\ \emph {et~al.}(2022{\natexlab{b}})\citenamefont
  {Ye}, \citenamefont {Zhao}, \citenamefont {Diroll}, \citenamefont
  {Ravichandran},\ and\ \citenamefont {Jaramillo}}]{ye_time-resolved_2022}%
  \BibitemOpen
  \bibfield  {author} {\bibinfo {author} {\bibfnamefont {K.}~\bibnamefont
  {Ye}}, \bibinfo {author} {\bibfnamefont {B.}~\bibnamefont {Zhao}}, \bibinfo
  {author} {\bibfnamefont {B.~T.}\ \bibnamefont {Diroll}}, \bibinfo {author}
  {\bibfnamefont {J.}~\bibnamefont {Ravichandran}},\ and\ \bibinfo {author}
  {\bibfnamefont {R.}~\bibnamefont {Jaramillo}},\ }\bibfield  {title} {\bibinfo
  {title} {Time-resolved photoluminescence studies of perovskite
  chalcogenides},\ }\href {https://doi.org/10.1039/D2FD00047D} {\bibfield
  {journal} {\bibinfo  {journal} {Faraday Discussions}\ }\textbf {\bibinfo
  {volume} {239}},\ \bibinfo {pages} {146} (\bibinfo {year}
  {2022}{\natexlab{b}})}\BibitemShut {NoStop}%
\bibitem [{\citenamefont {Pradhan}\ \emph {et~al.}(2023)\citenamefont
  {Pradhan}, \citenamefont {Uible}, \citenamefont {Agarwal}, \citenamefont
  {Turnley}, \citenamefont {Khandelwal}, \citenamefont {Peterson},
  \citenamefont {Blach}, \citenamefont {Swope}, \citenamefont {Huang},
  \citenamefont {Bart},\ and\ \citenamefont {Agrawal}}]{pradhan_angew_2023}%
  \BibitemOpen
  \bibfield  {author} {\bibinfo {author} {\bibfnamefont {A.}~\bibnamefont
  {Pradhan}}, \bibinfo {author} {\bibfnamefont {M.}~\bibnamefont {Uible}},
  \bibinfo {author} {\bibfnamefont {S.}~\bibnamefont {Agarwal}}, \bibinfo
  {author} {\bibfnamefont {J.}~\bibnamefont {Turnley}}, \bibinfo {author}
  {\bibfnamefont {S.}~\bibnamefont {Khandelwal}}, \bibinfo {author}
  {\bibfnamefont {J.}~\bibnamefont {Peterson}}, \bibinfo {author}
  {\bibfnamefont {D.}~\bibnamefont {Blach}}, \bibinfo {author} {\bibfnamefont
  {R.}~\bibnamefont {Swope}}, \bibinfo {author} {\bibfnamefont
  {L.}~\bibnamefont {Huang}}, \bibinfo {author} {\bibfnamefont
  {S.}~\bibnamefont {Bart}},\ and\ \bibinfo {author} {\bibfnamefont
  {R.}~\bibnamefont {Agrawal}},\ }\bibfield  {title} {\bibinfo {title}
  {{Synthesis} of {BaZrs3} and {BaHfS3} {Chalcogenide} {Perovskite} {Films}
  {Using} {Single}‐{Phase}},\ }\href {https://doi.org/10.1002/anie.202301049}
  {\bibfield  {journal} {\bibinfo  {journal} {Angewandte Chemie International
  Edition}\ }\textbf {\bibinfo {volume} {62}},\ \bibinfo {pages} {e202301049}
  (\bibinfo {year} {2023})}\BibitemShut {NoStop}%
\bibitem [{\citenamefont {Yang}\ \emph {et~al.}(2022)\citenamefont {Yang},
  \citenamefont {Jess}, \citenamefont {Fai},\ and\ \citenamefont
  {Hages}}]{yang_low-temperature_2022}%
  \BibitemOpen
  \bibfield  {author} {\bibinfo {author} {\bibfnamefont {R.}~\bibnamefont
  {Yang}}, \bibinfo {author} {\bibfnamefont {A.~D.}\ \bibnamefont {Jess}},
  \bibinfo {author} {\bibfnamefont {C.}~\bibnamefont {Fai}},\ and\ \bibinfo
  {author} {\bibfnamefont {C.~J.}\ \bibnamefont {Hages}},\ }\bibfield  {title}
  {\bibinfo {title} {Low-{Temperature}, {Solution}-{Based} {Synthesis} of
  {Luminescent} {Chalcogenide} {Perovskite} {Bazrs} $_{\textrm{3}}$
  {Nanoparticles}},\ }\href {https://doi.org/10.1021/jacs.2c06168} {\bibfield
  {journal} {\bibinfo  {journal} {Journal of the American Chemical Society}\
  }\textbf {\bibinfo {volume} {144}},\ \bibinfo {pages} {15928} (\bibinfo
  {year} {2022})}\BibitemShut {NoStop}%
\bibitem [{\citenamefont {Sharma}\ \emph
  {et~al.}(2020{\natexlab{a}})\citenamefont {Sharma}, \citenamefont {Dai},
  \citenamefont {Gao}, \citenamefont {Brenner}, \citenamefont {Yadgarov},
  \citenamefont {Zhang}, \citenamefont {Rakita}, \citenamefont {Korobko},
  \citenamefont {Rappe},\ and\ \citenamefont {Yaffe}}]{SharmaMAPI1}%
  \BibitemOpen
  \bibfield  {author} {\bibinfo {author} {\bibfnamefont {R.}~\bibnamefont
  {Sharma}}, \bibinfo {author} {\bibfnamefont {Z.}~\bibnamefont {Dai}},
  \bibinfo {author} {\bibfnamefont {L.}~\bibnamefont {Gao}}, \bibinfo {author}
  {\bibfnamefont {T.~M.}\ \bibnamefont {Brenner}}, \bibinfo {author}
  {\bibfnamefont {L.}~\bibnamefont {Yadgarov}}, \bibinfo {author}
  {\bibfnamefont {J.}~\bibnamefont {Zhang}}, \bibinfo {author} {\bibfnamefont
  {Y.}~\bibnamefont {Rakita}}, \bibinfo {author} {\bibfnamefont
  {R.}~\bibnamefont {Korobko}}, \bibinfo {author} {\bibfnamefont {A.~M.}\
  \bibnamefont {Rappe}},\ and\ \bibinfo {author} {\bibfnamefont
  {O.}~\bibnamefont {Yaffe}},\ }\bibfield  {title} {\bibinfo {title}
  {{Elucidating the atomistic origin of anharmonicity in tetragonal
  CH$_3$NH$_3$PbI$_3$ with Raman scattering}},\ }\href
  {http://arxiv.org/abs/2001.04306} {\bibfield  {journal} {\bibinfo  {journal}
  {Physical Review Materials}\ } (\bibinfo {year}
  {2020}{\natexlab{a}})}\BibitemShut {NoStop}%
\bibitem [{\citenamefont {Yaffe}\ \emph {et~al.}(2017)\citenamefont {Yaffe},
  \citenamefont {Guo}, \citenamefont {Tan}, \citenamefont {Egger},
  \citenamefont {Hull}, \citenamefont {Stoumpos}, \citenamefont {Zheng},
  \citenamefont {Heinz}, \citenamefont {Kronik}, \citenamefont {Kanatzidis},
  \citenamefont {Owen}, \citenamefont {Rappe}, \citenamefont {Pimenta},\ and\
  \citenamefont {Brus}}]{YaffePRL2017}%
  \BibitemOpen
  \bibfield  {author} {\bibinfo {author} {\bibfnamefont {O.}~\bibnamefont
  {Yaffe}}, \bibinfo {author} {\bibfnamefont {Y.}~\bibnamefont {Guo}}, \bibinfo
  {author} {\bibfnamefont {L.~Z.}\ \bibnamefont {Tan}}, \bibinfo {author}
  {\bibfnamefont {D.~A.}\ \bibnamefont {Egger}}, \bibinfo {author}
  {\bibfnamefont {T.~D.}\ \bibnamefont {Hull}}, \bibinfo {author}
  {\bibfnamefont {C.~C.}\ \bibnamefont {Stoumpos}}, \bibinfo {author}
  {\bibfnamefont {F.}~\bibnamefont {Zheng}}, \bibinfo {author} {\bibfnamefont
  {T.~F.}\ \bibnamefont {Heinz}}, \bibinfo {author} {\bibfnamefont
  {L.}~\bibnamefont {Kronik}}, \bibinfo {author} {\bibfnamefont {M.~G.}\
  \bibnamefont {Kanatzidis}}, \bibinfo {author} {\bibfnamefont {J.~S.}\
  \bibnamefont {Owen}}, \bibinfo {author} {\bibfnamefont {A.~M.}\ \bibnamefont
  {Rappe}}, \bibinfo {author} {\bibfnamefont {M.~A.}\ \bibnamefont {Pimenta}},\
  and\ \bibinfo {author} {\bibfnamefont {L.~E.}\ \bibnamefont {Brus}},\
  }\bibfield  {title} {\bibinfo {title} {{Local Polar Fluctuations in Lead
  Halide Perovskite Crystals}},\ }\href
  {https://doi.org/10.1103/PhysRevLett.118.136001} {\bibfield  {journal}
  {\bibinfo  {journal} {Physical Review Letters}\ }\textbf {\bibinfo {volume}
  {118}},\ \bibinfo {pages} {1} (\bibinfo {year} {2017})}\BibitemShut {NoStop}%
\bibitem [{\citenamefont {Guo}\ \emph {et~al.}(2017)\citenamefont {Guo},
  \citenamefont {Yaffe}, \citenamefont {Paley}, \citenamefont {Beecher},
  \citenamefont {Hull}, \citenamefont {Szpak}, \citenamefont {Owen},
  \citenamefont {Brus},\ and\ \citenamefont {Pimenta}}]{Guo2017b}%
  \BibitemOpen
  \bibfield  {author} {\bibinfo {author} {\bibfnamefont {Y.}~\bibnamefont
  {Guo}}, \bibinfo {author} {\bibfnamefont {O.}~\bibnamefont {Yaffe}}, \bibinfo
  {author} {\bibfnamefont {D.~W.}\ \bibnamefont {Paley}}, \bibinfo {author}
  {\bibfnamefont {A.~N.}\ \bibnamefont {Beecher}}, \bibinfo {author}
  {\bibfnamefont {T.~D.}\ \bibnamefont {Hull}}, \bibinfo {author}
  {\bibfnamefont {G.}~\bibnamefont {Szpak}}, \bibinfo {author} {\bibfnamefont
  {J.~S.}\ \bibnamefont {Owen}}, \bibinfo {author} {\bibfnamefont {L.~E.}\
  \bibnamefont {Brus}},\ and\ \bibinfo {author} {\bibfnamefont {M.~A.}\
  \bibnamefont {Pimenta}},\ }\bibfield  {title} {\bibinfo {title} {{Interplay
  between organic cations and inorganic framework and incommensurability in
  hybrid lead-halide perovskite CH$_{3}$NH$ _{3}$PbBr$_{3}$}},\ }\href
  {https://doi.org/10.1103/PhysRevMaterials.1.042401} {\bibfield  {journal}
  {\bibinfo  {journal} {Phys. Rev. Mater.}\ }\textbf {\bibinfo {volume} {1}},\
  \bibinfo {pages} {042401} (\bibinfo {year} {2017})}\BibitemShut {NoStop}%
\bibitem [{\citenamefont {Mayers}\ \emph {et~al.}(2018)\citenamefont {Mayers},
  \citenamefont {Tan}, \citenamefont {Egger}, \citenamefont {Rappe},\ and\
  \citenamefont {Reichman}}]{Mayers2018}%
  \BibitemOpen
  \bibfield  {author} {\bibinfo {author} {\bibfnamefont {M.}~\bibnamefont
  {Mayers}}, \bibinfo {author} {\bibfnamefont {L.~Z.}\ \bibnamefont {Tan}},
  \bibinfo {author} {\bibfnamefont {D.~A.}\ \bibnamefont {Egger}}, \bibinfo
  {author} {\bibfnamefont {A.~M.}\ \bibnamefont {Rappe}},\ and\ \bibinfo
  {author} {\bibfnamefont {D.~R.}\ \bibnamefont {Reichman}},\ }\bibfield
  {title} {\bibinfo {title} {{How Lattice and Charge Fluctuations Control
  Carrier Dynamics in Halide Perovskites}},\ }\href
  {https://doi.org/10.1021/acs.nanolett.8b04276} {\bibfield  {journal}
  {\bibinfo  {journal} {Nano Letters}\ }\textbf {\bibinfo {volume} {18}},\
  \bibinfo {pages} {acs.nanolett.8b04276} (\bibinfo {year} {2018})}\BibitemShut
  {NoStop}%
\bibitem [{\citenamefont {Whalley}\ \emph {et~al.}(2016)\citenamefont
  {Whalley}, \citenamefont {Skelton}, \citenamefont {Frost},\ and\
  \citenamefont {Walsh}}]{Whalley2016b}%
  \BibitemOpen
  \bibfield  {author} {\bibinfo {author} {\bibfnamefont {L.~D.}\ \bibnamefont
  {Whalley}}, \bibinfo {author} {\bibfnamefont {J.~M.}\ \bibnamefont
  {Skelton}}, \bibinfo {author} {\bibfnamefont {J.~M.}\ \bibnamefont {Frost}},\
  and\ \bibinfo {author} {\bibfnamefont {A.}~\bibnamefont {Walsh}},\ }\bibfield
   {title} {\bibinfo {title} {{Phonon anharmonicity, lifetimes, and thermal
  transport in CH$ _{3} $NH$ _{3} $PbI$ _{3} $ from many-body perturbation
  theory}},\ }\href {https://doi.org/10.1103/PhysRevB.94.220301} {\bibfield
  {journal} {\bibinfo  {journal} {Phys. Rev. B}\ }\textbf {\bibinfo {volume}
  {94}},\ \bibinfo {pages} {220301} (\bibinfo {year} {2016})}\BibitemShut
  {NoStop}%
\bibitem [{\citenamefont {Egger}\ \emph {et~al.}(2018)\citenamefont {Egger},
  \citenamefont {Bera}, \citenamefont {Cahen}, \citenamefont {Hodes},
  \citenamefont {Kirchartz}, \citenamefont {Kronik}, \citenamefont {Lovrincic},
  \citenamefont {Rappe}, \citenamefont {Reichman},\ and\ \citenamefont
  {Yaffe}}]{Egger2018}%
  \BibitemOpen
  \bibfield  {author} {\bibinfo {author} {\bibfnamefont {D.~A.}\ \bibnamefont
  {Egger}}, \bibinfo {author} {\bibfnamefont {A.}~\bibnamefont {Bera}},
  \bibinfo {author} {\bibfnamefont {D.}~\bibnamefont {Cahen}}, \bibinfo
  {author} {\bibfnamefont {G.}~\bibnamefont {Hodes}}, \bibinfo {author}
  {\bibfnamefont {T.}~\bibnamefont {Kirchartz}}, \bibinfo {author}
  {\bibfnamefont {L.}~\bibnamefont {Kronik}}, \bibinfo {author} {\bibfnamefont
  {R.}~\bibnamefont {Lovrincic}}, \bibinfo {author} {\bibfnamefont {A.~M.}\
  \bibnamefont {Rappe}}, \bibinfo {author} {\bibfnamefont {D.~R.}\ \bibnamefont
  {Reichman}},\ and\ \bibinfo {author} {\bibfnamefont {O.}~\bibnamefont
  {Yaffe}},\ }\bibfield  {title} {\bibinfo {title} {{What Remains Unexplained
  about the Properties of Halide Perovskites?}},\ }\href
  {https://doi.org/10.1002/adma.201800691} {\bibfield  {journal} {\bibinfo
  {journal} {Adv. Mater.}\ }\textbf {\bibinfo {volume} {30}},\ \bibinfo {pages}
  {1800691} (\bibinfo {year} {2018})}\BibitemShut {NoStop}%
\bibitem [{\citenamefont {Panzer}\ \emph {et~al.}(2017)\citenamefont {Panzer},
  \citenamefont {Li}, \citenamefont {Meier}, \citenamefont {K{\"{o}}hler},\
  and\ \citenamefont {Huettner}}]{Panzer2017}%
  \BibitemOpen
  \bibfield  {author} {\bibinfo {author} {\bibfnamefont {F.}~\bibnamefont
  {Panzer}}, \bibinfo {author} {\bibfnamefont {C.}~\bibnamefont {Li}}, \bibinfo
  {author} {\bibfnamefont {T.}~\bibnamefont {Meier}}, \bibinfo {author}
  {\bibfnamefont {A.}~\bibnamefont {K{\"{o}}hler}},\ and\ \bibinfo {author}
  {\bibfnamefont {S.}~\bibnamefont {Huettner}},\ }\bibfield  {title} {\bibinfo
  {title} {{Impact of Structural Dynamics on the Optical Properties of
  Methylammonium Lead Iodide Perovskites}},\ }\href
  {https://doi.org/10.1002/aenm.201700286} {\bibfield  {journal} {\bibinfo
  {journal} {Adv. Energy Mater.}\ }\textbf {\bibinfo {volume} {7}},\ \bibinfo
  {pages} {1700286} (\bibinfo {year} {2017})}\BibitemShut {NoStop}%
\bibitem [{\citenamefont {Seidl}\ \emph {et~al.}(2023)\citenamefont {Seidl},
  \citenamefont {Zhu}, \citenamefont {Reuveni}, \citenamefont {Aharon},
  \citenamefont {Gehrmann}, \citenamefont {Caicedo-D{\'{a}}vila}, \citenamefont
  {Yaffe},\ and\ \citenamefont {Egger}}]{Seidl2023}%
  \BibitemOpen
  \bibfield  {author} {\bibinfo {author} {\bibfnamefont {S.~A.}\ \bibnamefont
  {Seidl}}, \bibinfo {author} {\bibfnamefont {X.}~\bibnamefont {Zhu}}, \bibinfo
  {author} {\bibfnamefont {G.}~\bibnamefont {Reuveni}}, \bibinfo {author}
  {\bibfnamefont {S.}~\bibnamefont {Aharon}}, \bibinfo {author} {\bibfnamefont
  {C.}~\bibnamefont {Gehrmann}}, \bibinfo {author} {\bibfnamefont
  {S.}~\bibnamefont {Caicedo-D{\'{a}}vila}}, \bibinfo {author} {\bibfnamefont
  {O.}~\bibnamefont {Yaffe}},\ and\ \bibinfo {author} {\bibfnamefont {D.~A.}\
  \bibnamefont {Egger}},\ }\bibfield  {title} {\bibinfo {title} {{Anharmonic
  Fluctuations Govern the Band Gap of Halide Perovskites}},\ }\href
  {https://doi.org/10.48550/arXiv.2303.01603} {\bibfield  {journal} {\bibinfo
  {journal} {ArXiv}\ ,\ \bibinfo {pages} {1}} (\bibinfo {year}
  {2023})}\BibitemShut {NoStop}%
\bibitem [{\citenamefont {Gehrmann}\ and\ \citenamefont
  {Egger}(2019)}]{Gehrmann2019}%
  \BibitemOpen
  \bibfield  {author} {\bibinfo {author} {\bibfnamefont {C.}~\bibnamefont
  {Gehrmann}}\ and\ \bibinfo {author} {\bibfnamefont {D.~A.}\ \bibnamefont
  {Egger}},\ }\bibfield  {title} {\bibinfo {title} {{Dynamic shortening of
  disorder potentials in anharmonic halide perovskites}},\ }\bibfield
  {journal} {\bibinfo  {journal} {Nature Communications}\ }\textbf {\bibinfo
  {volume} {10}},\ \href {https://doi.org/10.1038/s41467-019-11087-y}
  {10.1038/s41467-019-11087-y} (\bibinfo {year} {2019})\BibitemShut {NoStop}%
\bibitem [{\citenamefont {Schilcher}\ \emph {et~al.}(2021)\citenamefont
  {Schilcher}, \citenamefont {Robinson}, \citenamefont {Abramovitch},
  \citenamefont {Tan}, \citenamefont {Rappe}, \citenamefont {Reichman},\ and\
  \citenamefont {Egger}}]{Schilcher2021}%
  \BibitemOpen
  \bibfield  {author} {\bibinfo {author} {\bibfnamefont {M.~J.}\ \bibnamefont
  {Schilcher}}, \bibinfo {author} {\bibfnamefont {P.~J.}\ \bibnamefont
  {Robinson}}, \bibinfo {author} {\bibfnamefont {D.~J.}\ \bibnamefont
  {Abramovitch}}, \bibinfo {author} {\bibfnamefont {L.~Z.}\ \bibnamefont
  {Tan}}, \bibinfo {author} {\bibfnamefont {A.~M.}\ \bibnamefont {Rappe}},
  \bibinfo {author} {\bibfnamefont {D.~R.}\ \bibnamefont {Reichman}},\ and\
  \bibinfo {author} {\bibfnamefont {D.~A.}\ \bibnamefont {Egger}},\ }\bibfield
  {title} {\bibinfo {title} {{The Significance of Polarons and Dynamic Disorder
  in Halide Perovskites}},\ }\href
  {https://doi.org/10.1021/acsenergylett.1c00506} {\bibfield  {journal}
  {\bibinfo  {journal} {ACS Energy Letters}\ }\textbf {\bibinfo {volume} {6}},\
  \bibinfo {pages} {2162} (\bibinfo {year} {2021})}\BibitemShut {NoStop}%
\bibitem [{\citenamefont {Niu}\ \emph {et~al.}(2019)\citenamefont {Niu},
  \citenamefont {Zhao}, \citenamefont {Ye}, \citenamefont {Bianco},
  \citenamefont {Zhou}, \citenamefont {McConney}, \citenamefont {Settens},
  \citenamefont {Haiges}, \citenamefont {Jaramillo},\ and\ \citenamefont
  {Ravichandran}}]{Niu2019}%
  \BibitemOpen
  \bibfield  {author} {\bibinfo {author} {\bibfnamefont {S.}~\bibnamefont
  {Niu}}, \bibinfo {author} {\bibfnamefont {B.}~\bibnamefont {Zhao}}, \bibinfo
  {author} {\bibfnamefont {K.}~\bibnamefont {Ye}}, \bibinfo {author}
  {\bibfnamefont {E.}~\bibnamefont {Bianco}}, \bibinfo {author} {\bibfnamefont
  {J.}~\bibnamefont {Zhou}}, \bibinfo {author} {\bibfnamefont {M.~E.}\
  \bibnamefont {McConney}}, \bibinfo {author} {\bibfnamefont {C.}~\bibnamefont
  {Settens}}, \bibinfo {author} {\bibfnamefont {R.}~\bibnamefont {Haiges}},
  \bibinfo {author} {\bibfnamefont {R.}~\bibnamefont {Jaramillo}},\ and\
  \bibinfo {author} {\bibfnamefont {J.}~\bibnamefont {Ravichandran}},\
  }\bibfield  {title} {\bibinfo {title} {{Crystal growth and structural
  analysis of perovskite chalcogenide BaZrS$_3$ and Ruddlesden-Popper phase
  Ba$_3$Zr2S$_7$}},\ }\href {https://doi.org/10.1557/jmr.2019.348} {\bibfield
  {journal} {\bibinfo  {journal} {Journal of Materials Research}\ }\textbf
  {\bibinfo {volume} {34}},\ \bibinfo {pages} {3819} (\bibinfo {year}
  {2019})}\BibitemShut {NoStop}%
\bibitem [{\citenamefont {Lanigan-Atkins}\ \emph {et~al.}(2021)\citenamefont
  {Lanigan-Atkins}, \citenamefont {He}, \citenamefont {Krogstad}, \citenamefont
  {Pajerowski}, \citenamefont {Abernathy}, \citenamefont {Xu}, \citenamefont
  {Xu}, \citenamefont {Chung}, \citenamefont {Kanatzidis}, \citenamefont
  {Rosenkranz}, \citenamefont {Osborn},\ and\ \citenamefont
  {Delaire}}]{Lanigan-Atkins2021}%
  \BibitemOpen
  \bibfield  {author} {\bibinfo {author} {\bibfnamefont {T.}~\bibnamefont
  {Lanigan-Atkins}}, \bibinfo {author} {\bibfnamefont {X.}~\bibnamefont {He}},
  \bibinfo {author} {\bibfnamefont {M.~J.}\ \bibnamefont {Krogstad}}, \bibinfo
  {author} {\bibfnamefont {D.~M.}\ \bibnamefont {Pajerowski}}, \bibinfo
  {author} {\bibfnamefont {D.~L.}\ \bibnamefont {Abernathy}}, \bibinfo {author}
  {\bibfnamefont {G.~N.}\ \bibnamefont {Xu}}, \bibinfo {author} {\bibfnamefont
  {Z.}~\bibnamefont {Xu}}, \bibinfo {author} {\bibfnamefont {D.~Y.}\
  \bibnamefont {Chung}}, \bibinfo {author} {\bibfnamefont {M.~G.}\ \bibnamefont
  {Kanatzidis}}, \bibinfo {author} {\bibfnamefont {S.}~\bibnamefont
  {Rosenkranz}}, \bibinfo {author} {\bibfnamefont {R.}~\bibnamefont {Osborn}},\
  and\ \bibinfo {author} {\bibfnamefont {O.}~\bibnamefont {Delaire}},\
  }\bibfield  {title} {\bibinfo {title} {{Two-dimensional overdamped
  fluctuations of the soft perovskite lattice in CsPbBr$_3$}},\ }\href
  {https://doi.org/10.1038/s41563-021-00947-y} {\bibfield  {journal} {\bibinfo
  {journal} {Nature Materials}\ }\textbf {\bibinfo {volume} {20}},\ \bibinfo
  {pages} {977} (\bibinfo {year} {2021})}\BibitemShut {NoStop}%
\bibitem [{\citenamefont {Das}\ \emph {et~al.}(2020)\citenamefont {Das},
  \citenamefont {Aguilera}, \citenamefont {Rau},\ and\ \citenamefont
  {Kirchartz}}]{das_what_2020}%
  \BibitemOpen
  \bibfield  {author} {\bibinfo {author} {\bibfnamefont {B.}~\bibnamefont
  {Das}}, \bibinfo {author} {\bibfnamefont {I.}~\bibnamefont {Aguilera}},
  \bibinfo {author} {\bibfnamefont {U.}~\bibnamefont {Rau}},\ and\ \bibinfo
  {author} {\bibfnamefont {T.}~\bibnamefont {Kirchartz}},\ }\bibfield  {title}
  {\bibinfo {title} {What is a deep defect? {Combining}
  {Shockley}-{Read}-{Hall} statistics with multiphonon recombination theory},\
  }\href {https://doi.org/10.1103/PhysRevMaterials.4.024602} {\bibfield
  {journal} {\bibinfo  {journal} {Physical Review Materials}\ }\textbf
  {\bibinfo {volume} {4}},\ \bibinfo {pages} {024602} (\bibinfo {year}
  {2020})}\BibitemShut {NoStop}%
\bibitem [{\citenamefont {Cahen}\ \emph {et~al.}(2021)\citenamefont {Cahen},
  \citenamefont {Kronik},\ and\ \citenamefont {Hodes}}]{cahen_are_2021}%
  \BibitemOpen
  \bibfield  {author} {\bibinfo {author} {\bibfnamefont {D.}~\bibnamefont
  {Cahen}}, \bibinfo {author} {\bibfnamefont {L.}~\bibnamefont {Kronik}},\ and\
  \bibinfo {author} {\bibfnamefont {G.}~\bibnamefont {Hodes}},\ }\bibfield
  {title} {\bibinfo {title} {Are {Defects} in {Lead}-{Halide} {Perovskites}
  {Healed}, {Tolerated}, or {Both}?},\ }\href
  {https://doi.org/10.1021/acsenergylett.1c02027} {\bibfield  {journal}
  {\bibinfo  {journal} {ACS Energy Letters}\ }\textbf {\bibinfo {volume} {6}},\
  \bibinfo {pages} {4108} (\bibinfo {year} {2021})}\BibitemShut {NoStop}%
\bibitem [{\citenamefont {Rakita}\ \emph {et~al.}(2016)\citenamefont {Rakita},
  \citenamefont {Kedem}, \citenamefont {Gupta}, \citenamefont {Sadhanala},
  \citenamefont {Kalchenko}, \citenamefont {B{\"{o}}hm}, \citenamefont
  {Kulbak}, \citenamefont {Friend}, \citenamefont {Cahen},\ and\ \citenamefont
  {Hodes}}]{Rakita2016}%
  \BibitemOpen
  \bibfield  {author} {\bibinfo {author} {\bibfnamefont {Y.}~\bibnamefont
  {Rakita}}, \bibinfo {author} {\bibfnamefont {N.}~\bibnamefont {Kedem}},
  \bibinfo {author} {\bibfnamefont {S.}~\bibnamefont {Gupta}}, \bibinfo
  {author} {\bibfnamefont {A.}~\bibnamefont {Sadhanala}}, \bibinfo {author}
  {\bibfnamefont {V.}~\bibnamefont {Kalchenko}}, \bibinfo {author}
  {\bibfnamefont {M.~L.}\ \bibnamefont {B{\"{o}}hm}}, \bibinfo {author}
  {\bibfnamefont {M.}~\bibnamefont {Kulbak}}, \bibinfo {author} {\bibfnamefont
  {R.~H.}\ \bibnamefont {Friend}}, \bibinfo {author} {\bibfnamefont
  {D.}~\bibnamefont {Cahen}},\ and\ \bibinfo {author} {\bibfnamefont
  {G.}~\bibnamefont {Hodes}},\ }\bibfield  {title} {\bibinfo {title}
  {{Low-Temperature Solution-Grown CsPbBr$_3$ Single Crystals and Their
  Characterization}},\ }\href {https://doi.org/10.1021/acs.cgd.6b00764}
  {\bibfield  {journal} {\bibinfo  {journal} {Crystal Growth and Design}\
  }\textbf {\bibinfo {volume} {16}},\ \bibinfo {pages} {5717} (\bibinfo {year}
  {2016})}\BibitemShut {NoStop}%
\bibitem [{\citenamefont {Nishigaki}\ \emph {et~al.}(2020)\citenamefont
  {Nishigaki}, \citenamefont {Nagai}, \citenamefont {Nishiwaki}, \citenamefont
  {Aizawa}, \citenamefont {Kozawa}, \citenamefont {Hanzawa}, \citenamefont
  {Kato}, \citenamefont {Sai}, \citenamefont {Hiramatsu}, \citenamefont
  {Hosono},\ and\ \citenamefont {Fujiwara}}]{Nishigaki2020}%
  \BibitemOpen
  \bibfield  {author} {\bibinfo {author} {\bibfnamefont {Y.}~\bibnamefont
  {Nishigaki}}, \bibinfo {author} {\bibfnamefont {T.}~\bibnamefont {Nagai}},
  \bibinfo {author} {\bibfnamefont {M.}~\bibnamefont {Nishiwaki}}, \bibinfo
  {author} {\bibfnamefont {T.}~\bibnamefont {Aizawa}}, \bibinfo {author}
  {\bibfnamefont {M.}~\bibnamefont {Kozawa}}, \bibinfo {author} {\bibfnamefont
  {K.}~\bibnamefont {Hanzawa}}, \bibinfo {author} {\bibfnamefont
  {Y.}~\bibnamefont {Kato}}, \bibinfo {author} {\bibfnamefont {H.}~\bibnamefont
  {Sai}}, \bibinfo {author} {\bibfnamefont {H.}~\bibnamefont {Hiramatsu}},
  \bibinfo {author} {\bibfnamefont {H.}~\bibnamefont {Hosono}},\ and\ \bibinfo
  {author} {\bibfnamefont {H.}~\bibnamefont {Fujiwara}},\ }\bibfield  {title}
  {\bibinfo {title} {{Extraordinary Strong Band-Edge Absorption in Distorted
  Chalcogenide Perovskites}},\ }\href {https://doi.org/10.1002/solr.201900555}
  {\bibfield  {journal} {\bibinfo  {journal} {Solar RRL}\ }\textbf {\bibinfo
  {volume} {4}},\ \bibinfo {pages} {1} (\bibinfo {year} {2020})}\BibitemShut
  {NoStop}%
\bibitem [{\citenamefont {Guo}\ \emph {et~al.}(2019)\citenamefont {Guo},
  \citenamefont {Yaffe}, \citenamefont {Hull}, \citenamefont {Owen},
  \citenamefont {Reichman},\ and\ \citenamefont {Brus}}]{Guo2018}%
  \BibitemOpen
  \bibfield  {author} {\bibinfo {author} {\bibfnamefont {Y.}~\bibnamefont
  {Guo}}, \bibinfo {author} {\bibfnamefont {O.}~\bibnamefont {Yaffe}}, \bibinfo
  {author} {\bibfnamefont {T.~D.}\ \bibnamefont {Hull}}, \bibinfo {author}
  {\bibfnamefont {J.~S.}\ \bibnamefont {Owen}}, \bibinfo {author}
  {\bibfnamefont {D.~R.}\ \bibnamefont {Reichman}},\ and\ \bibinfo {author}
  {\bibfnamefont {L.~E.}\ \bibnamefont {Brus}},\ }\bibfield  {title} {\bibinfo
  {title} {{Dynamic Emission Stokes Shift and Liquid-Like Dielectric Solvation
  of Band Edge Carriers in Lead-Halide Perovskites}},\ }\href
  {https://doi.org/10.1038/s41467-019-09057-5} {\bibfield  {journal} {\bibinfo
  {journal} {Nature Communications}\ }\textbf {\bibinfo {volume} {10}},\
  \bibinfo {pages} {1175} (\bibinfo {year} {2019})}\BibitemShut {NoStop}%
\bibitem [{\citenamefont {Menahem}\ \emph {et~al.}(2021)\citenamefont
  {Menahem}, \citenamefont {Dai}, \citenamefont {Aharon}, \citenamefont
  {Sharma}, \citenamefont {Asher}, \citenamefont {Diskin-Posner}, \citenamefont
  {Korobko}, \citenamefont {Rappe},\ and\ \citenamefont {Yaffe}}]{Menahem2021}%
  \BibitemOpen
  \bibfield  {author} {\bibinfo {author} {\bibfnamefont {M.}~\bibnamefont
  {Menahem}}, \bibinfo {author} {\bibfnamefont {Z.}~\bibnamefont {Dai}},
  \bibinfo {author} {\bibfnamefont {S.}~\bibnamefont {Aharon}}, \bibinfo
  {author} {\bibfnamefont {R.}~\bibnamefont {Sharma}}, \bibinfo {author}
  {\bibfnamefont {M.}~\bibnamefont {Asher}}, \bibinfo {author} {\bibfnamefont
  {Y.}~\bibnamefont {Diskin-Posner}}, \bibinfo {author} {\bibfnamefont
  {R.}~\bibnamefont {Korobko}}, \bibinfo {author} {\bibfnamefont {A.~M.}\
  \bibnamefont {Rappe}},\ and\ \bibinfo {author} {\bibfnamefont
  {O.}~\bibnamefont {Yaffe}},\ }\bibfield  {title} {\bibinfo {title} {{Strongly
  Anharmonic Octahedral Tilting in Two-Dimensional Hybrid Halide
  Perovskites}},\ }\href {https://doi.org/10.1021/acsnano.1c02022} {\bibfield
  {journal} {\bibinfo  {journal} {ACS Nano}\ }\textbf {\bibinfo {volume}
  {15}},\ \bibinfo {pages} {10153} (\bibinfo {year} {2021})}\BibitemShut
  {NoStop}%
\bibitem [{\citenamefont {Sharma}\ \emph
  {et~al.}(2020{\natexlab{b}})\citenamefont {Sharma}, \citenamefont {Menahem},
  \citenamefont {Dai}, \citenamefont {Gao}, \citenamefont {Korobko},
  \citenamefont {Pinkas}, \citenamefont {Rappe}, \citenamefont {Yaffe},
  \citenamefont {Brenner}, \citenamefont {Yadgarov}, \citenamefont {Zhang},
  \citenamefont {Rakita}, \citenamefont {Korobko}, \citenamefont {Pinkas},
  \citenamefont {Rappe},\ and\ \citenamefont {Yaffe}}]{SharmaMAPI2}%
  \BibitemOpen
  \bibfield  {author} {\bibinfo {author} {\bibfnamefont {R.}~\bibnamefont
  {Sharma}}, \bibinfo {author} {\bibfnamefont {M.}~\bibnamefont {Menahem}},
  \bibinfo {author} {\bibfnamefont {Z.}~\bibnamefont {Dai}}, \bibinfo {author}
  {\bibfnamefont {L.}~\bibnamefont {Gao}}, \bibinfo {author} {\bibfnamefont
  {R.}~\bibnamefont {Korobko}}, \bibinfo {author} {\bibfnamefont
  {I.}~\bibnamefont {Pinkas}}, \bibinfo {author} {\bibfnamefont {A.~M.}\
  \bibnamefont {Rappe}}, \bibinfo {author} {\bibfnamefont {O.}~\bibnamefont
  {Yaffe}}, \bibinfo {author} {\bibfnamefont {T.~M.}\ \bibnamefont {Brenner}},
  \bibinfo {author} {\bibfnamefont {L.}~\bibnamefont {Yadgarov}}, \bibinfo
  {author} {\bibfnamefont {J.}~\bibnamefont {Zhang}}, \bibinfo {author}
  {\bibfnamefont {Y.}~\bibnamefont {Rakita}}, \bibinfo {author} {\bibfnamefont
  {R.}~\bibnamefont {Korobko}}, \bibinfo {author} {\bibfnamefont
  {I.}~\bibnamefont {Pinkas}}, \bibinfo {author} {\bibfnamefont {A.~M.}\
  \bibnamefont {Rappe}},\ and\ \bibinfo {author} {\bibfnamefont
  {O.}~\bibnamefont {Yaffe}},\ }\bibfield  {title} {\bibinfo {title} {{Lattice
  mode symmetry analysis of the orthorhombic phase of methylammonium lead
  iodide using polarized Raman}},\ }\href
  {https://doi.org/10.1103/PhysRevMaterials.4.051601} {\bibfield  {journal}
  {\bibinfo  {journal} {Physical Review Materials}\ }\textbf {\bibinfo {volume}
  {4}},\ \bibinfo {pages} {051601} (\bibinfo {year}
  {2020}{\natexlab{b}})}\BibitemShut {NoStop}%
\bibitem [{\citenamefont {Asher}\ \emph {et~al.}(2020)\citenamefont {Asher},
  \citenamefont {Angerer}, \citenamefont {Korobko}, \citenamefont
  {Diskin-Posner}, \citenamefont {Egger},\ and\ \citenamefont
  {Yaffe}}]{Asher2020}%
  \BibitemOpen
  \bibfield  {author} {\bibinfo {author} {\bibfnamefont {M.}~\bibnamefont
  {Asher}}, \bibinfo {author} {\bibfnamefont {D.}~\bibnamefont {Angerer}},
  \bibinfo {author} {\bibfnamefont {R.}~\bibnamefont {Korobko}}, \bibinfo
  {author} {\bibfnamefont {Y.}~\bibnamefont {Diskin-Posner}}, \bibinfo {author}
  {\bibfnamefont {D.~A.}\ \bibnamefont {Egger}},\ and\ \bibinfo {author}
  {\bibfnamefont {O.}~\bibnamefont {Yaffe}},\ }\bibfield  {title} {\bibinfo
  {title} {{Anharmonic Lattice Vibrations in Small-Molecule Organic
  Semiconductors}},\ }\bibfield  {journal} {\bibinfo  {journal} {Advanced
  Materials}\ }\textbf {\bibinfo {volume} {32}},\ \href
  {https://doi.org/10.1002/adma.201908028} {10.1002/adma.201908028} (\bibinfo
  {year} {2020})\BibitemShut {NoStop}%
\bibitem [{\citenamefont {Cohen}\ \emph {et~al.}(2022)\citenamefont {Cohen},
  \citenamefont {Brenner}, \citenamefont {Klarbring}, \citenamefont {Sharma},
  \citenamefont {Fabini}, \citenamefont {Korobko}, \citenamefont {Nayak},
  \citenamefont {Hellman},\ and\ \citenamefont {Yaffe}}]{Cohen2022}%
  \BibitemOpen
  \bibfield  {author} {\bibinfo {author} {\bibfnamefont {A.}~\bibnamefont
  {Cohen}}, \bibinfo {author} {\bibfnamefont {T.~M.}\ \bibnamefont {Brenner}},
  \bibinfo {author} {\bibfnamefont {J.}~\bibnamefont {Klarbring}}, \bibinfo
  {author} {\bibfnamefont {R.}~\bibnamefont {Sharma}}, \bibinfo {author}
  {\bibfnamefont {D.~H.}\ \bibnamefont {Fabini}}, \bibinfo {author}
  {\bibfnamefont {R.}~\bibnamefont {Korobko}}, \bibinfo {author} {\bibfnamefont
  {P.~K.}\ \bibnamefont {Nayak}}, \bibinfo {author} {\bibfnamefont
  {O.}~\bibnamefont {Hellman}},\ and\ \bibinfo {author} {\bibfnamefont
  {O.}~\bibnamefont {Yaffe}},\ }\bibfield  {title} {\bibinfo {title} {Diverging
  expressions of anharmonicity in halide perovskites},\ }\href
  {https://doi.org/https://doi.org/10.1002/adma.202107932} {\bibfield
  {journal} {\bibinfo  {journal} {Advanced Materials}\ }\textbf {\bibinfo
  {volume} {34}},\ \bibinfo {pages} {2107932} (\bibinfo {year}
  {2022})}\BibitemShut {NoStop}%
\bibitem [{\citenamefont {Asher}\ \emph {et~al.}(2023)\citenamefont {Asher},
  \citenamefont {Bardini}, \citenamefont {Catalano}, \citenamefont {Jouclas},
  \citenamefont {Schweicher}, \citenamefont {Liu}, \citenamefont {Korobko},
  \citenamefont {Cohen}, \citenamefont {Geerts}, \citenamefont {Beljonne},\
  and\ \citenamefont {Yaffe}}]{Asher2023}%
  \BibitemOpen
  \bibfield  {author} {\bibinfo {author} {\bibfnamefont {M.}~\bibnamefont
  {Asher}}, \bibinfo {author} {\bibfnamefont {M.}~\bibnamefont {Bardini}},
  \bibinfo {author} {\bibfnamefont {L.}~\bibnamefont {Catalano}}, \bibinfo
  {author} {\bibfnamefont {R.}~\bibnamefont {Jouclas}}, \bibinfo {author}
  {\bibfnamefont {G.}~\bibnamefont {Schweicher}}, \bibinfo {author}
  {\bibfnamefont {J.}~\bibnamefont {Liu}}, \bibinfo {author} {\bibfnamefont
  {R.}~\bibnamefont {Korobko}}, \bibinfo {author} {\bibfnamefont
  {A.}~\bibnamefont {Cohen}}, \bibinfo {author} {\bibfnamefont
  {Y.}~\bibnamefont {Geerts}}, \bibinfo {author} {\bibfnamefont
  {D.}~\bibnamefont {Beljonne}},\ and\ \bibinfo {author} {\bibfnamefont
  {O.}~\bibnamefont {Yaffe}},\ }\bibfield  {title} {\bibinfo {title}
  {{Mechanistic View on the Order-Disorder Phase Transition in Amphidynamic
  Crystals}},\ }\href {https://doi.org/10.1021/acs.jpclett.2c03316} {\bibfield
  {journal} {\bibinfo  {journal} {The Journal of Physical Chemistry Letters}\
  }\textbf {\bibinfo {volume} {14}},\ \bibinfo {pages} {1570} (\bibinfo {year}
  {2023})}\BibitemShut {NoStop}%
\bibitem [{SM()}]{SM}%
  \BibitemOpen
  \href@noop {} {\bibinfo {title} {{See Supplemental Material at
  \url{https://link.aps.org/supplemental/} for additional Raman spectra and
  fitting, derivations of the equations for the numerical simulation and
  additional results, and fit results to the model.}}}\BibitemShut {Stop}%
\bibitem [{\citenamefont {Ermolaev}\ \emph {et~al.}(2023)\citenamefont
  {Ermolaev}, \citenamefont {Pushkarev}, \citenamefont {Zhizhchenko},
  \citenamefont {Kuchmizhak}, \citenamefont {Iorsh}, \citenamefont {Kruglov},
  \citenamefont {Mazitov}, \citenamefont {Ishteev}, \citenamefont
  {Konstantinova}, \citenamefont {Saranin}, \citenamefont {Slavich},
  \citenamefont {Stosic}, \citenamefont {Zhukova}, \citenamefont {Tselikov},
  \citenamefont {Di~Carlo}, \citenamefont {Arsenin}, \citenamefont {Novoselov},
  \citenamefont {Makarov},\ and\ \citenamefont {Volkov}}]{ermolaev_giant_2023}%
  \BibitemOpen
  \bibfield  {author} {\bibinfo {author} {\bibfnamefont {G.}~\bibnamefont
  {Ermolaev}}, \bibinfo {author} {\bibfnamefont {A.~P.}\ \bibnamefont
  {Pushkarev}}, \bibinfo {author} {\bibfnamefont {A.}~\bibnamefont
  {Zhizhchenko}}, \bibinfo {author} {\bibfnamefont {A.~A.}\ \bibnamefont
  {Kuchmizhak}}, \bibinfo {author} {\bibfnamefont {I.}~\bibnamefont {Iorsh}},
  \bibinfo {author} {\bibfnamefont {I.}~\bibnamefont {Kruglov}}, \bibinfo
  {author} {\bibfnamefont {A.}~\bibnamefont {Mazitov}}, \bibinfo {author}
  {\bibfnamefont {A.}~\bibnamefont {Ishteev}}, \bibinfo {author} {\bibfnamefont
  {K.}~\bibnamefont {Konstantinova}}, \bibinfo {author} {\bibfnamefont
  {D.}~\bibnamefont {Saranin}}, \bibinfo {author} {\bibfnamefont
  {A.}~\bibnamefont {Slavich}}, \bibinfo {author} {\bibfnamefont
  {D.}~\bibnamefont {Stosic}}, \bibinfo {author} {\bibfnamefont {E.~S.}\
  \bibnamefont {Zhukova}}, \bibinfo {author} {\bibfnamefont {G.}~\bibnamefont
  {Tselikov}}, \bibinfo {author} {\bibfnamefont {A.}~\bibnamefont {Di~Carlo}},
  \bibinfo {author} {\bibfnamefont {A.}~\bibnamefont {Arsenin}}, \bibinfo
  {author} {\bibfnamefont {K.~S.}\ \bibnamefont {Novoselov}}, \bibinfo {author}
  {\bibfnamefont {S.~V.}\ \bibnamefont {Makarov}},\ and\ \bibinfo {author}
  {\bibfnamefont {V.~S.}\ \bibnamefont {Volkov}},\ }\bibfield  {title}
  {\bibinfo {title} {Giant and {Tunable} {Excitonic} {Optical} {Anisotropy} in
  {Single}-{Crystal} {Halide} {Perovskites}},\ }\href
  {https://doi.org/10.1021/acs.nanolett.2c04792} {\bibfield  {journal}
  {\bibinfo  {journal} {Nano Letters}\ }\textbf {\bibinfo {volume} {23}},\
  \bibinfo {pages} {2570} (\bibinfo {year} {2023})}\BibitemShut {NoStop}%
\bibitem [{\citenamefont {Bimberg}\ \emph {et~al.}(1971)\citenamefont
  {Bimberg}, \citenamefont {Sondergeld},\ and\ \citenamefont
  {Grobe}}]{Bimberg1971}%
  \BibitemOpen
  \bibfield  {author} {\bibinfo {author} {\bibfnamefont {D.}~\bibnamefont
  {Bimberg}}, \bibinfo {author} {\bibfnamefont {M.}~\bibnamefont
  {Sondergeld}},\ and\ \bibinfo {author} {\bibfnamefont {E.}~\bibnamefont
  {Grobe}},\ }\bibfield  {title} {\bibinfo {title} {{Thermal dissociation of
  excitons bounds to neutral acceptors in high-purity GaAs}},\ }\href
  {https://doi.org/10.1103/PhysRevB.4.3451} {\bibfield  {journal} {\bibinfo
  {journal} {Physical Review B}\ }\textbf {\bibinfo {volume} {4}},\ \bibinfo
  {pages} {3451} (\bibinfo {year} {1971})}\BibitemShut {NoStop}%
\bibitem [{\citenamefont {Lambkin}\ \emph {et~al.}(1994)\citenamefont
  {Lambkin}, \citenamefont {Considine}, \citenamefont {Walsh}, \citenamefont
  {O'Connor}, \citenamefont {McDonagh},\ and\ \citenamefont
  {Glynn}}]{Lambkin1994}%
  \BibitemOpen
  \bibfield  {author} {\bibinfo {author} {\bibfnamefont {J.~D.}\ \bibnamefont
  {Lambkin}}, \bibinfo {author} {\bibfnamefont {L.}~\bibnamefont {Considine}},
  \bibinfo {author} {\bibfnamefont {S.}~\bibnamefont {Walsh}}, \bibinfo
  {author} {\bibfnamefont {G.~M.}\ \bibnamefont {O'Connor}}, \bibinfo {author}
  {\bibfnamefont {C.~J.}\ \bibnamefont {McDonagh}},\ and\ \bibinfo {author}
  {\bibfnamefont {T.~J.}\ \bibnamefont {Glynn}},\ }\bibfield  {title} {\bibinfo
  {title} {{Temperature dependence of the photoluminescence intensity of
  ordered and disordered In$_{0.48}$Ga$_{0.52}$P}},\ }\href
  {https://doi.org/10.1063/1.113078} {\bibfield  {journal} {\bibinfo  {journal}
  {Applied Physics Letters}\ }\textbf {\bibinfo {volume} {65}},\ \bibinfo
  {pages} {73} (\bibinfo {year} {1994})}\BibitemShut {NoStop}%
\bibitem [{\citenamefont {Li}\ \emph {et~al.}(2019)\citenamefont {Li},
  \citenamefont {Niu}, \citenamefont {Zhao}, \citenamefont {Haiges},
  \citenamefont {Zhang}, \citenamefont {Ravichandran},\ and\ \citenamefont
  {Janotti}}]{li_band_2019}%
  \BibitemOpen
  \bibfield  {author} {\bibinfo {author} {\bibfnamefont {W.}~\bibnamefont
  {Li}}, \bibinfo {author} {\bibfnamefont {S.}~\bibnamefont {Niu}}, \bibinfo
  {author} {\bibfnamefont {B.}~\bibnamefont {Zhao}}, \bibinfo {author}
  {\bibfnamefont {R.}~\bibnamefont {Haiges}}, \bibinfo {author} {\bibfnamefont
  {Z.}~\bibnamefont {Zhang}}, \bibinfo {author} {\bibfnamefont
  {J.}~\bibnamefont {Ravichandran}},\ and\ \bibinfo {author} {\bibfnamefont
  {A.}~\bibnamefont {Janotti}},\ }\bibfield  {title} {\bibinfo {title} {Band
  gap evolution in {Ruddlesden}-{Popper} phases},\ }\href
  {https://doi.org/10.1103/PhysRevMaterials.3.101601} {\bibfield  {journal}
  {\bibinfo  {journal} {Physical Review Materials}\ }\textbf {\bibinfo {volume}
  {3}},\ \bibinfo {pages} {101601} (\bibinfo {year} {2019})}\BibitemShut
  {NoStop}%
\bibitem [{\citenamefont {Wolter}\ \emph {et~al.}(2021)\citenamefont {Wolter},
  \citenamefont {Carron}, \citenamefont {Avancini}, \citenamefont {Bissig},
  \citenamefont {Weiss}, \citenamefont {Nishiwaki}, \citenamefont {Feurer},
  \citenamefont {Buecheler}, \citenamefont {Jackson}, \citenamefont {Witte},\
  and\ \citenamefont {Siebentritt}}]{wolter_progress_2021}%
  \BibitemOpen
  \bibfield  {author} {\bibinfo {author} {\bibfnamefont {M.}~\bibnamefont
  {Wolter}}, \bibinfo {author} {\bibfnamefont {R.}~\bibnamefont {Carron}},
  \bibinfo {author} {\bibfnamefont {E.}~\bibnamefont {Avancini}}, \bibinfo
  {author} {\bibfnamefont {B.}~\bibnamefont {Bissig}}, \bibinfo {author}
  {\bibfnamefont {T.}~\bibnamefont {Weiss}}, \bibinfo {author} {\bibfnamefont
  {S.}~\bibnamefont {Nishiwaki}}, \bibinfo {author} {\bibfnamefont
  {T.}~\bibnamefont {Feurer}}, \bibinfo {author} {\bibfnamefont
  {S.}~\bibnamefont {Buecheler}}, \bibinfo {author} {\bibfnamefont
  {P.}~\bibnamefont {Jackson}}, \bibinfo {author} {\bibfnamefont
  {W.}~\bibnamefont {Witte}},\ and\ \bibinfo {author} {\bibfnamefont
  {S.}~\bibnamefont {Siebentritt}},\ }\bibfield  {title} {\bibinfo {title}
  {Progress in {Photovoltaics} - 2021 - {Wolter} - {How} band tail
  recombination influences the open‐circuit voltage of solar cells.pdf},\
  }\href {https://doi.org/10.1002/pip.3449} {\bibfield  {journal} {\bibinfo
  {journal} {Progress in Photovoltaics: Research and Applications}\ }\textbf
  {\bibinfo {volume} {30}},\ \bibinfo {pages} {702} (\bibinfo {year}
  {2021})}\BibitemShut {NoStop}%
\bibitem [{\citenamefont {Wang}\ \emph {et~al.}(2020)\citenamefont {Wang},
  \citenamefont {Yu},\ and\ \citenamefont {Xu}}]{wang_determination_2020}%
  \BibitemOpen
  \bibfield  {author} {\bibinfo {author} {\bibfnamefont {X.}~\bibnamefont
  {Wang}}, \bibinfo {author} {\bibfnamefont {D.}~\bibnamefont {Yu}},\ and\
  \bibinfo {author} {\bibfnamefont {S.}~\bibnamefont {Xu}},\ }\bibfield
  {title} {\bibinfo {title} {Determination of absorption coefficients and
  {Urbach} tail depth of {Zno} below the bandgap with two-photon
  photoluminescence},\ }\href {https://doi.org/10.1364/OE.391534} {\bibfield
  {journal} {\bibinfo  {journal} {Optics Express}\ }\textbf {\bibinfo {volume}
  {28}},\ \bibinfo {pages} {13817} (\bibinfo {year} {2020})}\BibitemShut
  {NoStop}%
\bibitem [{\citenamefont {Falsini}\ \emph {et~al.}(2022)\citenamefont
  {Falsini}, \citenamefont {Roini}, \citenamefont {Ristori}, \citenamefont
  {Calisi}, \citenamefont {Biccari},\ and\ \citenamefont
  {Vinattieri}}]{falsini_analysis_2022}%
  \BibitemOpen
  \bibfield  {author} {\bibinfo {author} {\bibfnamefont {N.}~\bibnamefont
  {Falsini}}, \bibinfo {author} {\bibfnamefont {G.}~\bibnamefont {Roini}},
  \bibinfo {author} {\bibfnamefont {A.}~\bibnamefont {Ristori}}, \bibinfo
  {author} {\bibfnamefont {N.}~\bibnamefont {Calisi}}, \bibinfo {author}
  {\bibfnamefont {F.}~\bibnamefont {Biccari}},\ and\ \bibinfo {author}
  {\bibfnamefont {A.}~\bibnamefont {Vinattieri}},\ }\bibfield  {title}
  {\bibinfo {title} {Analysis of the {Urbach} tail in cesium lead halide
  perovskites},\ }\href {https://doi.org/10.1063/5.0076712} {\bibfield
  {journal} {\bibinfo  {journal} {Journal of Applied Physics}\ }\textbf
  {\bibinfo {volume} {131}},\ \bibinfo {pages} {010902} (\bibinfo {year}
  {2022})}\BibitemShut {NoStop}%
\bibitem [{\citenamefont {Knoop}\ \emph {et~al.}(2024)\citenamefont {Knoop},
  \citenamefont {Shulumba}, \citenamefont {Castellano}, \citenamefont
  {Batista}, \citenamefont {Farris}, \citenamefont {Verstraete}, \citenamefont
  {Heine}, \citenamefont {Broido}, \citenamefont {Kim}, \citenamefont
  {Klarbring}, \citenamefont {Abrikosov}, \citenamefont {Simak},\ and\
  \citenamefont {Hellman}}]{Knoop.2024}%
  \BibitemOpen
  \bibfield  {author} {\bibinfo {author} {\bibfnamefont {F.}~\bibnamefont
  {Knoop}}, \bibinfo {author} {\bibfnamefont {N.}~\bibnamefont {Shulumba}},
  \bibinfo {author} {\bibfnamefont {A.}~\bibnamefont {Castellano}}, \bibinfo
  {author} {\bibfnamefont {J.~P.~A.}\ \bibnamefont {Batista}}, \bibinfo
  {author} {\bibfnamefont {R.}~\bibnamefont {Farris}}, \bibinfo {author}
  {\bibfnamefont {M.~J.}\ \bibnamefont {Verstraete}}, \bibinfo {author}
  {\bibfnamefont {M.}~\bibnamefont {Heine}}, \bibinfo {author} {\bibfnamefont
  {D.}~\bibnamefont {Broido}}, \bibinfo {author} {\bibfnamefont {D.~S.}\
  \bibnamefont {Kim}}, \bibinfo {author} {\bibfnamefont {J.}~\bibnamefont
  {Klarbring}}, \bibinfo {author} {\bibfnamefont {I.~A.}\ \bibnamefont
  {Abrikosov}}, \bibinfo {author} {\bibfnamefont {S.~I.}\ \bibnamefont
  {Simak}},\ and\ \bibinfo {author} {\bibfnamefont {O.}~\bibnamefont
  {Hellman}},\ }\bibfield  {title} {\bibinfo {title} {{TDEP: Temperature
  Dependent Effective Potentials}},\ }\href
  {https://doi.org/10.21105/joss.06150} {\bibfield  {journal} {\bibinfo
  {journal} {Journal of Open Source Software}\ }\textbf {\bibinfo {volume}
  {9}},\ \bibinfo {pages} {6150} (\bibinfo {year} {2024})}\BibitemShut
  {NoStop}%
\bibitem [{\citenamefont {Benshalom}\ \emph {et~al.}(2022)\citenamefont
  {Benshalom}, \citenamefont {Reuveni}, \citenamefont {Korobko}, \citenamefont
  {Yaffe},\ and\ \citenamefont {Hellman}}]{Benshalom2022}%
  \BibitemOpen
  \bibfield  {author} {\bibinfo {author} {\bibfnamefont {N.}~\bibnamefont
  {Benshalom}}, \bibinfo {author} {\bibfnamefont {G.}~\bibnamefont {Reuveni}},
  \bibinfo {author} {\bibfnamefont {R.}~\bibnamefont {Korobko}}, \bibinfo
  {author} {\bibfnamefont {O.}~\bibnamefont {Yaffe}},\ and\ \bibinfo {author}
  {\bibfnamefont {O.}~\bibnamefont {Hellman}},\ }\bibfield  {title} {\bibinfo
  {title} {{Dielectric response of rock-salt crystals at finite temperatures
  from first principles}},\ }\href
  {https://doi.org/10.1103/PhysRevMaterials.6.033607} {\bibfield  {journal}
  {\bibinfo  {journal} {Physical Review Materials}\ }\textbf {\bibinfo {volume}
  {6}},\ \bibinfo {pages} {033607} (\bibinfo {year} {2022})}\BibitemShut
  {NoStop}%
\bibitem [{\citenamefont {Roekeghem}\ \emph {et~al.}(2021)\citenamefont
  {Roekeghem}, \citenamefont {Carrete},\ and\ \citenamefont
  {Mingo}}]{Roekeghem.2021}%
  \BibitemOpen
  \bibfield  {author} {\bibinfo {author} {\bibfnamefont {A.~v.}\ \bibnamefont
  {Roekeghem}}, \bibinfo {author} {\bibfnamefont {J.}~\bibnamefont {Carrete}},\
  and\ \bibinfo {author} {\bibfnamefont {N.}~\bibnamefont {Mingo}},\ }\bibfield
   {title} {\bibinfo {title} {{Quantum Self-Consistent Ab-Initio Lattice
  Dynamics}},\ }\href {https://doi.org/10.1016/j.cpc.2021.107945} {\bibfield
  {journal} {\bibinfo  {journal} {Computer Physics Communications}\ }\textbf
  {\bibinfo {volume} {263}},\ \bibinfo {pages} {107945} (\bibinfo {year}
  {2021})}\BibitemShut {NoStop}%
\bibitem [{\citenamefont {Giannozzi}\ \emph {et~al.}(2020)\citenamefont
  {Giannozzi}, \citenamefont {Baseggio}, \citenamefont {Bonfà}, \citenamefont
  {Brunato}, \citenamefont {Car}, \citenamefont {Carnimeo}, \citenamefont
  {Cavazzoni}, \citenamefont {Gironcoli}, \citenamefont {Delugas},
  \citenamefont {Ruffino}, \citenamefont {Ferretti}, \citenamefont {Marzari},
  \citenamefont {Timrov}, \citenamefont {Urru},\ and\ \citenamefont
  {Baroni}}]{Giannozzi.2020}%
  \BibitemOpen
  \bibfield  {author} {\bibinfo {author} {\bibfnamefont {P.}~\bibnamefont
  {Giannozzi}}, \bibinfo {author} {\bibfnamefont {O.}~\bibnamefont {Baseggio}},
  \bibinfo {author} {\bibfnamefont {P.}~\bibnamefont {Bonfà}}, \bibinfo
  {author} {\bibfnamefont {D.}~\bibnamefont {Brunato}}, \bibinfo {author}
  {\bibfnamefont {R.}~\bibnamefont {Car}}, \bibinfo {author} {\bibfnamefont
  {I.}~\bibnamefont {Carnimeo}}, \bibinfo {author} {\bibfnamefont
  {C.}~\bibnamefont {Cavazzoni}}, \bibinfo {author} {\bibfnamefont {S.~d.}\
  \bibnamefont {Gironcoli}}, \bibinfo {author} {\bibfnamefont {P.}~\bibnamefont
  {Delugas}}, \bibinfo {author} {\bibfnamefont {F.~F.}\ \bibnamefont
  {Ruffino}}, \bibinfo {author} {\bibfnamefont {A.}~\bibnamefont {Ferretti}},
  \bibinfo {author} {\bibfnamefont {N.}~\bibnamefont {Marzari}}, \bibinfo
  {author} {\bibfnamefont {I.}~\bibnamefont {Timrov}}, \bibinfo {author}
  {\bibfnamefont {A.}~\bibnamefont {Urru}},\ and\ \bibinfo {author}
  {\bibfnamefont {S.}~\bibnamefont {Baroni}},\ }\bibfield  {title} {\bibinfo
  {title} {{Quantum ESPRESSO toward the exascale}},\ }\href
  {https://doi.org/10.1063/5.0005082} {\bibfield  {journal} {\bibinfo
  {journal} {The Journal of Chemical Physics}\ }\textbf {\bibinfo {volume}
  {152}},\ \bibinfo {pages} {154105} (\bibinfo {year} {2020})}\BibitemShut
  {NoStop}%
\bibitem [{\citenamefont {Bianco}\ \emph {et~al.}(2017)\citenamefont {Bianco},
  \citenamefont {Errea}, \citenamefont {Paulatto}, \citenamefont {Calandra},\
  and\ \citenamefont {Mauri}}]{Bianco.2017}%
  \BibitemOpen
  \bibfield  {author} {\bibinfo {author} {\bibfnamefont {R.}~\bibnamefont
  {Bianco}}, \bibinfo {author} {\bibfnamefont {I.}~\bibnamefont {Errea}},
  \bibinfo {author} {\bibfnamefont {L.}~\bibnamefont {Paulatto}}, \bibinfo
  {author} {\bibfnamefont {M.}~\bibnamefont {Calandra}},\ and\ \bibinfo
  {author} {\bibfnamefont {F.}~\bibnamefont {Mauri}},\ }\bibfield  {title}
  {\bibinfo {title} {{Second-order structural phase transitions, free energy
  curvature, and temperature-dependent anharmonic phonons in the
  self-consistent harmonic approximation: Theory and stochastic
  implementation}},\ }\href {https://doi.org/10.1103/physrevb.96.014111}
  {\bibfield  {journal} {\bibinfo  {journal} {Physical Review B}\ }\textbf
  {\bibinfo {volume} {96}},\ \bibinfo {pages} {014111} (\bibinfo {year}
  {2017})}\BibitemShut {NoStop}%
\bibitem [{\citenamefont {Shulumba}\ \emph {et~al.}(2017)\citenamefont
  {Shulumba}, \citenamefont {Hellman},\ and\ \citenamefont
  {Minnich}}]{Shulumba.2017}%
  \BibitemOpen
  \bibfield  {author} {\bibinfo {author} {\bibfnamefont {N.}~\bibnamefont
  {Shulumba}}, \bibinfo {author} {\bibfnamefont {O.}~\bibnamefont {Hellman}},\
  and\ \bibinfo {author} {\bibfnamefont {A.~J.}\ \bibnamefont {Minnich}},\
  }\bibfield  {title} {\bibinfo {title} {{Intrinsic localized mode and low
  thermal conductivity of PbSe}},\ }\href
  {https://doi.org/10.1103/physrevb.95.014302} {\bibfield  {journal} {\bibinfo
  {journal} {Physical Review B}\ }\textbf {\bibinfo {volume} {95}},\ \bibinfo
  {pages} {014302} (\bibinfo {year} {2017})}\BibitemShut {NoStop}%
\bibitem [{\citenamefont {Giannozzi}\ \emph {et~al.}(2017)\citenamefont
  {Giannozzi}, \citenamefont {Andreussi}, \citenamefont {Brumme}, \citenamefont
  {Bunau}, \citenamefont {Nardelli}, \citenamefont {Calandra}, \citenamefont
  {Car}, \citenamefont {Cavazzoni}, \citenamefont {Ceresoli}, \citenamefont
  {Cococcioni}, \citenamefont {Colonna}, \citenamefont {Carnimeo},
  \citenamefont {Corso}, \citenamefont {Gironcoli}, \citenamefont {Delugas},
  \citenamefont {Jr}, \citenamefont {Ferretti}, \citenamefont {Floris},
  \citenamefont {Fratesi}, \citenamefont {Fugallo}, \citenamefont {Gebauer},
  \citenamefont {Gerstmann}, \citenamefont {Giustino}, \citenamefont {Gorni},
  \citenamefont {Jia}, \citenamefont {Kawamura}, \citenamefont {Ko},
  \citenamefont {Kokalj}, \citenamefont {Küçükbenli}, \citenamefont
  {Lazzeri}, \citenamefont {Marsili}, \citenamefont {Marzari}, \citenamefont
  {Mauri}, \citenamefont {Nguyen}, \citenamefont {Nguyen}, \citenamefont
  {Otero-de-la Roza}, \citenamefont {Paulatto}, \citenamefont {Poncé},
  \citenamefont {Rocca}, \citenamefont {Sabatini}, \citenamefont {Santra},
  \citenamefont {Schlipf}, \citenamefont {Seitsonen}, \citenamefont {Smogunov},
  \citenamefont {Timrov}, \citenamefont {Thonhauser}, \citenamefont {Umari},
  \citenamefont {Vast}, \citenamefont {Wu},\ and\ \citenamefont
  {Baroni}}]{Giannozzi.2017}%
  \BibitemOpen
  \bibfield  {author} {\bibinfo {author} {\bibfnamefont {P.}~\bibnamefont
  {Giannozzi}}, \bibinfo {author} {\bibfnamefont {O.}~\bibnamefont
  {Andreussi}}, \bibinfo {author} {\bibfnamefont {T.}~\bibnamefont {Brumme}},
  \bibinfo {author} {\bibfnamefont {O.}~\bibnamefont {Bunau}}, \bibinfo
  {author} {\bibfnamefont {M.~B.}\ \bibnamefont {Nardelli}}, \bibinfo {author}
  {\bibfnamefont {M.}~\bibnamefont {Calandra}}, \bibinfo {author}
  {\bibfnamefont {R.}~\bibnamefont {Car}}, \bibinfo {author} {\bibfnamefont
  {C.}~\bibnamefont {Cavazzoni}}, \bibinfo {author} {\bibfnamefont
  {D.}~\bibnamefont {Ceresoli}}, \bibinfo {author} {\bibfnamefont
  {M.}~\bibnamefont {Cococcioni}}, \bibinfo {author} {\bibfnamefont
  {N.}~\bibnamefont {Colonna}}, \bibinfo {author} {\bibfnamefont
  {I.}~\bibnamefont {Carnimeo}}, \bibinfo {author} {\bibfnamefont {A.~D.}\
  \bibnamefont {Corso}}, \bibinfo {author} {\bibfnamefont {S.~d.}\ \bibnamefont
  {Gironcoli}}, \bibinfo {author} {\bibfnamefont {P.}~\bibnamefont {Delugas}},
  \bibinfo {author} {\bibfnamefont {R.~A.~D.}\ \bibnamefont {Jr}}, \bibinfo
  {author} {\bibfnamefont {A.}~\bibnamefont {Ferretti}}, \bibinfo {author}
  {\bibfnamefont {A.}~\bibnamefont {Floris}}, \bibinfo {author} {\bibfnamefont
  {G.}~\bibnamefont {Fratesi}}, \bibinfo {author} {\bibfnamefont
  {G.}~\bibnamefont {Fugallo}}, \bibinfo {author} {\bibfnamefont
  {R.}~\bibnamefont {Gebauer}}, \bibinfo {author} {\bibfnamefont
  {U.}~\bibnamefont {Gerstmann}}, \bibinfo {author} {\bibfnamefont
  {F.}~\bibnamefont {Giustino}}, \bibinfo {author} {\bibfnamefont
  {T.}~\bibnamefont {Gorni}}, \bibinfo {author} {\bibfnamefont
  {J.}~\bibnamefont {Jia}}, \bibinfo {author} {\bibfnamefont {M.}~\bibnamefont
  {Kawamura}}, \bibinfo {author} {\bibfnamefont {H.-Y.}\ \bibnamefont {Ko}},
  \bibinfo {author} {\bibfnamefont {A.}~\bibnamefont {Kokalj}}, \bibinfo
  {author} {\bibfnamefont {E.}~\bibnamefont {Küçükbenli}}, \bibinfo {author}
  {\bibfnamefont {M.}~\bibnamefont {Lazzeri}}, \bibinfo {author} {\bibfnamefont
  {M.}~\bibnamefont {Marsili}}, \bibinfo {author} {\bibfnamefont
  {N.}~\bibnamefont {Marzari}}, \bibinfo {author} {\bibfnamefont
  {F.}~\bibnamefont {Mauri}}, \bibinfo {author} {\bibfnamefont {N.~L.}\
  \bibnamefont {Nguyen}}, \bibinfo {author} {\bibfnamefont {H.-V.}\
  \bibnamefont {Nguyen}}, \bibinfo {author} {\bibfnamefont {A.}~\bibnamefont
  {Otero-de-la Roza}}, \bibinfo {author} {\bibfnamefont {L.}~\bibnamefont
  {Paulatto}}, \bibinfo {author} {\bibfnamefont {S.}~\bibnamefont {Poncé}},
  \bibinfo {author} {\bibfnamefont {D.}~\bibnamefont {Rocca}}, \bibinfo
  {author} {\bibfnamefont {R.}~\bibnamefont {Sabatini}}, \bibinfo {author}
  {\bibfnamefont {B.}~\bibnamefont {Santra}}, \bibinfo {author} {\bibfnamefont
  {M.}~\bibnamefont {Schlipf}}, \bibinfo {author} {\bibfnamefont {A.~P.}\
  \bibnamefont {Seitsonen}}, \bibinfo {author} {\bibfnamefont {A.}~\bibnamefont
  {Smogunov}}, \bibinfo {author} {\bibfnamefont {I.}~\bibnamefont {Timrov}},
  \bibinfo {author} {\bibfnamefont {T.}~\bibnamefont {Thonhauser}}, \bibinfo
  {author} {\bibfnamefont {P.}~\bibnamefont {Umari}}, \bibinfo {author}
  {\bibfnamefont {N.}~\bibnamefont {Vast}}, \bibinfo {author} {\bibfnamefont
  {X.}~\bibnamefont {Wu}},\ and\ \bibinfo {author} {\bibfnamefont
  {S.}~\bibnamefont {Baroni}},\ }\bibfield  {title} {\bibinfo {title}
  {{Advanced capabilities for materials modelling with Quantum ESPRESSO}},\
  }\href {https://doi.org/10.1088/1361-648x/aa8f79} {\bibfield  {journal}
  {\bibinfo  {journal} {Journal of Physics: Condensed Matter}\ }\textbf
  {\bibinfo {volume} {29}},\ \bibinfo {pages} {465901} (\bibinfo {year}
  {2017})}\BibitemShut {NoStop}%
\bibitem [{\citenamefont {Schlipf}\ and\ \citenamefont
  {Gygi}(2015)}]{Schlipf.2015ake}%
  \BibitemOpen
  \bibfield  {author} {\bibinfo {author} {\bibfnamefont {M.}~\bibnamefont
  {Schlipf}}\ and\ \bibinfo {author} {\bibfnamefont {F.}~\bibnamefont {Gygi}},\
  }\bibfield  {title} {\bibinfo {title} {{Optimization algorithm for the
  generation of ONCV pseudopotentials}},\ }\href
  {https://doi.org/10.1016/j.cpc.2015.05.011} {\bibfield  {journal} {\bibinfo
  {journal} {Computer Physics Communications}\ }\textbf {\bibinfo {volume}
  {196}},\ \bibinfo {pages} {36} (\bibinfo {year} {2015})}\BibitemShut
  {NoStop}%
\bibitem [{\citenamefont {Hellman}\ and\ \citenamefont
  {Abrikosov}(2013)}]{Hellman.2013oi5}%
  \BibitemOpen
  \bibfield  {author} {\bibinfo {author} {\bibfnamefont {O.}~\bibnamefont
  {Hellman}}\ and\ \bibinfo {author} {\bibfnamefont {I.~A.}\ \bibnamefont
  {Abrikosov}},\ }\bibfield  {title} {\bibinfo {title} {{Temperature-dependent
  effective third-order interatomic force constants from first principles}},\
  }\href {https://doi.org/10.1103/physrevb.88.144301} {\bibfield  {journal}
  {\bibinfo  {journal} {Physical Review B}\ }\textbf {\bibinfo {volume} {88}},\
  \bibinfo {pages} {144301} (\bibinfo {year} {2013})}\BibitemShut {NoStop}%
\bibitem [{\citenamefont {Hamann}(2013)}]{Hamann.2013}%
  \BibitemOpen
  \bibfield  {author} {\bibinfo {author} {\bibfnamefont {D.~R.}\ \bibnamefont
  {Hamann}},\ }\bibfield  {title} {\bibinfo {title} {{Optimized norm-conserving
  Vanderbilt pseudopotentials}},\ }\href
  {https://doi.org/10.1103/physrevb.88.085117} {\bibfield  {journal} {\bibinfo
  {journal} {Physical Review B}\ }\textbf {\bibinfo {volume} {88}},\ \bibinfo
  {pages} {085117} (\bibinfo {year} {2013})}\BibitemShut {NoStop}%
\bibitem [{\citenamefont {Jain}\ \emph {et~al.}(2013)\citenamefont {Jain},
  \citenamefont {Ong}, \citenamefont {Hautier}, \citenamefont {Chen},
  \citenamefont {Richards}, \citenamefont {Dacek}, \citenamefont {Cholia},
  \citenamefont {Gunter}, \citenamefont {Skinner}, \citenamefont {Ceder},\ and\
  \citenamefont {Persson}}]{Jain.2013}%
  \BibitemOpen
  \bibfield  {author} {\bibinfo {author} {\bibfnamefont {A.}~\bibnamefont
  {Jain}}, \bibinfo {author} {\bibfnamefont {S.~P.}\ \bibnamefont {Ong}},
  \bibinfo {author} {\bibfnamefont {G.}~\bibnamefont {Hautier}}, \bibinfo
  {author} {\bibfnamefont {W.}~\bibnamefont {Chen}}, \bibinfo {author}
  {\bibfnamefont {W.~D.}\ \bibnamefont {Richards}}, \bibinfo {author}
  {\bibfnamefont {S.}~\bibnamefont {Dacek}}, \bibinfo {author} {\bibfnamefont
  {S.}~\bibnamefont {Cholia}}, \bibinfo {author} {\bibfnamefont
  {D.}~\bibnamefont {Gunter}}, \bibinfo {author} {\bibfnamefont
  {D.}~\bibnamefont {Skinner}}, \bibinfo {author} {\bibfnamefont
  {G.}~\bibnamefont {Ceder}},\ and\ \bibinfo {author} {\bibfnamefont {K.~A.}\
  \bibnamefont {Persson}},\ }\bibfield  {title} {\bibinfo {title} {{Commentary:
  The Materials Project: A materials genome approach to accelerating materials
  innovation}},\ }\href {https://doi.org/10.1063/1.4812323} {\bibfield
  {journal} {\bibinfo  {journal} {APL Materials}\ }\textbf {\bibinfo {volume}
  {1}},\ \bibinfo {pages} {011002} (\bibinfo {year} {2013})}\BibitemShut
  {NoStop}%
\bibitem [{\citenamefont {Giannozzi}\ \emph {et~al.}(2009)\citenamefont
  {Giannozzi}, \citenamefont {Baroni}, \citenamefont {Bonini}, \citenamefont
  {Calandra}, \citenamefont {Car}, \citenamefont {Cavazzoni}, \citenamefont
  {Ceresoli}, \citenamefont {Chiarotti}, \citenamefont {Cococcioni},
  \citenamefont {Dabo}, \citenamefont {Corso}, \citenamefont {Gironcoli},
  \citenamefont {Fabris}, \citenamefont {Fratesi}, \citenamefont {Gebauer},
  \citenamefont {Gerstmann}, \citenamefont {Gougoussis}, \citenamefont
  {Kokalj}, \citenamefont {Lazzeri}, \citenamefont {Martin-Samos},
  \citenamefont {Marzari}, \citenamefont {Mauri}, \citenamefont {Mazzarello},
  \citenamefont {Paolini}, \citenamefont {Pasquarello}, \citenamefont
  {Paulatto}, \citenamefont {Sbraccia}, \citenamefont {Scandolo}, \citenamefont
  {Sclauzero}, \citenamefont {Seitsonen}, \citenamefont {Smogunov},
  \citenamefont {Umari},\ and\ \citenamefont {Wentzcovitch}}]{Giannozzi.2009}%
  \BibitemOpen
  \bibfield  {author} {\bibinfo {author} {\bibfnamefont {P.}~\bibnamefont
  {Giannozzi}}, \bibinfo {author} {\bibfnamefont {S.}~\bibnamefont {Baroni}},
  \bibinfo {author} {\bibfnamefont {N.}~\bibnamefont {Bonini}}, \bibinfo
  {author} {\bibfnamefont {M.}~\bibnamefont {Calandra}}, \bibinfo {author}
  {\bibfnamefont {R.}~\bibnamefont {Car}}, \bibinfo {author} {\bibfnamefont
  {C.}~\bibnamefont {Cavazzoni}}, \bibinfo {author} {\bibfnamefont
  {D.}~\bibnamefont {Ceresoli}}, \bibinfo {author} {\bibfnamefont {G.~L.}\
  \bibnamefont {Chiarotti}}, \bibinfo {author} {\bibfnamefont {M.}~\bibnamefont
  {Cococcioni}}, \bibinfo {author} {\bibfnamefont {I.}~\bibnamefont {Dabo}},
  \bibinfo {author} {\bibfnamefont {A.~D.}\ \bibnamefont {Corso}}, \bibinfo
  {author} {\bibfnamefont {S.~d.}\ \bibnamefont {Gironcoli}}, \bibinfo {author}
  {\bibfnamefont {S.}~\bibnamefont {Fabris}}, \bibinfo {author} {\bibfnamefont
  {G.}~\bibnamefont {Fratesi}}, \bibinfo {author} {\bibfnamefont
  {R.}~\bibnamefont {Gebauer}}, \bibinfo {author} {\bibfnamefont
  {U.}~\bibnamefont {Gerstmann}}, \bibinfo {author} {\bibfnamefont
  {C.}~\bibnamefont {Gougoussis}}, \bibinfo {author} {\bibfnamefont
  {A.}~\bibnamefont {Kokalj}}, \bibinfo {author} {\bibfnamefont
  {M.}~\bibnamefont {Lazzeri}}, \bibinfo {author} {\bibfnamefont
  {L.}~\bibnamefont {Martin-Samos}}, \bibinfo {author} {\bibfnamefont
  {N.}~\bibnamefont {Marzari}}, \bibinfo {author} {\bibfnamefont
  {F.}~\bibnamefont {Mauri}}, \bibinfo {author} {\bibfnamefont
  {R.}~\bibnamefont {Mazzarello}}, \bibinfo {author} {\bibfnamefont
  {S.}~\bibnamefont {Paolini}}, \bibinfo {author} {\bibfnamefont
  {A.}~\bibnamefont {Pasquarello}}, \bibinfo {author} {\bibfnamefont
  {L.}~\bibnamefont {Paulatto}}, \bibinfo {author} {\bibfnamefont
  {C.}~\bibnamefont {Sbraccia}}, \bibinfo {author} {\bibfnamefont
  {S.}~\bibnamefont {Scandolo}}, \bibinfo {author} {\bibfnamefont
  {G.}~\bibnamefont {Sclauzero}}, \bibinfo {author} {\bibfnamefont {A.~P.}\
  \bibnamefont {Seitsonen}}, \bibinfo {author} {\bibfnamefont {A.}~\bibnamefont
  {Smogunov}}, \bibinfo {author} {\bibfnamefont {P.}~\bibnamefont {Umari}},\
  and\ \bibinfo {author} {\bibfnamefont {R.~M.}\ \bibnamefont {Wentzcovitch}},\
  }\bibfield  {title} {\bibinfo {title} {{QUANTUM ESPRESSO: a modular and
  open-source software project for quantum simulations of materials}},\ }\href
  {https://doi.org/10.1088/0953-8984/21/39/395502} {\bibfield  {journal}
  {\bibinfo  {journal} {Journal of Physics: Condensed Matter}\ }\textbf
  {\bibinfo {volume} {21}},\ \bibinfo {pages} {395502} (\bibinfo {year}
  {2009})}\BibitemShut {NoStop}%
\bibitem [{\citenamefont {Blum}\ \emph {et~al.}(2009)\citenamefont {Blum},
  \citenamefont {Gehrke}, \citenamefont {Hanke}, \citenamefont {Havu},
  \citenamefont {Havu}, \citenamefont {Ren}, \citenamefont {Reuter},\ and\
  \citenamefont {Scheffler}}]{Blum.2009}%
  \BibitemOpen
  \bibfield  {author} {\bibinfo {author} {\bibfnamefont {V.}~\bibnamefont
  {Blum}}, \bibinfo {author} {\bibfnamefont {R.}~\bibnamefont {Gehrke}},
  \bibinfo {author} {\bibfnamefont {F.}~\bibnamefont {Hanke}}, \bibinfo
  {author} {\bibfnamefont {P.}~\bibnamefont {Havu}}, \bibinfo {author}
  {\bibfnamefont {V.}~\bibnamefont {Havu}}, \bibinfo {author} {\bibfnamefont
  {X.}~\bibnamefont {Ren}}, \bibinfo {author} {\bibfnamefont {K.}~\bibnamefont
  {Reuter}},\ and\ \bibinfo {author} {\bibfnamefont {M.}~\bibnamefont
  {Scheffler}},\ }\bibfield  {title} {\bibinfo {title} {{Ab initio molecular
  simulations with numeric atom-centered orbitals}},\ }\href
  {https://doi.org/10.1016/j.cpc.2009.06.022} {\bibfield  {journal} {\bibinfo
  {journal} {Computer Physics Communications}\ }\textbf {\bibinfo {volume}
  {180}},\ \bibinfo {pages} {2175} (\bibinfo {year} {2009})}\BibitemShut
  {NoStop}%
\bibitem [{\citenamefont {Mattsson}\ \emph {et~al.}(2008)\citenamefont
  {Mattsson}, \citenamefont {Armiento}, \citenamefont {Paier}, \citenamefont
  {Kresse}, \citenamefont {Wills},\ and\ \citenamefont
  {Mattsson}}]{Mattsson.2008}%
  \BibitemOpen
  \bibfield  {author} {\bibinfo {author} {\bibfnamefont {A.~E.}\ \bibnamefont
  {Mattsson}}, \bibinfo {author} {\bibfnamefont {R.}~\bibnamefont {Armiento}},
  \bibinfo {author} {\bibfnamefont {J.}~\bibnamefont {Paier}}, \bibinfo
  {author} {\bibfnamefont {G.}~\bibnamefont {Kresse}}, \bibinfo {author}
  {\bibfnamefont {J.~M.}\ \bibnamefont {Wills}},\ and\ \bibinfo {author}
  {\bibfnamefont {T.~R.}\ \bibnamefont {Mattsson}},\ }\bibfield  {title}
  {\bibinfo {title} {{The AM05 density functional applied to solids}},\ }\href
  {https://doi.org/10.1063/1.2835596} {\bibfield  {journal} {\bibinfo
  {journal} {The Journal of Chemical Physics}\ }\textbf {\bibinfo {volume}
  {128}},\ \bibinfo {pages} {084714} (\bibinfo {year} {2008})}\BibitemShut
  {NoStop}%
\bibitem [{\citenamefont {Armiento}\ and\ \citenamefont
  {Mattsson}(2005)}]{Armiento.2005}%
  \BibitemOpen
  \bibfield  {author} {\bibinfo {author} {\bibfnamefont {R.}~\bibnamefont
  {Armiento}}\ and\ \bibinfo {author} {\bibfnamefont {A.~E.}\ \bibnamefont
  {Mattsson}},\ }\bibfield  {title} {\bibinfo {title} {{Functional designed to
  include surface effects in self-consistent density functional theory}},\
  }\href {https://doi.org/10.1103/physrevb.72.085108} {\bibfield  {journal}
  {\bibinfo  {journal} {Physical Review B}\ }\textbf {\bibinfo {volume} {72}},\
  \bibinfo {pages} {085108} (\bibinfo {year} {2005})}\BibitemShut {NoStop}%
\bibitem [{\citenamefont {Perdew}\ \emph {et~al.}(1996)\citenamefont {Perdew},
  \citenamefont {Burke},\ and\ \citenamefont {Ernzerhof}}]{Perdew.1996}%
  \BibitemOpen
  \bibfield  {author} {\bibinfo {author} {\bibfnamefont {J.~P.}\ \bibnamefont
  {Perdew}}, \bibinfo {author} {\bibfnamefont {K.}~\bibnamefont {Burke}},\ and\
  \bibinfo {author} {\bibfnamefont {M.}~\bibnamefont {Ernzerhof}},\ }\bibfield
  {title} {\bibinfo {title} {{Generalized Gradient Approximation Made
  Simple}},\ }\href {https://doi.org/10.1103/physrevlett.77.3865} {\bibfield
  {journal} {\bibinfo  {journal} {Physical Review Letters}\ }\textbf {\bibinfo
  {volume} {77}},\ \bibinfo {pages} {3865} (\bibinfo {year} {1996})},\ \bibinfo
  {note} {pBE}\BibitemShut {NoStop}%
\bibitem [{\citenamefont {Fox}(2010)}]{FoxOPOS}%
  \BibitemOpen
  \bibfield  {author} {\bibinfo {author} {\bibfnamefont {M.}~\bibnamefont
  {Fox}},\ }\bibinfo {title} {{Phonons}},\ in\ \href
  {https://doi.org/10.1007/BF02751482} {\emph {\bibinfo {booktitle} {{Optical
  Properties of Solids}}}}\ (\bibinfo  {publisher} {Oxford University Press},\
  \bibinfo {address} {New York, USA},\ \bibinfo {year} {2010})\ pp.\ \bibinfo
  {pages} {273--276},\ \bibinfo {edition} {2nd}\ ed.\BibitemShut {Stop}%
\bibitem [{\citenamefont {Yazdani}\ \emph {et~al.}(2024)\citenamefont
  {Yazdani}, \citenamefont {Bodnarchuk}, \citenamefont {Bertolotti},
  \citenamefont {Masciocchi}, \citenamefont {Fureraj}, \citenamefont
  {Guzelturk}, \citenamefont {Cotts}, \citenamefont {Zajac}, \citenamefont
  {Rain{\`{o}}}, \citenamefont {Jansen}, \citenamefont {Boehme}, \citenamefont
  {Yarema}, \citenamefont {Lin}, \citenamefont {Kozina}, \citenamefont {Reid},
  \citenamefont {Shen}, \citenamefont {Weathersby}, \citenamefont {Wang},
  \citenamefont {Vauthey}, \citenamefont {Guagliardi}, \citenamefont
  {Kovalenko}, \citenamefont {Wood},\ and\ \citenamefont
  {Lindenberg}}]{yazdani_coupling_2024}%
  \BibitemOpen
  \bibfield  {author} {\bibinfo {author} {\bibfnamefont {N.}~\bibnamefont
  {Yazdani}}, \bibinfo {author} {\bibfnamefont {M.~I.}\ \bibnamefont
  {Bodnarchuk}}, \bibinfo {author} {\bibfnamefont {F.}~\bibnamefont
  {Bertolotti}}, \bibinfo {author} {\bibfnamefont {N.}~\bibnamefont
  {Masciocchi}}, \bibinfo {author} {\bibfnamefont {I.}~\bibnamefont {Fureraj}},
  \bibinfo {author} {\bibfnamefont {B.}~\bibnamefont {Guzelturk}}, \bibinfo
  {author} {\bibfnamefont {B.~L.}\ \bibnamefont {Cotts}}, \bibinfo {author}
  {\bibfnamefont {M.}~\bibnamefont {Zajac}}, \bibinfo {author} {\bibfnamefont
  {G.}~\bibnamefont {Rain{\`{o}}}}, \bibinfo {author} {\bibfnamefont
  {M.}~\bibnamefont {Jansen}}, \bibinfo {author} {\bibfnamefont {S.~C.}\
  \bibnamefont {Boehme}}, \bibinfo {author} {\bibfnamefont {M.}~\bibnamefont
  {Yarema}}, \bibinfo {author} {\bibfnamefont {M.~F.}\ \bibnamefont {Lin}},
  \bibinfo {author} {\bibfnamefont {M.}~\bibnamefont {Kozina}}, \bibinfo
  {author} {\bibfnamefont {A.}~\bibnamefont {Reid}}, \bibinfo {author}
  {\bibfnamefont {X.}~\bibnamefont {Shen}}, \bibinfo {author} {\bibfnamefont
  {S.}~\bibnamefont {Weathersby}}, \bibinfo {author} {\bibfnamefont
  {X.}~\bibnamefont {Wang}}, \bibinfo {author} {\bibfnamefont {E.}~\bibnamefont
  {Vauthey}}, \bibinfo {author} {\bibfnamefont {A.}~\bibnamefont {Guagliardi}},
  \bibinfo {author} {\bibfnamefont {M.~V.}\ \bibnamefont {Kovalenko}}, \bibinfo
  {author} {\bibfnamefont {V.}~\bibnamefont {Wood}},\ and\ \bibinfo {author}
  {\bibfnamefont {A.~M.}\ \bibnamefont {Lindenberg}},\ }\bibfield  {title}
  {\bibinfo {title} {{Coupling to octahedral tilts in halide perovskite
  nanocrystals induces phonon-mediated attractive interactions between
  excitons}},\ }\href {https://doi.org/10.1038/s41567-023-02253-7} {\bibfield
  {journal} {\bibinfo  {journal} {Nature Physics}\ }\textbf {\bibinfo {volume}
  {20}},\ \bibinfo {pages} {47} (\bibinfo {year} {2024})}\BibitemShut {NoStop}%
\bibitem [{\citenamefont {Gross}\ \emph {et~al.}(2017)\citenamefont {Gross},
  \citenamefont {Sun}, \citenamefont {Perera}, \citenamefont {Hui},
  \citenamefont {Wei}, \citenamefont {Zhang}, \citenamefont {Zeng},\ and\
  \citenamefont {Weinstein}}]{Gross2017}%
  \BibitemOpen
  \bibfield  {author} {\bibinfo {author} {\bibfnamefont {N.}~\bibnamefont
  {Gross}}, \bibinfo {author} {\bibfnamefont {Y.-Y.}\ \bibnamefont {Sun}},
  \bibinfo {author} {\bibfnamefont {S.}~\bibnamefont {Perera}}, \bibinfo
  {author} {\bibfnamefont {H.}~\bibnamefont {Hui}}, \bibinfo {author}
  {\bibfnamefont {X.}~\bibnamefont {Wei}}, \bibinfo {author} {\bibfnamefont
  {S.}~\bibnamefont {Zhang}}, \bibinfo {author} {\bibfnamefont
  {H.}~\bibnamefont {Zeng}},\ and\ \bibinfo {author} {\bibfnamefont {B.~A.}\
  \bibnamefont {Weinstein}},\ }\bibfield  {title} {\bibinfo {title} {{Stability
  and Band-Gap Tuning of the Chalcogenide Perovskite \BZS\ in Raman and Optical
  Investigations at High Pressures}},\ }\href
  {https://doi.org/10.1103/PhysRevApplied.8.044014} {\bibfield  {journal}
  {\bibinfo  {journal} {Physical Review Applied}\ }\textbf {\bibinfo {volume}
  {8}},\ \bibinfo {pages} {044014} (\bibinfo {year} {2017})}\BibitemShut
  {NoStop}%
\bibitem [{\citenamefont {Xu}\ \emph {et~al.}(2022)\citenamefont {Xu},
  \citenamefont {Fan}, \citenamefont {Tian}, \citenamefont {Ye}, \citenamefont
  {Zhang}, \citenamefont {Tian}, \citenamefont {Han},\ and\ \citenamefont
  {Shi}}]{Xu2022}%
  \BibitemOpen
  \bibfield  {author} {\bibinfo {author} {\bibfnamefont {J.}~\bibnamefont
  {Xu}}, \bibinfo {author} {\bibfnamefont {Y.}~\bibnamefont {Fan}}, \bibinfo
  {author} {\bibfnamefont {W.}~\bibnamefont {Tian}}, \bibinfo {author}
  {\bibfnamefont {L.}~\bibnamefont {Ye}}, \bibinfo {author} {\bibfnamefont
  {Y.}~\bibnamefont {Zhang}}, \bibinfo {author} {\bibfnamefont
  {Y.}~\bibnamefont {Tian}}, \bibinfo {author} {\bibfnamefont {Y.}~\bibnamefont
  {Han}},\ and\ \bibinfo {author} {\bibfnamefont {Z.}~\bibnamefont {Shi}},\
  }\bibfield  {title} {\bibinfo {title} {{Enhancing the optical absorption of
  chalcogenide perovskite BaZrS$_3$ by optimizing the synthesis and
  post-processing conditions}},\ }\href
  {https://doi.org/10.1016/j.jssc.2021.122872} {\bibfield  {journal} {\bibinfo
  {journal} {Journal of Solid State Chemistry}\ }\textbf {\bibinfo {volume}
  {307}},\ \bibinfo {pages} {122872} (\bibinfo {year} {2022})}\BibitemShut
  {NoStop}%
\bibitem [{\citenamefont {Yu}\ \emph {et~al.}(2021)\citenamefont {Yu},
  \citenamefont {Wei}, \citenamefont {Zheng}, \citenamefont {Hui},
  \citenamefont {Bian}, \citenamefont {Dhole}, \citenamefont {Seo},
  \citenamefont {Sun}, \citenamefont {Jia}, \citenamefont {Zhang},
  \citenamefont {Yang},\ and\ \citenamefont {Zeng}}]{Yu2021}%
  \BibitemOpen
  \bibfield  {author} {\bibinfo {author} {\bibfnamefont {Z.}~\bibnamefont
  {Yu}}, \bibinfo {author} {\bibfnamefont {X.}~\bibnamefont {Wei}}, \bibinfo
  {author} {\bibfnamefont {Y.}~\bibnamefont {Zheng}}, \bibinfo {author}
  {\bibfnamefont {H.}~\bibnamefont {Hui}}, \bibinfo {author} {\bibfnamefont
  {M.}~\bibnamefont {Bian}}, \bibinfo {author} {\bibfnamefont {S.}~\bibnamefont
  {Dhole}}, \bibinfo {author} {\bibfnamefont {J.-H.}\ \bibnamefont {Seo}},
  \bibinfo {author} {\bibfnamefont {Y.-Y.}\ \bibnamefont {Sun}}, \bibinfo
  {author} {\bibfnamefont {Q.}~\bibnamefont {Jia}}, \bibinfo {author}
  {\bibfnamefont {S.}~\bibnamefont {Zhang}}, \bibinfo {author} {\bibfnamefont
  {S.}~\bibnamefont {Yang}},\ and\ \bibinfo {author} {\bibfnamefont
  {H.}~\bibnamefont {Zeng}},\ }\bibfield  {title} {\bibinfo {title}
  {{Chalcogenide perovskite BaZrS$_3$ thin-film electronic and optoelectronic
  devices by low temperature processing}},\ }\href
  {https://doi.org/10.1016/j.nanoen.2021.105959} {\bibfield  {journal}
  {\bibinfo  {journal} {Nano Energy}\ }\textbf {\bibinfo {volume} {85}},\
  \bibinfo {pages} {105959} (\bibinfo {year} {2021})}\BibitemShut {NoStop}%
\bibitem [{\citenamefont {Cardona}\ and\ \citenamefont
  {Guntherodt}(1982)}]{Cardona1982}%
  \BibitemOpen
  \bibfield  {author} {\bibinfo {author} {\bibfnamefont {M.}~\bibnamefont
  {Cardona}}\ and\ \bibinfo {author} {\bibfnamefont {G.}~\bibnamefont
  {Guntherodt}},\ }\href {https://doi.org/10.1016/0030-3992(77)90116-5} {\emph
  {\bibinfo {title} {{Light Scattering in Solids II - Basic Concepts and
  Instrumentation}}}}\ (\bibinfo  {publisher} {Springer-Verlag},\ \bibinfo
  {year} {1982})\ pp.\ \bibinfo {pages} {1--252}\BibitemShut {NoStop}%
\bibitem [{\citenamefont {Menahem}\ \emph {et~al.}(2023)\citenamefont
  {Menahem}, \citenamefont {Benshalom}, \citenamefont {Asher}, \citenamefont
  {Aharon}, \citenamefont {Korobko}, \citenamefont {Hellman},\ and\
  \citenamefont {Yaffe}}]{Menahem2023}%
  \BibitemOpen
  \bibfield  {author} {\bibinfo {author} {\bibfnamefont {M.}~\bibnamefont
  {Menahem}}, \bibinfo {author} {\bibfnamefont {N.}~\bibnamefont {Benshalom}},
  \bibinfo {author} {\bibfnamefont {M.}~\bibnamefont {Asher}}, \bibinfo
  {author} {\bibfnamefont {S.}~\bibnamefont {Aharon}}, \bibinfo {author}
  {\bibfnamefont {R.}~\bibnamefont {Korobko}}, \bibinfo {author} {\bibfnamefont
  {O.}~\bibnamefont {Hellman}},\ and\ \bibinfo {author} {\bibfnamefont
  {O.}~\bibnamefont {Yaffe}},\ }\bibfield  {title} {\bibinfo {title} {{Disorder
  origin of Raman scattering in perovskite single crystals}},\ }\href
  {https://doi.org/10.1103/PhysRevMaterials.7.044602} {\bibfield  {journal}
  {\bibinfo  {journal} {Physical Review Materials}\ }\textbf {\bibinfo {volume}
  {7}},\ \bibinfo {pages} {044602} (\bibinfo {year} {2023})}\BibitemShut
  {NoStop}%
\bibitem [{\citenamefont {Perry}\ \emph {et~al.}(1967)\citenamefont {Perry},
  \citenamefont {Fertel},\ and\ \citenamefont {McNelly}}]{Perry1967}%
  \BibitemOpen
  \bibfield  {author} {\bibinfo {author} {\bibfnamefont {C.~H.}\ \bibnamefont
  {Perry}}, \bibinfo {author} {\bibfnamefont {J.~H.}\ \bibnamefont {Fertel}},\
  and\ \bibinfo {author} {\bibfnamefont {T.~F.}\ \bibnamefont {McNelly}},\
  }\bibfield  {title} {\bibinfo {title} {{Temperature dependence of the Raman
  spectrum of SrTiO$_3$ and KTaO$_3$}},\ }\href
  {https://doi.org/10.1063/1.1712142} {\bibfield  {journal} {\bibinfo
  {journal} {The Journal of Chemical Physics}\ }\textbf {\bibinfo {volume}
  {47}},\ \bibinfo {pages} {1619} (\bibinfo {year} {1967})}\BibitemShut
  {NoStop}%
\bibitem [{\citenamefont {McMillan}\ and\ \citenamefont
  {Ross}(1988)}]{McMillan1988}%
  \BibitemOpen
  \bibfield  {author} {\bibinfo {author} {\bibfnamefont {P.}~\bibnamefont
  {McMillan}}\ and\ \bibinfo {author} {\bibfnamefont {N.}~\bibnamefont
  {Ross}},\ }\bibfield  {title} {\bibinfo {title} {{The Raman spectra of
  several orthorhombic calcium oxide perovskites}},\ }\href
  {https://doi.org/10.1007/BF00201326} {\bibfield  {journal} {\bibinfo
  {journal} {Physics and Chemistry of Minerals}\ }\textbf {\bibinfo {volume}
  {16}},\ \bibinfo {pages} {21} (\bibinfo {year} {1988})}\BibitemShut {NoStop}%
\bibitem [{\citenamefont {Ouillon}\ \emph {et~al.}(2002)\citenamefont
  {Ouillon}, \citenamefont {Pinan-Lucarre}, \citenamefont {Ranson},
  \citenamefont {Pruzan}, \citenamefont {Mishra}, \citenamefont {Ranjan},\ and\
  \citenamefont {Pandey}}]{ROuillon2002}%
  \BibitemOpen
  \bibfield  {author} {\bibinfo {author} {\bibfnamefont {R.}~\bibnamefont
  {Ouillon}}, \bibinfo {author} {\bibfnamefont {J.-P.}\ \bibnamefont
  {Pinan-Lucarre}}, \bibinfo {author} {\bibfnamefont {P.}~\bibnamefont
  {Ranson}}, \bibinfo {author} {\bibfnamefont {P.}~\bibnamefont {Pruzan}},
  \bibinfo {author} {\bibfnamefont {S.~K.}\ \bibnamefont {Mishra}}, \bibinfo
  {author} {\bibfnamefont {R.}~\bibnamefont {Ranjan}},\ and\ \bibinfo {author}
  {\bibfnamefont {D.}~\bibnamefont {Pandey}},\ }\bibfield  {title} {\bibinfo
  {title} {{A Raman scattering study of the phase transitions in SrTiO$_3$ and
  in the mixed system (Sr$_{1-x}$Ca$_x$)TiO$_3$ at ambient pressure from T =
  300~K down to 8~K}},\ }\href {https://doi.org/10.1088/0953-8984/14/8/333}
  {\bibfield  {journal} {\bibinfo  {journal} {Journal of Physics: Condensed
  Matter}\ }\textbf {\bibinfo {volume} {14}},\ \bibinfo {pages} {2079}
  (\bibinfo {year} {2002})}\BibitemShut {NoStop}%
\bibitem [{\citenamefont {Reuveni}\ \emph {et~al.}(2023)\citenamefont
  {Reuveni}, \citenamefont {Diskin-Posner}, \citenamefont {Gehrmann},
  \citenamefont {Godse}, \citenamefont {Gkikas}, \citenamefont {Buchine},
  \citenamefont {Aharon}, \citenamefont {Korobko}, \citenamefont {Stoumpos},
  \citenamefont {Egger},\ and\ \citenamefont {Yaffe}}]{Reuveni2023}%
  \BibitemOpen
  \bibfield  {author} {\bibinfo {author} {\bibfnamefont {G.}~\bibnamefont
  {Reuveni}}, \bibinfo {author} {\bibfnamefont {Y.}~\bibnamefont
  {Diskin-Posner}}, \bibinfo {author} {\bibfnamefont {C.}~\bibnamefont
  {Gehrmann}}, \bibinfo {author} {\bibfnamefont {S.}~\bibnamefont {Godse}},
  \bibinfo {author} {\bibfnamefont {G.~G.}\ \bibnamefont {Gkikas}}, \bibinfo
  {author} {\bibfnamefont {I.}~\bibnamefont {Buchine}}, \bibinfo {author}
  {\bibfnamefont {S.}~\bibnamefont {Aharon}}, \bibinfo {author} {\bibfnamefont
  {R.}~\bibnamefont {Korobko}}, \bibinfo {author} {\bibfnamefont {C.~C.}\
  \bibnamefont {Stoumpos}}, \bibinfo {author} {\bibfnamefont {D.~A.}\
  \bibnamefont {Egger}},\ and\ \bibinfo {author} {\bibfnamefont
  {O.}~\bibnamefont {Yaffe}},\ }\bibfield  {title} {\bibinfo {title} {{Static
  and Dynamic Disorder in Formamidinium Lead Bromide Single Crystals}},\ }\href
  {https://doi.org/10.1021/acs.jpclett.2c03337} {\bibfield  {journal} {\bibinfo
   {journal} {The Journal of Physical Chemistry Letters}\ }\textbf {\bibinfo
  {volume} {14}},\ \bibinfo {pages} {1288} (\bibinfo {year}
  {2023})}\BibitemShut {NoStop}%
\bibitem [{\citenamefont {Born}\ and\ \citenamefont
  {Bradburn}(1947)}]{Born1947}%
  \BibitemOpen
  \bibfield  {author} {\bibinfo {author} {\bibfnamefont {M.}~\bibnamefont
  {Born}}\ and\ \bibinfo {author} {\bibfnamefont {M.}~\bibnamefont
  {Bradburn}},\ }\bibfield  {title} {\bibinfo {title} {{The theory of the Raman
  effect in crystals, in particular rocksalt.}},\ }\href
  {https://doi.org/10.1098/rspa.1947.0002} {\bibfield  {journal} {\bibinfo
  {journal} {Proceedings of the Royal Society A}\ }\textbf {\bibinfo {volume}
  {188}},\ \bibinfo {pages} {161} (\bibinfo {year} {1947})}\BibitemShut
  {NoStop}%
\bibitem [{\citenamefont {{Y. Yu}}\ and\ \citenamefont
  {Cardona}(2010)}]{Y.Yu2010}%
  \BibitemOpen
  \bibfield  {author} {\bibinfo {author} {\bibfnamefont {P.}~\bibnamefont {{Y.
  Yu}}}\ and\ \bibinfo {author} {\bibfnamefont {M.}~\bibnamefont {Cardona}},\
  }\href@noop {} {\emph {\bibinfo {title} {{Fundamentals of Semiconductors,
  Physics and Materials Properties}}}},\ \bibinfo {edition} {4th}\ ed.\
  (\bibinfo  {publisher} {Springer},\ \bibinfo {address} {Berlin/Heidelberg},\
  \bibinfo {year} {2010})\ p.\ \bibinfo {pages} {775}\BibitemShut {NoStop}%
\bibitem [{\citenamefont {Kwok}(1968)}]{Kwok1968}%
  \BibitemOpen
  \bibfield  {author} {\bibinfo {author} {\bibfnamefont {P.~C.}\ \bibnamefont
  {Kwok}},\ }\bibfield  {title} {\bibinfo {title} {{Green's Function Method in
  Lattice Dynamics}},\ }in\ \href
  {https://doi.org/10.1016/S0081-1947(08)60219-2} {\emph {\bibinfo {booktitle}
  {Solid State Phys.}}}\ (\bibinfo  {publisher} {Academic Press},\ \bibinfo
  {address} {Cambridge, MA, USA},\ \bibinfo {year} {1968})\ pp.\ \bibinfo
  {pages} {213--303}\BibitemShut {NoStop}%
\bibitem [{\citenamefont {Maradudin}\ \emph {et~al.}(1971)\citenamefont
  {Maradudin}, \citenamefont {Montroll}, \citenamefont {Weiss},\ and\
  \citenamefont {Ipatova}}]{Maradudin1971}%
  \BibitemOpen
  \bibfield  {author} {\bibinfo {author} {\bibfnamefont {A.~A.}\ \bibnamefont
  {Maradudin}}, \bibinfo {author} {\bibfnamefont {E.~W.}\ \bibnamefont
  {Montroll}}, \bibinfo {author} {\bibfnamefont {G.~H.}\ \bibnamefont
  {Weiss}},\ and\ \bibinfo {author} {\bibfnamefont {I.~P.}\ \bibnamefont
  {Ipatova}},\ }\bibinfo {title} {{Theory of Lattice Dynamics in the Harmonic
  Approximation}}\ (\bibinfo  {publisher} {Academic Press},\ \bibinfo {address}
  {New York, USA},\ \bibinfo {year} {1971})\ pp.\ \bibinfo {pages} {1--67},\
  \bibinfo {edition} {2nd}\ ed.\BibitemShut {Stop}%
\bibitem [{\citenamefont {Safran}\ \emph {et~al.}(1977)\citenamefont {Safran},
  \citenamefont {Dresselhaus},\ and\ \citenamefont {Lax}}]{Safran1977}%
  \BibitemOpen
  \bibfield  {author} {\bibinfo {author} {\bibfnamefont {S.~A.}\ \bibnamefont
  {Safran}}, \bibinfo {author} {\bibfnamefont {G.}~\bibnamefont
  {Dresselhaus}},\ and\ \bibinfo {author} {\bibfnamefont {B.}~\bibnamefont
  {Lax}},\ }\bibfield  {title} {\bibinfo {title} {{Theory of spin-disorder
  Raman scattering in magnetic semiconductors}},\ }\href
  {https://doi.org/https://doi.org/10.1103/PhysRevB.16.2749} {\bibfield
  {journal} {\bibinfo  {journal} {Phys. Rev. B}\ }\textbf {\bibinfo {volume}
  {16}},\ \bibinfo {pages} {2749} (\bibinfo {year} {1977})}\BibitemShut
  {NoStop}%
\bibitem [{\citenamefont {Haro}\ \emph {et~al.}(1986)\citenamefont {Haro},
  \citenamefont {Balkanski}, \citenamefont {Wallis},\ and\ \citenamefont
  {Wanser}}]{Haro1986}%
  \BibitemOpen
  \bibfield  {author} {\bibinfo {author} {\bibfnamefont {E.}~\bibnamefont
  {Haro}}, \bibinfo {author} {\bibfnamefont {M.}~\bibnamefont {Balkanski}},
  \bibinfo {author} {\bibfnamefont {R.~F.}\ \bibnamefont {Wallis}},\ and\
  \bibinfo {author} {\bibfnamefont {K.~H.}\ \bibnamefont {Wanser}},\ }\bibfield
   {title} {\bibinfo {title} {{Theory of the anharmonic damping and shift of
  the Raman mode in silicon}},\ }\href
  {https://doi.org/10.1103/PhysRevB.34.5358} {\bibfield  {journal} {\bibinfo
  {journal} {Physical Review B}\ }\textbf {\bibinfo {volume} {34}},\ \bibinfo
  {pages} {5358} (\bibinfo {year} {1986})}\BibitemShut {NoStop}%
\bibitem [{\citenamefont {Men{\'{e}}ndez}\ and\ \citenamefont
  {Cardona}(1984)}]{Menendez1984}%
  \BibitemOpen
  \bibfield  {author} {\bibinfo {author} {\bibfnamefont {J.}~\bibnamefont
  {Men{\'{e}}ndez}}\ and\ \bibinfo {author} {\bibfnamefont {M.}~\bibnamefont
  {Cardona}},\ }\bibfield  {title} {\bibinfo {title} {{Temperature dependence
  of the first-order Raman scattering by phonons in Si, Ge, and $\alpha$ -- Sn
  : Anharmonic effects}},\ }\href {https://doi.org/10.1103/PhysRevB.29.2051}
  {\bibfield  {journal} {\bibinfo  {journal} {Physical Review B}\ }\textbf
  {\bibinfo {volume} {29}},\ \bibinfo {pages} {2051} (\bibinfo {year}
  {1984})}\BibitemShut {NoStop}%
\bibitem [{\citenamefont {Kirchartz}\ \emph {et~al.}(2018)\citenamefont
  {Kirchartz}, \citenamefont {Markvart}, \citenamefont {Rau},\ and\
  \citenamefont {Egger}}]{Kirchartz2018}%
  \BibitemOpen
  \bibfield  {author} {\bibinfo {author} {\bibfnamefont {T.}~\bibnamefont
  {Kirchartz}}, \bibinfo {author} {\bibfnamefont {T.}~\bibnamefont {Markvart}},
  \bibinfo {author} {\bibfnamefont {U.}~\bibnamefont {Rau}},\ and\ \bibinfo
  {author} {\bibfnamefont {D.~A.}\ \bibnamefont {Egger}},\ }\bibfield  {title}
  {\bibinfo {title} {{Impact of Small Phonon Energies on the Charge-Carrier
  Lifetimes in Metal-Halide Perovskites}},\ }\href
  {https://doi.org/10.1021/acs.jpclett.7b03414} {\bibfield  {journal} {\bibinfo
   {journal} {Journal of Physical Chemistry Letters}\ }\textbf {\bibinfo
  {volume} {9}},\ \bibinfo {pages} {939} (\bibinfo {year} {2018})}\BibitemShut
  {NoStop}%
\bibitem [{\citenamefont {Lang}\ and\ \citenamefont
  {Logan}(1977)}]{lang_large-lattice-relaxation_1977}%
  \BibitemOpen
  \bibfield  {author} {\bibinfo {author} {\bibfnamefont {D.~V.}\ \bibnamefont
  {Lang}}\ and\ \bibinfo {author} {\bibfnamefont {R.~A.}\ \bibnamefont
  {Logan}},\ }\bibfield  {title} {\bibinfo {title}
  {Large-{Lattice}-{Relaxation} {Model} for {Persistent} {Photoconductivity} in
  {Compound} {Semiconductors}},\ }\href
  {https://doi.org/10.1103/PhysRevLett.39.635} {\bibfield  {journal} {\bibinfo
  {journal} {Physical Review Letters}\ }\textbf {\bibinfo {volume} {39}},\
  \bibinfo {pages} {635} (\bibinfo {year} {1977})}\BibitemShut {NoStop}%
\bibitem [{\citenamefont {Mooney}(1990)}]{mooney_deep_1990}%
  \BibitemOpen
  \bibfield  {author} {\bibinfo {author} {\bibfnamefont {P.~M.}\ \bibnamefont
  {Mooney}},\ }\bibfield  {title} {\bibinfo {title} {Deep donor levels ({DX}
  centers) in {III}‐{V} semiconductors},\ }\href
  {https://doi.org/10.1063/1.345628} {\bibfield  {journal} {\bibinfo  {journal}
  {Journal of Applied Physics}\ }\textbf {\bibinfo {volume} {67}},\ \bibinfo
  {pages} {R1} (\bibinfo {year} {1990})}\BibitemShut {NoStop}%
\bibitem [{\citenamefont {Yin}\ \emph {et~al.}(2018)\citenamefont {Yin},
  \citenamefont {Akey},\ and\ \citenamefont {Jaramillo}}]{yin_large_2018}%
  \BibitemOpen
  \bibfield  {author} {\bibinfo {author} {\bibfnamefont {H.}~\bibnamefont
  {Yin}}, \bibinfo {author} {\bibfnamefont {A.}~\bibnamefont {Akey}},\ and\
  \bibinfo {author} {\bibfnamefont {R.}~\bibnamefont {Jaramillo}},\ }\bibfield
  {title} {\bibinfo {title} {Large and persistent photoconductivity due to
  hole-hole correlation in {CdS}},\ }\href
  {https://doi.org/10.1103/PhysRevMaterials.2.084602} {\bibfield  {journal}
  {\bibinfo  {journal} {Physical Review Materials}\ }\textbf {\bibinfo {volume}
  {2}},\ \bibinfo {pages} {084602} (\bibinfo {year} {2018})}\BibitemShut
  {NoStop}%
\bibitem [{\citenamefont {Yin}\ \emph {et~al.}(2021)\citenamefont {Yin},
  \citenamefont {Kumar}, \citenamefont {LeBeau},\ and\ \citenamefont
  {Jaramillo}}]{yin_defect-level_2021}%
  \BibitemOpen
  \bibfield  {author} {\bibinfo {author} {\bibfnamefont {H.}~\bibnamefont
  {Yin}}, \bibinfo {author} {\bibfnamefont {A.}~\bibnamefont {Kumar}}, \bibinfo
  {author} {\bibfnamefont {J.~M.}\ \bibnamefont {LeBeau}},\ and\ \bibinfo
  {author} {\bibfnamefont {R.}~\bibnamefont {Jaramillo}},\ }\bibfield  {title}
  {\bibinfo {title} {Defect-{Level} {Switching} for {Highly} {Nonlinear} and
  {Hysteretic} {Electronic} {Devices}},\ }\href
  {https://doi.org/10.1103/PhysRevApplied.15.014014} {\bibfield  {journal}
  {\bibinfo  {journal} {Physical Review Applied}\ }\textbf {\bibinfo {volume}
  {15}},\ \bibinfo {pages} {014014} (\bibinfo {year} {2021})}\BibitemShut
  {NoStop}%
\bibitem [{\citenamefont {Liu}\ \emph {et~al.}(2023)\citenamefont {Liu},
  \citenamefont {Peters}, \citenamefont {Bayikadi}, \citenamefont {Klepov},
  \citenamefont {Pan}, \citenamefont {Pandey}, \citenamefont {Kanatzidis},\
  and\ \citenamefont {Wessels}}]{Liu2023}%
  \BibitemOpen
  \bibfield  {author} {\bibinfo {author} {\bibfnamefont {Z.}~\bibnamefont
  {Liu}}, \bibinfo {author} {\bibfnamefont {J.~A.}\ \bibnamefont {Peters}},
  \bibinfo {author} {\bibfnamefont {K.~S.}\ \bibnamefont {Bayikadi}}, \bibinfo
  {author} {\bibfnamefont {V.}~\bibnamefont {Klepov}}, \bibinfo {author}
  {\bibfnamefont {L.}~\bibnamefont {Pan}}, \bibinfo {author} {\bibfnamefont
  {I.~R.}\ \bibnamefont {Pandey}}, \bibinfo {author} {\bibfnamefont {M.~G.}\
  \bibnamefont {Kanatzidis}},\ and\ \bibinfo {author} {\bibfnamefont {B.~W.}\
  \bibnamefont {Wessels}},\ }\bibfield  {title} {\bibinfo {title} {{Defects of
  perovskite semiconductor CsPbBr3 investigated via photoluminescence and
  thermally stimulated current spectroscopies}},\ }\href
  {https://doi.org/10.1063/5.0177809} {\bibfield  {journal} {\bibinfo
  {journal} {Journal of Applied Physics}\ }\textbf {\bibinfo {volume} {134}},\
  \bibinfo {pages} {245101} (\bibinfo {year} {2023})}\BibitemShut {NoStop}%
\bibitem [{\citenamefont {Cohen}\ \emph {et~al.}(2019)\citenamefont {Cohen},
  \citenamefont {Egger}, \citenamefont {Rappe},\ and\ \citenamefont
  {Kronik}}]{Cohen2019}%
  \BibitemOpen
  \bibfield  {author} {\bibinfo {author} {\bibfnamefont {A.~V.}\ \bibnamefont
  {Cohen}}, \bibinfo {author} {\bibfnamefont {D.~A.}\ \bibnamefont {Egger}},
  \bibinfo {author} {\bibfnamefont {A.~M.}\ \bibnamefont {Rappe}},\ and\
  \bibinfo {author} {\bibfnamefont {L.}~\bibnamefont {Kronik}},\ }\bibfield
  {title} {\bibinfo {title} {{Breakdown of the Static Picture of Defect
  Energetics in Halide Perovskites: The Case of the Br Vacancy in CsPbBr3}},\
  }\href {https://doi.org/10.1021/acs.jpclett.9b01855} {\bibfield  {journal}
  {\bibinfo  {journal} {Journal of Physical Chemistry Letters}\ }\textbf
  {\bibinfo {volume} {10}},\ \bibinfo {pages} {4490} (\bibinfo {year}
  {2019})}\BibitemShut {NoStop}%
\bibitem [{\citenamefont {Kang}\ and\ \citenamefont {Wang}(2017)}]{Kang2017}%
  \BibitemOpen
  \bibfield  {author} {\bibinfo {author} {\bibfnamefont {J.}~\bibnamefont
  {Kang}}\ and\ \bibinfo {author} {\bibfnamefont {L.~W.}\ \bibnamefont
  {Wang}},\ }\bibfield  {title} {\bibinfo {title} {{High Defect Tolerance in
  Lead Halide Perovskite CsPbBr3}},\ }\href
  {https://doi.org/10.1021/acs.jpclett.6b02800} {\bibfield  {journal} {\bibinfo
   {journal} {Journal of Physical Chemistry Letters}\ }\textbf {\bibinfo
  {volume} {8}},\ \bibinfo {pages} {489} (\bibinfo {year} {2017})}\BibitemShut
  {NoStop}%
\bibitem [{\citenamefont {Sebastian}\ \emph {et~al.}(2015)\citenamefont
  {Sebastian}, \citenamefont {Peters}, \citenamefont {Stoumpos}, \citenamefont
  {Im}, \citenamefont {Kostina}, \citenamefont {Liu}, \citenamefont
  {Kanatzidis}, \citenamefont {Freeman},\ and\ \citenamefont
  {Wessels}}]{Sebastian2015}%
  \BibitemOpen
  \bibfield  {author} {\bibinfo {author} {\bibfnamefont {M.}~\bibnamefont
  {Sebastian}}, \bibinfo {author} {\bibfnamefont {J.~A.}\ \bibnamefont
  {Peters}}, \bibinfo {author} {\bibfnamefont {C.~C.}\ \bibnamefont
  {Stoumpos}}, \bibinfo {author} {\bibfnamefont {J.}~\bibnamefont {Im}},
  \bibinfo {author} {\bibfnamefont {S.~S.}\ \bibnamefont {Kostina}}, \bibinfo
  {author} {\bibfnamefont {Z.}~\bibnamefont {Liu}}, \bibinfo {author}
  {\bibfnamefont {M.~G.}\ \bibnamefont {Kanatzidis}}, \bibinfo {author}
  {\bibfnamefont {A.~J.}\ \bibnamefont {Freeman}},\ and\ \bibinfo {author}
  {\bibfnamefont {B.~W.}\ \bibnamefont {Wessels}},\ }\bibfield  {title}
  {\bibinfo {title} {{Excitonic emissions and above-band-gap luminescence in
  the single-crystal perovskite semiconductors CsPbB r3 and CsPbC l3}},\ }\href
  {https://doi.org/10.1103/PhysRevB.92.235210} {\bibfield  {journal} {\bibinfo
  {journal} {Physical Review B - Condensed Matter and Materials Physics}\
  }\textbf {\bibinfo {volume} {92}},\ \bibinfo {pages} {235210} (\bibinfo
  {year} {2015})}\BibitemShut {NoStop}%
\bibitem [{\citenamefont {Zhao}\ \emph {et~al.}(2023)\citenamefont {Zhao},
  \citenamefont {Chen}, \citenamefont {Ahsan}, \citenamefont {Hou},
  \citenamefont {Hoglund}, \citenamefont {Singh}, \citenamefont {Zhao},
  \citenamefont {Htoon}, \citenamefont {Krayev}, \citenamefont
  {Shanmugasundaram}, \citenamefont {Hopkins}, \citenamefont {Seidel},
  \citenamefont {Kapadia},\ and\ \citenamefont
  {Ravichandran}}]{zhao_photoconductive_2023}%
  \BibitemOpen
  \bibfield  {author} {\bibinfo {author} {\bibfnamefont {B.}~\bibnamefont
  {Zhao}}, \bibinfo {author} {\bibfnamefont {H.}~\bibnamefont {Chen}}, \bibinfo
  {author} {\bibfnamefont {R.}~\bibnamefont {Ahsan}}, \bibinfo {author}
  {\bibfnamefont {F.}~\bibnamefont {Hou}}, \bibinfo {author} {\bibfnamefont
  {E.~R.}\ \bibnamefont {Hoglund}}, \bibinfo {author} {\bibfnamefont
  {S.}~\bibnamefont {Singh}}, \bibinfo {author} {\bibfnamefont
  {H.}~\bibnamefont {Zhao}}, \bibinfo {author} {\bibfnamefont {H.}~\bibnamefont
  {Htoon}}, \bibinfo {author} {\bibfnamefont {A.}~\bibnamefont {Krayev}},
  \bibinfo {author} {\bibfnamefont {M.}~\bibnamefont {Shanmugasundaram}},
  \bibinfo {author} {\bibfnamefont {P.~E.}\ \bibnamefont {Hopkins}}, \bibinfo
  {author} {\bibfnamefont {J.}~\bibnamefont {Seidel}}, \bibinfo {author}
  {\bibfnamefont {R.}~\bibnamefont {Kapadia}},\ and\ \bibinfo {author}
  {\bibfnamefont {J.}~\bibnamefont {Ravichandran}},\ }\href
  {https://doi.org/10.48550/arXiv.2310.03198} {\bibinfo {title}
  {Photoconductive {Effects} in {Single} {Crystals} of {BaZrS}\$\_3\$}}
  (\bibinfo {year} {2023})\BibitemShut {NoStop}%
\bibitem [{\citenamefont {Emin}(2018)}]{emin_barrier_2018}%
  \BibitemOpen
  \bibfield  {author} {\bibinfo {author} {\bibfnamefont {D.}~\bibnamefont
  {Emin}},\ }\bibfield  {title} {\bibinfo {title} {Barrier to recombination of
  oppositely charged large polarons},\ }\href
  {https://doi.org/10.1063/1.5019834} {\bibfield  {journal} {\bibinfo
  {journal} {Journal of Applied Physics}\ }\textbf {\bibinfo {volume} {123}},\
  \bibinfo {pages} {055105} (\bibinfo {year} {2018})}\BibitemShut {NoStop}%
\bibitem [{\citenamefont {Franchini}\ \emph {et~al.}(2021)\citenamefont
  {Franchini}, \citenamefont {Reticcioli}, \citenamefont {Setvin},\ and\
  \citenamefont {Diebold}}]{franchini_polarons_2021}%
  \BibitemOpen
  \bibfield  {author} {\bibinfo {author} {\bibfnamefont {C.}~\bibnamefont
  {Franchini}}, \bibinfo {author} {\bibfnamefont {M.}~\bibnamefont
  {Reticcioli}}, \bibinfo {author} {\bibfnamefont {M.}~\bibnamefont {Setvin}},\
  and\ \bibinfo {author} {\bibfnamefont {U.}~\bibnamefont {Diebold}},\
  }\bibfield  {title} {\bibinfo {title} {Polarons in materials},\ }\href
  {https://doi.org/10.1038/s41578-021-00289-w} {\bibfield  {journal} {\bibinfo
  {journal} {Nature Reviews Materials}\ }\textbf {\bibinfo {volume} {6}},\
  \bibinfo {pages} {560} (\bibinfo {year} {2021})}\BibitemShut {NoStop}%
\bibitem [{\citenamefont {Brenner}\ \emph {et~al.}(2020)\citenamefont
  {Brenner}, \citenamefont {Gehrmann}, \citenamefont {Korobko}, \citenamefont
  {Livneh}, \citenamefont {Egger},\ and\ \citenamefont {Yaffe}}]{Brenner2020}%
  \BibitemOpen
  \bibfield  {author} {\bibinfo {author} {\bibfnamefont {T.~M.}\ \bibnamefont
  {Brenner}}, \bibinfo {author} {\bibfnamefont {C.}~\bibnamefont {Gehrmann}},
  \bibinfo {author} {\bibfnamefont {R.}~\bibnamefont {Korobko}}, \bibinfo
  {author} {\bibfnamefont {T.}~\bibnamefont {Livneh}}, \bibinfo {author}
  {\bibfnamefont {D.~A.}\ \bibnamefont {Egger}},\ and\ \bibinfo {author}
  {\bibfnamefont {O.}~\bibnamefont {Yaffe}},\ }\bibfield  {title} {\bibinfo
  {title} {{Anharmonic host-lattice dynamics enable fast ion conduction in
  superionic AgI}},\ }\href {https://doi.org/10.1103/PhysRevMaterials.4.115402}
  {\bibfield  {journal} {\bibinfo  {journal} {Physical Review Materials}\
  }\textbf {\bibinfo {volume} {4}},\ \bibinfo {pages} {115402} (\bibinfo {year}
  {2020})}\BibitemShut {NoStop}%
\bibitem [{\citenamefont {Limmer}\ and\ \citenamefont
  {Ginsberg}(2020)}]{LimmerJCP2020}%
  \BibitemOpen
  \bibfield  {author} {\bibinfo {author} {\bibfnamefont {D.~T.}\ \bibnamefont
  {Limmer}}\ and\ \bibinfo {author} {\bibfnamefont {N.~S.}\ \bibnamefont
  {Ginsberg}},\ }\bibfield  {title} {\bibinfo {title} {{Photoinduced phase
  separation in the lead halides is a polaronic effect}},\ }\bibfield
  {journal} {\bibinfo  {journal} {The Journal of Chemical Physics}\ }\textbf
  {\bibinfo {volume} {152}},\ \href {https://doi.org/10.1063/1.5144291}
  {10.1063/1.5144291} (\bibinfo {year} {2020})\BibitemShut {NoStop}%
\bibitem [{\citenamefont {Zhao}\ \emph {et~al.}(2016)\citenamefont {Zhao},
  \citenamefont {Zhou}, \citenamefont {Zhou}, \citenamefont {Liu},
  \citenamefont {Yu},\ and\ \citenamefont {Zhao}}]{Zhao2016}%
  \BibitemOpen
  \bibfield  {author} {\bibinfo {author} {\bibfnamefont {Y.-C.}\ \bibnamefont
  {Zhao}}, \bibinfo {author} {\bibfnamefont {W.-K.}\ \bibnamefont {Zhou}},
  \bibinfo {author} {\bibfnamefont {X.}~\bibnamefont {Zhou}}, \bibinfo {author}
  {\bibfnamefont {K.-H.}\ \bibnamefont {Liu}}, \bibinfo {author} {\bibfnamefont
  {D.-P.}\ \bibnamefont {Yu}},\ and\ \bibinfo {author} {\bibfnamefont
  {Q.}~\bibnamefont {Zhao}},\ }\bibfield  {title} {\bibinfo {title}
  {{Quantification of light-enhanced ionic transport in lead iodide perovskite
  thin films and its solar cell applications}},\ }\bibfield  {journal}
  {\bibinfo  {journal} {Science {\&} Applications}\ }\textbf {\bibinfo {volume}
  {6}},\ \href {https://doi.org/10.1038/lsa.2016.243} {10.1038/lsa.2016.243}
  (\bibinfo {year} {2016})\BibitemShut {NoStop}%
\bibitem [{\citenamefont {Aharon}\ \emph {et~al.}(2022)\citenamefont {Aharon},
  \citenamefont {Ceratti}, \citenamefont {Jasti}, \citenamefont {Cremonesi},
  \citenamefont {Feldman}, \citenamefont {Potenza}, \citenamefont {Hodes},\
  and\ \citenamefont {Cahen}}]{Aharon2022}%
  \BibitemOpen
  \bibfield  {author} {\bibinfo {author} {\bibfnamefont {S.}~\bibnamefont
  {Aharon}}, \bibinfo {author} {\bibfnamefont {D.~R.}\ \bibnamefont {Ceratti}},
  \bibinfo {author} {\bibfnamefont {N.~P.}\ \bibnamefont {Jasti}}, \bibinfo
  {author} {\bibfnamefont {L.}~\bibnamefont {Cremonesi}}, \bibinfo {author}
  {\bibfnamefont {Y.}~\bibnamefont {Feldman}}, \bibinfo {author} {\bibfnamefont
  {M.~A.~C.}\ \bibnamefont {Potenza}}, \bibinfo {author} {\bibfnamefont
  {G.}~\bibnamefont {Hodes}},\ and\ \bibinfo {author} {\bibfnamefont
  {D.}~\bibnamefont {Cahen}},\ }\bibfield  {title} {\bibinfo {title} {{2D
  Pb‐Halide Perovskites Can Self‐Heal Photodamage Better than 3D Ones}},\
  }\bibfield  {journal} {\bibinfo  {journal} {Advanced Functional Materials}\
  }\textbf {\bibinfo {volume} {32}},\ \href
  {https://doi.org/10.1002/adfm.202113354} {10.1002/adfm.202113354} (\bibinfo
  {year} {2022})\BibitemShut {NoStop}%
\bibitem [{\citenamefont {Finkenauer}\ \emph {et~al.}(2022)\citenamefont
  {Finkenauer}, \citenamefont {Akriti}, \citenamefont {Ma},\ and\ \citenamefont
  {Dou}}]{Finkenauer2022}%
  \BibitemOpen
  \bibfield  {author} {\bibinfo {author} {\bibfnamefont {B.~P.}\ \bibnamefont
  {Finkenauer}}, \bibinfo {author} {\bibnamefont {Akriti}}, \bibinfo {author}
  {\bibfnamefont {K.}~\bibnamefont {Ma}},\ and\ \bibinfo {author}
  {\bibfnamefont {L.}~\bibnamefont {Dou}},\ }\bibfield  {title} {\bibinfo
  {title} {{Degradation and Self-Healing in Perovskite Solar Cells}},\ }\href
  {https://doi.org/10.1021/ACSAMI.2C01925} {\bibfield  {journal} {\bibinfo
  {journal} {ACS Applied Materials and Interfaces}\ }\textbf {\bibinfo {volume}
  {14}},\ \bibinfo {pages} {24073} (\bibinfo {year} {2022})}\BibitemShut
  {NoStop}%
\bibitem [{\citenamefont {Ceratti}\ \emph {et~al.}(2018)\citenamefont
  {Ceratti}, \citenamefont {Rakita}, \citenamefont {Cremonesi}, \citenamefont
  {Tenne}, \citenamefont {Kalchenko}, \citenamefont {Elbaum}, \citenamefont
  {Oron}, \citenamefont {Potenza}, \citenamefont {Hodes},\ and\ \citenamefont
  {Cahen}}]{Ceratti2018}%
  \BibitemOpen
  \bibfield  {author} {\bibinfo {author} {\bibfnamefont {D.~R.}\ \bibnamefont
  {Ceratti}}, \bibinfo {author} {\bibfnamefont {Y.}~\bibnamefont {Rakita}},
  \bibinfo {author} {\bibfnamefont {L.}~\bibnamefont {Cremonesi}}, \bibinfo
  {author} {\bibfnamefont {R.}~\bibnamefont {Tenne}}, \bibinfo {author}
  {\bibfnamefont {V.}~\bibnamefont {Kalchenko}}, \bibinfo {author}
  {\bibfnamefont {M.}~\bibnamefont {Elbaum}}, \bibinfo {author} {\bibfnamefont
  {D.}~\bibnamefont {Oron}}, \bibinfo {author} {\bibfnamefont {M.~A.~C.}\
  \bibnamefont {Potenza}}, \bibinfo {author} {\bibfnamefont {G.}~\bibnamefont
  {Hodes}},\ and\ \bibinfo {author} {\bibfnamefont {D.}~\bibnamefont {Cahen}},\
  }\bibfield  {title} {\bibinfo {title} {{Self‐Healing Inside APbBr 3 Halide
  Perovskite Crystals}},\ }\bibfield  {journal} {\bibinfo  {journal} {Advanced
  Materials}\ }\textbf {\bibinfo {volume} {30}},\ \href
  {https://doi.org/10.1002/adma.201706273} {10.1002/adma.201706273} (\bibinfo
  {year} {2018})\BibitemShut {NoStop}%
\bibitem [{\citenamefont {Parida}\ \emph {et~al.}(2023)\citenamefont {Parida},
  \citenamefont {Kumar}, \citenamefont {Cherf}, \citenamefont {Aharon},
  \citenamefont {Cahen},\ and\ \citenamefont {Eren}}]{Parida2023}%
  \BibitemOpen
  \bibfield  {author} {\bibinfo {author} {\bibfnamefont {S.}~\bibnamefont
  {Parida}}, \bibinfo {author} {\bibfnamefont {S.}~\bibnamefont {Kumar}},
  \bibinfo {author} {\bibfnamefont {S.}~\bibnamefont {Cherf}}, \bibinfo
  {author} {\bibfnamefont {S.}~\bibnamefont {Aharon}}, \bibinfo {author}
  {\bibfnamefont {D.}~\bibnamefont {Cahen}},\ and\ \bibinfo {author}
  {\bibfnamefont {B.}~\bibnamefont {Eren}},\ }\bibfield  {title} {\bibinfo
  {title} {{Self‐Healing and ‐Repair of Nanomechanical Damages in Lead
  Halide Perovskites}},\ }\bibfield  {journal} {\bibinfo  {journal} {Advanced
  Functional Materials}\ }\textbf {\bibinfo {volume} {33}},\ \href
  {https://doi.org/10.1002/adfm.202304278} {10.1002/adfm.202304278} (\bibinfo
  {year} {2023})\BibitemShut {NoStop}%
\bibitem [{\citenamefont {Hutter}\ \emph {et~al.}(2015)\citenamefont {Hutter},
  \citenamefont {Eperon}, \citenamefont {Stranks},\ and\ \citenamefont
  {Savenije}}]{Hutter2015}%
  \BibitemOpen
  \bibfield  {author} {\bibinfo {author} {\bibfnamefont {E.~M.}\ \bibnamefont
  {Hutter}}, \bibinfo {author} {\bibfnamefont {G.~E.}\ \bibnamefont {Eperon}},
  \bibinfo {author} {\bibfnamefont {S.~D.}\ \bibnamefont {Stranks}},\ and\
  \bibinfo {author} {\bibfnamefont {T.~J.}\ \bibnamefont {Savenije}},\
  }\bibfield  {title} {\bibinfo {title} {{Charge Carriers in Planar and
  Meso-Structured Organic–Inorganic Perovskites: Mobilities, Lifetimes, and
  Concentrations of Trap States}},\ }\href
  {https://doi.org/10.1021/acs.jpclett.5b01361} {\bibfield  {journal} {\bibinfo
   {journal} {The Journal of Physical Chemistry Letters}\ }\textbf {\bibinfo
  {volume} {6}},\ \bibinfo {pages} {3082} (\bibinfo {year} {2015})}\BibitemShut
  {NoStop}%
\bibitem [{\citenamefont {Bi}\ \emph {et~al.}(2016)\citenamefont {Bi},
  \citenamefont {Hutter}, \citenamefont {Fang}, \citenamefont {Dong},
  \citenamefont {Huang},\ and\ \citenamefont {Savenije}}]{Bi2016}%
  \BibitemOpen
  \bibfield  {author} {\bibinfo {author} {\bibfnamefont {Y.}~\bibnamefont
  {Bi}}, \bibinfo {author} {\bibfnamefont {E.~M.}\ \bibnamefont {Hutter}},
  \bibinfo {author} {\bibfnamefont {Y.}~\bibnamefont {Fang}}, \bibinfo {author}
  {\bibfnamefont {Q.}~\bibnamefont {Dong}}, \bibinfo {author} {\bibfnamefont
  {J.}~\bibnamefont {Huang}},\ and\ \bibinfo {author} {\bibfnamefont {T.~J.}\
  \bibnamefont {Savenije}},\ }\bibfield  {title} {\bibinfo {title} {{Charge
  Carrier Lifetimes Exceeding 15 $\mu$s in Methylammonium Lead Iodide Single
  Crystals}},\ }\href {https://doi.org/10.1021/acs.jpclett.6b00269} {\bibfield
  {journal} {\bibinfo  {journal} {J. Phys. Chem. Lett}\ }\textbf {\bibinfo
  {volume} {7}},\ \bibinfo {pages} {45} (\bibinfo {year} {2016})}\BibitemShut
  {NoStop}%
\bibitem [{\citenamefont {Xing}\ \emph {et~al.}(2014)\citenamefont {Xing},
  \citenamefont {Mathews}, \citenamefont {Lim}, \citenamefont {Yantara},
  \citenamefont {Liu}, \citenamefont {Sabba}, \citenamefont {Gr{\"{a}}tzel},
  \citenamefont {Mhaisalkar},\ and\ \citenamefont {Sum}}]{Xing2014}%
  \BibitemOpen
  \bibfield  {author} {\bibinfo {author} {\bibfnamefont {G.}~\bibnamefont
  {Xing}}, \bibinfo {author} {\bibfnamefont {N.}~\bibnamefont {Mathews}},
  \bibinfo {author} {\bibfnamefont {S.~S.}\ \bibnamefont {Lim}}, \bibinfo
  {author} {\bibfnamefont {N.}~\bibnamefont {Yantara}}, \bibinfo {author}
  {\bibfnamefont {X.}~\bibnamefont {Liu}}, \bibinfo {author} {\bibfnamefont
  {D.}~\bibnamefont {Sabba}}, \bibinfo {author} {\bibfnamefont
  {M.}~\bibnamefont {Gr{\"{a}}tzel}}, \bibinfo {author} {\bibfnamefont
  {S.}~\bibnamefont {Mhaisalkar}},\ and\ \bibinfo {author} {\bibfnamefont
  {T.~C.}\ \bibnamefont {Sum}},\ }\bibfield  {title} {\bibinfo {title}
  {{Low-temperature solution-processed wavelength-tunable perovskites for
  lasing}},\ }\href {https://doi.org/10.1038/nmat3911} {\bibfield  {journal}
  {\bibinfo  {journal} {Nature Materials}\ }\textbf {\bibinfo {volume} {13}},\
  \bibinfo {pages} {476} (\bibinfo {year} {2014})}\BibitemShut {NoStop}%
\bibitem [{\citenamefont {Lei}\ \emph {et~al.}(2021)\citenamefont {Lei},
  \citenamefont {Xu}, \citenamefont {Wang}, \citenamefont {Zhu}, \citenamefont
  {Jin}, \citenamefont {Lei}, \citenamefont {Xu}, \citenamefont {Wang},
  \citenamefont {Jin},\ and\ \citenamefont {Zhu}}]{Lei2021}%
  \BibitemOpen
  \bibfield  {author} {\bibinfo {author} {\bibfnamefont {Y.}~\bibnamefont
  {Lei}}, \bibinfo {author} {\bibfnamefont {Y.}~\bibnamefont {Xu}}, \bibinfo
  {author} {\bibfnamefont {M.}~\bibnamefont {Wang}}, \bibinfo {author}
  {\bibfnamefont {G.}~\bibnamefont {Zhu}}, \bibinfo {author} {\bibfnamefont
  {Z.}~\bibnamefont {Jin}}, \bibinfo {author} {\bibfnamefont {Y.}~\bibnamefont
  {Lei}}, \bibinfo {author} {\bibfnamefont {Y.}~\bibnamefont {Xu}}, \bibinfo
  {author} {\bibfnamefont {M.}~\bibnamefont {Wang}}, \bibinfo {author}
  {\bibfnamefont {Z.}~\bibnamefont {Jin}},\ and\ \bibinfo {author}
  {\bibfnamefont {G.}~\bibnamefont {Zhu}},\ }\bibfield  {title} {\bibinfo
  {title} {{Origin, Influence, and Countermeasures of Defects in Perovskite
  Solar Cells}},\ }\href {https://doi.org/10.1002/SMLL.202005495} {\bibfield
  {journal} {\bibinfo  {journal} {Small}\ }\textbf {\bibinfo {volume} {17}},\
  \bibinfo {pages} {2005495} (\bibinfo {year} {2021})}\BibitemShut {NoStop}%
\bibitem [{\citenamefont {{Le Corre}}\ \emph {et~al.}(2021)\citenamefont {{Le
  Corre}}, \citenamefont {Duijnstee}, \citenamefont {{El Tambouli}},
  \citenamefont {Ball}, \citenamefont {Snaith}, \citenamefont {Lim},\ and\
  \citenamefont {Koster}}]{LeCorre2021}%
  \BibitemOpen
  \bibfield  {author} {\bibinfo {author} {\bibfnamefont {V.~M.}\ \bibnamefont
  {{Le Corre}}}, \bibinfo {author} {\bibfnamefont {E.~A.}\ \bibnamefont
  {Duijnstee}}, \bibinfo {author} {\bibfnamefont {O.}~\bibnamefont {{El
  Tambouli}}}, \bibinfo {author} {\bibfnamefont {J.~M.}\ \bibnamefont {Ball}},
  \bibinfo {author} {\bibfnamefont {H.~J.}\ \bibnamefont {Snaith}}, \bibinfo
  {author} {\bibfnamefont {J.}~\bibnamefont {Lim}},\ and\ \bibinfo {author}
  {\bibfnamefont {L.~J.~A.}\ \bibnamefont {Koster}},\ }\bibfield  {title}
  {\bibinfo {title} {{Revealing Charge Carrier Mobility and Defect Densities in
  Metal Halide Perovskites via Space-Charge-Limited Current Measurements}},\
  }\href {https://doi.org/10.1021/acsenergylett.0c02599} {\bibfield  {journal}
  {\bibinfo  {journal} {ACS Energy Letters}\ }\textbf {\bibinfo {volume} {6}},\
  \bibinfo {pages} {1087} (\bibinfo {year} {2021})}\BibitemShut {NoStop}%
\bibitem [{\citenamefont {Herz}(2016)}]{Herz2016}%
  \BibitemOpen
  \bibfield  {author} {\bibinfo {author} {\bibfnamefont {L.~M.}\ \bibnamefont
  {Herz}},\ }\bibfield  {title} {\bibinfo {title} {{Charge-Carrier Dynamics in
  Organic-Inorganic Metal Halide Perovskites}},\ }\href
  {https://doi.org/10.1146/annurev-physchem-040215-112222} {\bibfield
  {journal} {\bibinfo  {journal} {Annual Review of Physical Chemistry}\
  }\textbf {\bibinfo {volume} {67}},\ \bibinfo {pages} {65} (\bibinfo {year}
  {2016})}\BibitemShut {NoStop}%
\bibitem [{\citenamefont {Rosenberg}\ \emph {et~al.}(2017)\citenamefont
  {Rosenberg}, \citenamefont {Legodi}, \citenamefont {Rakita}, \citenamefont
  {Cahen},\ and\ \citenamefont {Diale}}]{Rosenberg2017}%
  \BibitemOpen
  \bibfield  {author} {\bibinfo {author} {\bibfnamefont {J.~W.}\ \bibnamefont
  {Rosenberg}}, \bibinfo {author} {\bibfnamefont {M.~J.}\ \bibnamefont
  {Legodi}}, \bibinfo {author} {\bibfnamefont {Y.}~\bibnamefont {Rakita}},
  \bibinfo {author} {\bibfnamefont {D.}~\bibnamefont {Cahen}},\ and\ \bibinfo
  {author} {\bibfnamefont {M.}~\bibnamefont {Diale}},\ }\bibfield  {title}
  {\bibinfo {title} {{Laplace current deep level transient spectroscopy
  measurements of defect states in methylammonium lead bromide single
  crystals}},\ }\bibfield  {journal} {\bibinfo  {journal} {Journal of Applied
  Physics}\ }\textbf {\bibinfo {volume} {122}},\ \href
  {https://doi.org/10.1063/1.4995970} {10.1063/1.4995970} (\bibinfo {year}
  {2017})\BibitemShut {NoStop}%
\bibitem [{\citenamefont {Reichert}\ \emph {et~al.}(2020)\citenamefont
  {Reichert}, \citenamefont {An}, \citenamefont {Woo}, \citenamefont {Walsh},
  \citenamefont {Vaynzof},\ and\ \citenamefont
  {Deibel}}]{reichert_probing_2020}%
  \BibitemOpen
  \bibfield  {author} {\bibinfo {author} {\bibfnamefont {S.}~\bibnamefont
  {Reichert}}, \bibinfo {author} {\bibfnamefont {Q.}~\bibnamefont {An}},
  \bibinfo {author} {\bibfnamefont {Y.-W.}\ \bibnamefont {Woo}}, \bibinfo
  {author} {\bibfnamefont {A.}~\bibnamefont {Walsh}}, \bibinfo {author}
  {\bibfnamefont {Y.}~\bibnamefont {Vaynzof}},\ and\ \bibinfo {author}
  {\bibfnamefont {C.}~\bibnamefont {Deibel}},\ }\bibfield  {title} {\bibinfo
  {title} {Probing the ionic defect landscape in halide perovskite solar
  cells},\ }\href {https://doi.org/10.1038/s41467-020-19769-8} {\bibfield
  {journal} {\bibinfo  {journal} {Nature Communications}\ }\textbf {\bibinfo
  {volume} {11}},\ \bibinfo {pages} {6098} (\bibinfo {year}
  {2020})}\BibitemShut {NoStop}%
\end{thebibliography}

\begin{thebibliography}{31}%
\makeatletter
\providecommand \@ifxundefined [1]{%
 \@ifx{#1\undefined}
}%
\providecommand \@ifnum [1]{%
 \ifnum #1\expandafter \@firstoftwo
 \else \expandafter \@secondoftwo
 \fi
}%
\providecommand \@ifx [1]{%
 \ifx #1\expandafter \@firstoftwo
 \else \expandafter \@secondoftwo
 \fi
}%
\providecommand \natexlab [1]{#1}%
\providecommand \enquote  [1]{``#1''}%
\providecommand \bibnamefont  [1]{#1}%
\providecommand \bibfnamefont [1]{#1}%
\providecommand \citenamefont [1]{#1}%
\providecommand \href@noop [0]{\@secondoftwo}%
\providecommand \href [0]{\begingroup \@sanitize@url \@href}%
\providecommand \@href[1]{\@@startlink{#1}\@@href}%
\providecommand \@@href[1]{\endgroup#1\@@endlink}%
\providecommand \@sanitize@url [0]{\catcode `\\12\catcode `\$12\catcode
  `\&12\catcode `\#12\catcode `\^12\catcode `\_12\catcode `\%12\relax}%
\providecommand \@@startlink[1]{}%
\providecommand \@@endlink[0]{}%
\providecommand \url  [0]{\begingroup\@sanitize@url \@url }%
\providecommand \@url [1]{\endgroup\@href {#1}{\urlprefix }}%
\providecommand \urlprefix  [0]{URL }%
\providecommand \Eprint [0]{\href }%
\providecommand \doibase [0]{https://doi.org/}%
\providecommand \selectlanguage [0]{\@gobble}%
\providecommand \bibinfo  [0]{\@secondoftwo}%
\providecommand \bibfield  [0]{\@secondoftwo}%
\providecommand \translation [1]{[#1]}%
\providecommand \BibitemOpen [0]{}%
\providecommand \bibitemStop [0]{}%
\providecommand \bibitemNoStop [0]{.\EOS\space}%
\providecommand \EOS [0]{\spacefactor3000\relax}%
\providecommand \BibitemShut  [1]{\csname bibitem#1\endcsname}%
\let\auto@bib@innerbib\@empty
\bibitem [{Mai()}]{Main}%
  \BibitemOpen
  \href@noop {} {\bibinfo {title} {{See the main text of this supplemental
  material}}}\BibitemShut {NoStop}%
\bibitem [{\citenamefont {Bennett}\ \emph {et~al.}(2009)\citenamefont
  {Bennett}, \citenamefont {Grinberg},\ and\ \citenamefont
  {Rappe}}]{bennett_effect_2009}%
  \BibitemOpen
  \bibfield  {author} {\bibinfo {author} {\bibfnamefont {J.~W.}\ \bibnamefont
  {Bennett}}, \bibinfo {author} {\bibfnamefont {I.}~\bibnamefont {Grinberg}},\
  and\ \bibinfo {author} {\bibfnamefont {A.~M.}\ \bibnamefont {Rappe}},\
  }\bibfield  {title} {\bibinfo {title} {Effect of substituting of {S} for {O}:
  {The} sulfide perovskite {Bazrs} 3 investigated with density functional
  theory},\ }\href {https://doi.org/10.1103/PhysRevB.79.235115} {\bibfield
  {journal} {\bibinfo  {journal} {Physical Review B}\ }\textbf {\bibinfo
  {volume} {79}},\ \bibinfo {pages} {235115} (\bibinfo {year}
  {2009})}\BibitemShut {NoStop}%
\bibitem [{\citenamefont {Ermolaev}\ \emph {et~al.}(2023)\citenamefont
  {Ermolaev}, \citenamefont {Pushkarev}, \citenamefont {Zhizhchenko},
  \citenamefont {Kuchmizhak}, \citenamefont {Iorsh}, \citenamefont {Kruglov},
  \citenamefont {Mazitov}, \citenamefont {Ishteev}, \citenamefont
  {Konstantinova}, \citenamefont {Saranin}, \citenamefont {Slavich},
  \citenamefont {Stosic}, \citenamefont {Zhukova}, \citenamefont {Tselikov},
  \citenamefont {Di~Carlo}, \citenamefont {Arsenin}, \citenamefont {Novoselov},
  \citenamefont {Makarov},\ and\ \citenamefont {Volkov}}]{ermolaev_giant_2023}%
  \BibitemOpen
  \bibfield  {author} {\bibinfo {author} {\bibfnamefont {G.}~\bibnamefont
  {Ermolaev}}, \bibinfo {author} {\bibfnamefont {A.~P.}\ \bibnamefont
  {Pushkarev}}, \bibinfo {author} {\bibfnamefont {A.}~\bibnamefont
  {Zhizhchenko}}, \bibinfo {author} {\bibfnamefont {A.~A.}\ \bibnamefont
  {Kuchmizhak}}, \bibinfo {author} {\bibfnamefont {I.}~\bibnamefont {Iorsh}},
  \bibinfo {author} {\bibfnamefont {I.}~\bibnamefont {Kruglov}}, \bibinfo
  {author} {\bibfnamefont {A.}~\bibnamefont {Mazitov}}, \bibinfo {author}
  {\bibfnamefont {A.}~\bibnamefont {Ishteev}}, \bibinfo {author} {\bibfnamefont
  {K.}~\bibnamefont {Konstantinova}}, \bibinfo {author} {\bibfnamefont
  {D.}~\bibnamefont {Saranin}}, \bibinfo {author} {\bibfnamefont
  {A.}~\bibnamefont {Slavich}}, \bibinfo {author} {\bibfnamefont
  {D.}~\bibnamefont {Stosic}}, \bibinfo {author} {\bibfnamefont {E.~S.}\
  \bibnamefont {Zhukova}}, \bibinfo {author} {\bibfnamefont {G.}~\bibnamefont
  {Tselikov}}, \bibinfo {author} {\bibfnamefont {A.}~\bibnamefont {Di~Carlo}},
  \bibinfo {author} {\bibfnamefont {A.}~\bibnamefont {Arsenin}}, \bibinfo
  {author} {\bibfnamefont {K.~S.}\ \bibnamefont {Novoselov}}, \bibinfo {author}
  {\bibfnamefont {S.~V.}\ \bibnamefont {Makarov}},\ and\ \bibinfo {author}
  {\bibfnamefont {V.~S.}\ \bibnamefont {Volkov}},\ }\bibfield  {title}
  {\bibinfo {title} {Giant and {Tunable} {Excitonic} {Optical} {Anisotropy} in
  {Single}-{Crystal} {Halide} {Perovskites}},\ }\href
  {https://doi.org/10.1021/acs.nanolett.2c04792} {\bibfield  {journal}
  {\bibinfo  {journal} {Nano Letters}\ }\textbf {\bibinfo {volume} {23}},\
  \bibinfo {pages} {2570} (\bibinfo {year} {2023})}\BibitemShut {NoStop}%
\bibitem [{\citenamefont {Roekeghem}\ \emph {et~al.}(2021)\citenamefont
  {Roekeghem}, \citenamefont {Carrete},\ and\ \citenamefont
  {Mingo}}]{Roekeghem.2021}%
  \BibitemOpen
  \bibfield  {author} {\bibinfo {author} {\bibfnamefont {A.~v.}\ \bibnamefont
  {Roekeghem}}, \bibinfo {author} {\bibfnamefont {J.}~\bibnamefont {Carrete}},\
  and\ \bibinfo {author} {\bibfnamefont {N.}~\bibnamefont {Mingo}},\ }\bibfield
   {title} {\bibinfo {title} {{Quantum Self-Consistent Ab-Initio Lattice
  Dynamics}},\ }\href {https://doi.org/10.1016/j.cpc.2021.107945} {\bibfield
  {journal} {\bibinfo  {journal} {Computer Physics Communications}\ }\textbf
  {\bibinfo {volume} {263}},\ \bibinfo {pages} {107945} (\bibinfo {year}
  {2021})}\BibitemShut {NoStop}%
\bibitem [{\citenamefont {Bianco}\ \emph {et~al.}(2017)\citenamefont {Bianco},
  \citenamefont {Errea}, \citenamefont {Paulatto}, \citenamefont {Calandra},\
  and\ \citenamefont {Mauri}}]{Bianco.2017}%
  \BibitemOpen
  \bibfield  {author} {\bibinfo {author} {\bibfnamefont {R.}~\bibnamefont
  {Bianco}}, \bibinfo {author} {\bibfnamefont {I.}~\bibnamefont {Errea}},
  \bibinfo {author} {\bibfnamefont {L.}~\bibnamefont {Paulatto}}, \bibinfo
  {author} {\bibfnamefont {M.}~\bibnamefont {Calandra}},\ and\ \bibinfo
  {author} {\bibfnamefont {F.}~\bibnamefont {Mauri}},\ }\bibfield  {title}
  {\bibinfo {title} {{Second-order structural phase transitions, free energy
  curvature, and temperature-dependent anharmonic phonons in the
  self-consistent harmonic approximation: Theory and stochastic
  implementation}},\ }\href {https://doi.org/10.1103/physrevb.96.014111}
  {\bibfield  {journal} {\bibinfo  {journal} {Physical Review B}\ }\textbf
  {\bibinfo {volume} {96}},\ \bibinfo {pages} {014111} (\bibinfo {year}
  {2017})}\BibitemShut {NoStop}%
\bibitem [{\citenamefont {Benshalom}\ \emph {et~al.}(2022)\citenamefont
  {Benshalom}, \citenamefont {Reuveni}, \citenamefont {Korobko}, \citenamefont
  {Yaffe},\ and\ \citenamefont {Hellman}}]{Benshalom2022}%
  \BibitemOpen
  \bibfield  {author} {\bibinfo {author} {\bibfnamefont {N.}~\bibnamefont
  {Benshalom}}, \bibinfo {author} {\bibfnamefont {G.}~\bibnamefont {Reuveni}},
  \bibinfo {author} {\bibfnamefont {R.}~\bibnamefont {Korobko}}, \bibinfo
  {author} {\bibfnamefont {O.}~\bibnamefont {Yaffe}},\ and\ \bibinfo {author}
  {\bibfnamefont {O.}~\bibnamefont {Hellman}},\ }\bibfield  {title} {\bibinfo
  {title} {{Dielectric response of rock-salt crystals at finite temperatures
  from first principles}},\ }\href
  {https://doi.org/10.1103/PhysRevMaterials.6.033607} {\bibfield  {journal}
  {\bibinfo  {journal} {Physical Review Materials}\ }\textbf {\bibinfo {volume}
  {6}},\ \bibinfo {pages} {033607} (\bibinfo {year} {2022})}\BibitemShut
  {NoStop}%
\bibitem [{\citenamefont {Knoop}\ \emph {et~al.}(2024)\citenamefont {Knoop},
  \citenamefont {Shulumba}, \citenamefont {Castellano}, \citenamefont
  {Batista}, \citenamefont {Farris}, \citenamefont {Verstraete}, \citenamefont
  {Heine}, \citenamefont {Broido}, \citenamefont {Kim}, \citenamefont
  {Klarbring}, \citenamefont {Abrikosov}, \citenamefont {Simak},\ and\
  \citenamefont {Hellman}}]{Knoop.2024}%
  \BibitemOpen
  \bibfield  {author} {\bibinfo {author} {\bibfnamefont {F.}~\bibnamefont
  {Knoop}}, \bibinfo {author} {\bibfnamefont {N.}~\bibnamefont {Shulumba}},
  \bibinfo {author} {\bibfnamefont {A.}~\bibnamefont {Castellano}}, \bibinfo
  {author} {\bibfnamefont {J.~P.~A.}\ \bibnamefont {Batista}}, \bibinfo
  {author} {\bibfnamefont {R.}~\bibnamefont {Farris}}, \bibinfo {author}
  {\bibfnamefont {M.~J.}\ \bibnamefont {Verstraete}}, \bibinfo {author}
  {\bibfnamefont {M.}~\bibnamefont {Heine}}, \bibinfo {author} {\bibfnamefont
  {D.}~\bibnamefont {Broido}}, \bibinfo {author} {\bibfnamefont {D.~S.}\
  \bibnamefont {Kim}}, \bibinfo {author} {\bibfnamefont {J.}~\bibnamefont
  {Klarbring}}, \bibinfo {author} {\bibfnamefont {I.~A.}\ \bibnamefont
  {Abrikosov}}, \bibinfo {author} {\bibfnamefont {S.~I.}\ \bibnamefont
  {Simak}},\ and\ \bibinfo {author} {\bibfnamefont {O.}~\bibnamefont
  {Hellman}},\ }\bibfield  {title} {\bibinfo {title} {{TDEP: Temperature
  Dependent Effective Potentials}},\ }\href
  {https://doi.org/10.21105/joss.06150} {\bibfield  {journal} {\bibinfo
  {journal} {Journal of Open Source Software}\ }\textbf {\bibinfo {volume}
  {9}},\ \bibinfo {pages} {6150} (\bibinfo {year} {2024})}\BibitemShut
  {NoStop}%
\bibitem [{\citenamefont {Shulumba}\ \emph {et~al.}(2017)\citenamefont
  {Shulumba}, \citenamefont {Hellman},\ and\ \citenamefont
  {Minnich}}]{Shulumba.2017}%
  \BibitemOpen
  \bibfield  {author} {\bibinfo {author} {\bibfnamefont {N.}~\bibnamefont
  {Shulumba}}, \bibinfo {author} {\bibfnamefont {O.}~\bibnamefont {Hellman}},\
  and\ \bibinfo {author} {\bibfnamefont {A.~J.}\ \bibnamefont {Minnich}},\
  }\bibfield  {title} {\bibinfo {title} {{Intrinsic localized mode and low
  thermal conductivity of PbSe}},\ }\href
  {https://doi.org/10.1103/physrevb.95.014302} {\bibfield  {journal} {\bibinfo
  {journal} {Physical Review B}\ }\textbf {\bibinfo {volume} {95}},\ \bibinfo
  {pages} {014302} (\bibinfo {year} {2017})}\BibitemShut {NoStop}%
\bibitem [{\citenamefont {Hellman}\ and\ \citenamefont
  {Abrikosov}(2013)}]{Hellman.2013oi5}%
  \BibitemOpen
  \bibfield  {author} {\bibinfo {author} {\bibfnamefont {O.}~\bibnamefont
  {Hellman}}\ and\ \bibinfo {author} {\bibfnamefont {I.~A.}\ \bibnamefont
  {Abrikosov}},\ }\bibfield  {title} {\bibinfo {title} {{Temperature-dependent
  effective third-order interatomic force constants from first principles}},\
  }\href {https://doi.org/10.1103/physrevb.88.144301} {\bibfield  {journal}
  {\bibinfo  {journal} {Physical Review B}\ }\textbf {\bibinfo {volume} {88}},\
  \bibinfo {pages} {144301} (\bibinfo {year} {2013})}\BibitemShut {NoStop}%
\bibitem [{\citenamefont {Jain}\ \emph {et~al.}(2013)\citenamefont {Jain},
  \citenamefont {Ong}, \citenamefont {Hautier}, \citenamefont {Chen},
  \citenamefont {Richards}, \citenamefont {Dacek}, \citenamefont {Cholia},
  \citenamefont {Gunter}, \citenamefont {Skinner}, \citenamefont {Ceder},\ and\
  \citenamefont {Persson}}]{Jain.2013}%
  \BibitemOpen
  \bibfield  {author} {\bibinfo {author} {\bibfnamefont {A.}~\bibnamefont
  {Jain}}, \bibinfo {author} {\bibfnamefont {S.~P.}\ \bibnamefont {Ong}},
  \bibinfo {author} {\bibfnamefont {G.}~\bibnamefont {Hautier}}, \bibinfo
  {author} {\bibfnamefont {W.}~\bibnamefont {Chen}}, \bibinfo {author}
  {\bibfnamefont {W.~D.}\ \bibnamefont {Richards}}, \bibinfo {author}
  {\bibfnamefont {S.}~\bibnamefont {Dacek}}, \bibinfo {author} {\bibfnamefont
  {S.}~\bibnamefont {Cholia}}, \bibinfo {author} {\bibfnamefont
  {D.}~\bibnamefont {Gunter}}, \bibinfo {author} {\bibfnamefont
  {D.}~\bibnamefont {Skinner}}, \bibinfo {author} {\bibfnamefont
  {G.}~\bibnamefont {Ceder}},\ and\ \bibinfo {author} {\bibfnamefont {K.~A.}\
  \bibnamefont {Persson}},\ }\bibfield  {title} {\bibinfo {title} {{Commentary:
  The Materials Project: A materials genome approach to accelerating materials
  innovation}},\ }\href {https://doi.org/10.1063/1.4812323} {\bibfield
  {journal} {\bibinfo  {journal} {APL Materials}\ }\textbf {\bibinfo {volume}
  {1}},\ \bibinfo {pages} {011002} (\bibinfo {year} {2013})}\BibitemShut
  {NoStop}%
\bibitem [{\citenamefont {Blum}\ \emph {et~al.}(2009)\citenamefont {Blum},
  \citenamefont {Gehrke}, \citenamefont {Hanke}, \citenamefont {Havu},
  \citenamefont {Havu}, \citenamefont {Ren}, \citenamefont {Reuter},\ and\
  \citenamefont {Scheffler}}]{Blum.2009}%
  \BibitemOpen
  \bibfield  {author} {\bibinfo {author} {\bibfnamefont {V.}~\bibnamefont
  {Blum}}, \bibinfo {author} {\bibfnamefont {R.}~\bibnamefont {Gehrke}},
  \bibinfo {author} {\bibfnamefont {F.}~\bibnamefont {Hanke}}, \bibinfo
  {author} {\bibfnamefont {P.}~\bibnamefont {Havu}}, \bibinfo {author}
  {\bibfnamefont {V.}~\bibnamefont {Havu}}, \bibinfo {author} {\bibfnamefont
  {X.}~\bibnamefont {Ren}}, \bibinfo {author} {\bibfnamefont {K.}~\bibnamefont
  {Reuter}},\ and\ \bibinfo {author} {\bibfnamefont {M.}~\bibnamefont
  {Scheffler}},\ }\bibfield  {title} {\bibinfo {title} {{Ab initio molecular
  simulations with numeric atom-centered orbitals}},\ }\href
  {https://doi.org/10.1016/j.cpc.2009.06.022} {\bibfield  {journal} {\bibinfo
  {journal} {Computer Physics Communications}\ }\textbf {\bibinfo {volume}
  {180}},\ \bibinfo {pages} {2175} (\bibinfo {year} {2009})}\BibitemShut
  {NoStop}%
\bibitem [{\citenamefont {Armiento}\ and\ \citenamefont
  {Mattsson}(2005)}]{Armiento.2005}%
  \BibitemOpen
  \bibfield  {author} {\bibinfo {author} {\bibfnamefont {R.}~\bibnamefont
  {Armiento}}\ and\ \bibinfo {author} {\bibfnamefont {A.~E.}\ \bibnamefont
  {Mattsson}},\ }\bibfield  {title} {\bibinfo {title} {{Functional designed to
  include surface effects in self-consistent density functional theory}},\
  }\href {https://doi.org/10.1103/physrevb.72.085108} {\bibfield  {journal}
  {\bibinfo  {journal} {Physical Review B}\ }\textbf {\bibinfo {volume} {72}},\
  \bibinfo {pages} {085108} (\bibinfo {year} {2005})}\BibitemShut {NoStop}%
\bibitem [{\citenamefont {Mattsson}\ \emph {et~al.}(2008)\citenamefont
  {Mattsson}, \citenamefont {Armiento}, \citenamefont {Paier}, \citenamefont
  {Kresse}, \citenamefont {Wills},\ and\ \citenamefont
  {Mattsson}}]{Mattsson.2008}%
  \BibitemOpen
  \bibfield  {author} {\bibinfo {author} {\bibfnamefont {A.~E.}\ \bibnamefont
  {Mattsson}}, \bibinfo {author} {\bibfnamefont {R.}~\bibnamefont {Armiento}},
  \bibinfo {author} {\bibfnamefont {J.}~\bibnamefont {Paier}}, \bibinfo
  {author} {\bibfnamefont {G.}~\bibnamefont {Kresse}}, \bibinfo {author}
  {\bibfnamefont {J.~M.}\ \bibnamefont {Wills}},\ and\ \bibinfo {author}
  {\bibfnamefont {T.~R.}\ \bibnamefont {Mattsson}},\ }\bibfield  {title}
  {\bibinfo {title} {{The AM05 density functional applied to solids}},\ }\href
  {https://doi.org/10.1063/1.2835596} {\bibfield  {journal} {\bibinfo
  {journal} {The Journal of Chemical Physics}\ }\textbf {\bibinfo {volume}
  {128}},\ \bibinfo {pages} {084714} (\bibinfo {year} {2008})}\BibitemShut
  {NoStop}%
\bibitem [{\citenamefont {Giannozzi}\ \emph {et~al.}(2009)\citenamefont
  {Giannozzi}, \citenamefont {Baroni}, \citenamefont {Bonini}, \citenamefont
  {Calandra}, \citenamefont {Car}, \citenamefont {Cavazzoni}, \citenamefont
  {Ceresoli}, \citenamefont {Chiarotti}, \citenamefont {Cococcioni},
  \citenamefont {Dabo}, \citenamefont {Corso}, \citenamefont {Gironcoli},
  \citenamefont {Fabris}, \citenamefont {Fratesi}, \citenamefont {Gebauer},
  \citenamefont {Gerstmann}, \citenamefont {Gougoussis}, \citenamefont
  {Kokalj}, \citenamefont {Lazzeri}, \citenamefont {Martin-Samos},
  \citenamefont {Marzari}, \citenamefont {Mauri}, \citenamefont {Mazzarello},
  \citenamefont {Paolini}, \citenamefont {Pasquarello}, \citenamefont
  {Paulatto}, \citenamefont {Sbraccia}, \citenamefont {Scandolo}, \citenamefont
  {Sclauzero}, \citenamefont {Seitsonen}, \citenamefont {Smogunov},
  \citenamefont {Umari},\ and\ \citenamefont {Wentzcovitch}}]{Giannozzi.2009}%
  \BibitemOpen
  \bibfield  {author} {\bibinfo {author} {\bibfnamefont {P.}~\bibnamefont
  {Giannozzi}}, \bibinfo {author} {\bibfnamefont {S.}~\bibnamefont {Baroni}},
  \bibinfo {author} {\bibfnamefont {N.}~\bibnamefont {Bonini}}, \bibinfo
  {author} {\bibfnamefont {M.}~\bibnamefont {Calandra}}, \bibinfo {author}
  {\bibfnamefont {R.}~\bibnamefont {Car}}, \bibinfo {author} {\bibfnamefont
  {C.}~\bibnamefont {Cavazzoni}}, \bibinfo {author} {\bibfnamefont
  {D.}~\bibnamefont {Ceresoli}}, \bibinfo {author} {\bibfnamefont {G.~L.}\
  \bibnamefont {Chiarotti}}, \bibinfo {author} {\bibfnamefont {M.}~\bibnamefont
  {Cococcioni}}, \bibinfo {author} {\bibfnamefont {I.}~\bibnamefont {Dabo}},
  \bibinfo {author} {\bibfnamefont {A.~D.}\ \bibnamefont {Corso}}, \bibinfo
  {author} {\bibfnamefont {S.~d.}\ \bibnamefont {Gironcoli}}, \bibinfo {author}
  {\bibfnamefont {S.}~\bibnamefont {Fabris}}, \bibinfo {author} {\bibfnamefont
  {G.}~\bibnamefont {Fratesi}}, \bibinfo {author} {\bibfnamefont
  {R.}~\bibnamefont {Gebauer}}, \bibinfo {author} {\bibfnamefont
  {U.}~\bibnamefont {Gerstmann}}, \bibinfo {author} {\bibfnamefont
  {C.}~\bibnamefont {Gougoussis}}, \bibinfo {author} {\bibfnamefont
  {A.}~\bibnamefont {Kokalj}}, \bibinfo {author} {\bibfnamefont
  {M.}~\bibnamefont {Lazzeri}}, \bibinfo {author} {\bibfnamefont
  {L.}~\bibnamefont {Martin-Samos}}, \bibinfo {author} {\bibfnamefont
  {N.}~\bibnamefont {Marzari}}, \bibinfo {author} {\bibfnamefont
  {F.}~\bibnamefont {Mauri}}, \bibinfo {author} {\bibfnamefont
  {R.}~\bibnamefont {Mazzarello}}, \bibinfo {author} {\bibfnamefont
  {S.}~\bibnamefont {Paolini}}, \bibinfo {author} {\bibfnamefont
  {A.}~\bibnamefont {Pasquarello}}, \bibinfo {author} {\bibfnamefont
  {L.}~\bibnamefont {Paulatto}}, \bibinfo {author} {\bibfnamefont
  {C.}~\bibnamefont {Sbraccia}}, \bibinfo {author} {\bibfnamefont
  {S.}~\bibnamefont {Scandolo}}, \bibinfo {author} {\bibfnamefont
  {G.}~\bibnamefont {Sclauzero}}, \bibinfo {author} {\bibfnamefont {A.~P.}\
  \bibnamefont {Seitsonen}}, \bibinfo {author} {\bibfnamefont {A.}~\bibnamefont
  {Smogunov}}, \bibinfo {author} {\bibfnamefont {P.}~\bibnamefont {Umari}},\
  and\ \bibinfo {author} {\bibfnamefont {R.~M.}\ \bibnamefont {Wentzcovitch}},\
  }\bibfield  {title} {\bibinfo {title} {{QUANTUM ESPRESSO: a modular and
  open-source software project for quantum simulations of materials}},\ }\href
  {https://doi.org/10.1088/0953-8984/21/39/395502} {\bibfield  {journal}
  {\bibinfo  {journal} {Journal of Physics: Condensed Matter}\ }\textbf
  {\bibinfo {volume} {21}},\ \bibinfo {pages} {395502} (\bibinfo {year}
  {2009})}\BibitemShut {NoStop}%
\bibitem [{\citenamefont {Giannozzi}\ \emph {et~al.}(2017)\citenamefont
  {Giannozzi}, \citenamefont {Andreussi}, \citenamefont {Brumme}, \citenamefont
  {Bunau}, \citenamefont {Nardelli}, \citenamefont {Calandra}, \citenamefont
  {Car}, \citenamefont {Cavazzoni}, \citenamefont {Ceresoli}, \citenamefont
  {Cococcioni}, \citenamefont {Colonna}, \citenamefont {Carnimeo},
  \citenamefont {Corso}, \citenamefont {Gironcoli}, \citenamefont {Delugas},
  \citenamefont {Jr}, \citenamefont {Ferretti}, \citenamefont {Floris},
  \citenamefont {Fratesi}, \citenamefont {Fugallo}, \citenamefont {Gebauer},
  \citenamefont {Gerstmann}, \citenamefont {Giustino}, \citenamefont {Gorni},
  \citenamefont {Jia}, \citenamefont {Kawamura}, \citenamefont {Ko},
  \citenamefont {Kokalj}, \citenamefont {Küçükbenli}, \citenamefont
  {Lazzeri}, \citenamefont {Marsili}, \citenamefont {Marzari}, \citenamefont
  {Mauri}, \citenamefont {Nguyen}, \citenamefont {Nguyen}, \citenamefont
  {Otero-de-la Roza}, \citenamefont {Paulatto}, \citenamefont {Poncé},
  \citenamefont {Rocca}, \citenamefont {Sabatini}, \citenamefont {Santra},
  \citenamefont {Schlipf}, \citenamefont {Seitsonen}, \citenamefont {Smogunov},
  \citenamefont {Timrov}, \citenamefont {Thonhauser}, \citenamefont {Umari},
  \citenamefont {Vast}, \citenamefont {Wu},\ and\ \citenamefont
  {Baroni}}]{Giannozzi.2017}%
  \BibitemOpen
  \bibfield  {author} {\bibinfo {author} {\bibfnamefont {P.}~\bibnamefont
  {Giannozzi}}, \bibinfo {author} {\bibfnamefont {O.}~\bibnamefont
  {Andreussi}}, \bibinfo {author} {\bibfnamefont {T.}~\bibnamefont {Brumme}},
  \bibinfo {author} {\bibfnamefont {O.}~\bibnamefont {Bunau}}, \bibinfo
  {author} {\bibfnamefont {M.~B.}\ \bibnamefont {Nardelli}}, \bibinfo {author}
  {\bibfnamefont {M.}~\bibnamefont {Calandra}}, \bibinfo {author}
  {\bibfnamefont {R.}~\bibnamefont {Car}}, \bibinfo {author} {\bibfnamefont
  {C.}~\bibnamefont {Cavazzoni}}, \bibinfo {author} {\bibfnamefont
  {D.}~\bibnamefont {Ceresoli}}, \bibinfo {author} {\bibfnamefont
  {M.}~\bibnamefont {Cococcioni}}, \bibinfo {author} {\bibfnamefont
  {N.}~\bibnamefont {Colonna}}, \bibinfo {author} {\bibfnamefont
  {I.}~\bibnamefont {Carnimeo}}, \bibinfo {author} {\bibfnamefont {A.~D.}\
  \bibnamefont {Corso}}, \bibinfo {author} {\bibfnamefont {S.~d.}\ \bibnamefont
  {Gironcoli}}, \bibinfo {author} {\bibfnamefont {P.}~\bibnamefont {Delugas}},
  \bibinfo {author} {\bibfnamefont {R.~A.~D.}\ \bibnamefont {Jr}}, \bibinfo
  {author} {\bibfnamefont {A.}~\bibnamefont {Ferretti}}, \bibinfo {author}
  {\bibfnamefont {A.}~\bibnamefont {Floris}}, \bibinfo {author} {\bibfnamefont
  {G.}~\bibnamefont {Fratesi}}, \bibinfo {author} {\bibfnamefont
  {G.}~\bibnamefont {Fugallo}}, \bibinfo {author} {\bibfnamefont
  {R.}~\bibnamefont {Gebauer}}, \bibinfo {author} {\bibfnamefont
  {U.}~\bibnamefont {Gerstmann}}, \bibinfo {author} {\bibfnamefont
  {F.}~\bibnamefont {Giustino}}, \bibinfo {author} {\bibfnamefont
  {T.}~\bibnamefont {Gorni}}, \bibinfo {author} {\bibfnamefont
  {J.}~\bibnamefont {Jia}}, \bibinfo {author} {\bibfnamefont {M.}~\bibnamefont
  {Kawamura}}, \bibinfo {author} {\bibfnamefont {H.-Y.}\ \bibnamefont {Ko}},
  \bibinfo {author} {\bibfnamefont {A.}~\bibnamefont {Kokalj}}, \bibinfo
  {author} {\bibfnamefont {E.}~\bibnamefont {Küçükbenli}}, \bibinfo {author}
  {\bibfnamefont {M.}~\bibnamefont {Lazzeri}}, \bibinfo {author} {\bibfnamefont
  {M.}~\bibnamefont {Marsili}}, \bibinfo {author} {\bibfnamefont
  {N.}~\bibnamefont {Marzari}}, \bibinfo {author} {\bibfnamefont
  {F.}~\bibnamefont {Mauri}}, \bibinfo {author} {\bibfnamefont {N.~L.}\
  \bibnamefont {Nguyen}}, \bibinfo {author} {\bibfnamefont {H.-V.}\
  \bibnamefont {Nguyen}}, \bibinfo {author} {\bibfnamefont {A.}~\bibnamefont
  {Otero-de-la Roza}}, \bibinfo {author} {\bibfnamefont {L.}~\bibnamefont
  {Paulatto}}, \bibinfo {author} {\bibfnamefont {S.}~\bibnamefont {Poncé}},
  \bibinfo {author} {\bibfnamefont {D.}~\bibnamefont {Rocca}}, \bibinfo
  {author} {\bibfnamefont {R.}~\bibnamefont {Sabatini}}, \bibinfo {author}
  {\bibfnamefont {B.}~\bibnamefont {Santra}}, \bibinfo {author} {\bibfnamefont
  {M.}~\bibnamefont {Schlipf}}, \bibinfo {author} {\bibfnamefont {A.~P.}\
  \bibnamefont {Seitsonen}}, \bibinfo {author} {\bibfnamefont {A.}~\bibnamefont
  {Smogunov}}, \bibinfo {author} {\bibfnamefont {I.}~\bibnamefont {Timrov}},
  \bibinfo {author} {\bibfnamefont {T.}~\bibnamefont {Thonhauser}}, \bibinfo
  {author} {\bibfnamefont {P.}~\bibnamefont {Umari}}, \bibinfo {author}
  {\bibfnamefont {N.}~\bibnamefont {Vast}}, \bibinfo {author} {\bibfnamefont
  {X.}~\bibnamefont {Wu}},\ and\ \bibinfo {author} {\bibfnamefont
  {S.}~\bibnamefont {Baroni}},\ }\bibfield  {title} {\bibinfo {title}
  {{Advanced capabilities for materials modelling with Quantum ESPRESSO}},\
  }\href {https://doi.org/10.1088/1361-648x/aa8f79} {\bibfield  {journal}
  {\bibinfo  {journal} {Journal of Physics: Condensed Matter}\ }\textbf
  {\bibinfo {volume} {29}},\ \bibinfo {pages} {465901} (\bibinfo {year}
  {2017})}\BibitemShut {NoStop}%
\bibitem [{\citenamefont {Giannozzi}\ \emph {et~al.}(2020)\citenamefont
  {Giannozzi}, \citenamefont {Baseggio}, \citenamefont {Bonfà}, \citenamefont
  {Brunato}, \citenamefont {Car}, \citenamefont {Carnimeo}, \citenamefont
  {Cavazzoni}, \citenamefont {Gironcoli}, \citenamefont {Delugas},
  \citenamefont {Ruffino}, \citenamefont {Ferretti}, \citenamefont {Marzari},
  \citenamefont {Timrov}, \citenamefont {Urru},\ and\ \citenamefont
  {Baroni}}]{Giannozzi.2020}%
  \BibitemOpen
  \bibfield  {author} {\bibinfo {author} {\bibfnamefont {P.}~\bibnamefont
  {Giannozzi}}, \bibinfo {author} {\bibfnamefont {O.}~\bibnamefont {Baseggio}},
  \bibinfo {author} {\bibfnamefont {P.}~\bibnamefont {Bonfà}}, \bibinfo
  {author} {\bibfnamefont {D.}~\bibnamefont {Brunato}}, \bibinfo {author}
  {\bibfnamefont {R.}~\bibnamefont {Car}}, \bibinfo {author} {\bibfnamefont
  {I.}~\bibnamefont {Carnimeo}}, \bibinfo {author} {\bibfnamefont
  {C.}~\bibnamefont {Cavazzoni}}, \bibinfo {author} {\bibfnamefont {S.~d.}\
  \bibnamefont {Gironcoli}}, \bibinfo {author} {\bibfnamefont {P.}~\bibnamefont
  {Delugas}}, \bibinfo {author} {\bibfnamefont {F.~F.}\ \bibnamefont
  {Ruffino}}, \bibinfo {author} {\bibfnamefont {A.}~\bibnamefont {Ferretti}},
  \bibinfo {author} {\bibfnamefont {N.}~\bibnamefont {Marzari}}, \bibinfo
  {author} {\bibfnamefont {I.}~\bibnamefont {Timrov}}, \bibinfo {author}
  {\bibfnamefont {A.}~\bibnamefont {Urru}},\ and\ \bibinfo {author}
  {\bibfnamefont {S.}~\bibnamefont {Baroni}},\ }\bibfield  {title} {\bibinfo
  {title} {{Quantum ESPRESSO toward the exascale}},\ }\href
  {https://doi.org/10.1063/5.0005082} {\bibfield  {journal} {\bibinfo
  {journal} {The Journal of Chemical Physics}\ }\textbf {\bibinfo {volume}
  {152}},\ \bibinfo {pages} {154105} (\bibinfo {year} {2020})}\BibitemShut
  {NoStop}%
\bibitem [{\citenamefont {Perdew}\ \emph {et~al.}(1996)\citenamefont {Perdew},
  \citenamefont {Burke},\ and\ \citenamefont {Ernzerhof}}]{Perdew.1996}%
  \BibitemOpen
  \bibfield  {author} {\bibinfo {author} {\bibfnamefont {J.~P.}\ \bibnamefont
  {Perdew}}, \bibinfo {author} {\bibfnamefont {K.}~\bibnamefont {Burke}},\ and\
  \bibinfo {author} {\bibfnamefont {M.}~\bibnamefont {Ernzerhof}},\ }\bibfield
  {title} {\bibinfo {title} {{Generalized Gradient Approximation Made
  Simple}},\ }\href {https://doi.org/10.1103/physrevlett.77.3865} {\bibfield
  {journal} {\bibinfo  {journal} {Physical Review Letters}\ }\textbf {\bibinfo
  {volume} {77}},\ \bibinfo {pages} {3865} (\bibinfo {year} {1996})},\ \bibinfo
  {note} {pBE}\BibitemShut {NoStop}%
\bibitem [{\citenamefont {Hamann}(2013)}]{Hamann.2013}%
  \BibitemOpen
  \bibfield  {author} {\bibinfo {author} {\bibfnamefont {D.~R.}\ \bibnamefont
  {Hamann}},\ }\bibfield  {title} {\bibinfo {title} {{Optimized norm-conserving
  Vanderbilt pseudopotentials}},\ }\href
  {https://doi.org/10.1103/physrevb.88.085117} {\bibfield  {journal} {\bibinfo
  {journal} {Physical Review B}\ }\textbf {\bibinfo {volume} {88}},\ \bibinfo
  {pages} {085117} (\bibinfo {year} {2013})}\BibitemShut {NoStop}%
\bibitem [{\citenamefont {Schlipf}\ and\ \citenamefont
  {Gygi}(2015)}]{Schlipf.2015ake}%
  \BibitemOpen
  \bibfield  {author} {\bibinfo {author} {\bibfnamefont {M.}~\bibnamefont
  {Schlipf}}\ and\ \bibinfo {author} {\bibfnamefont {F.}~\bibnamefont {Gygi}},\
  }\bibfield  {title} {\bibinfo {title} {{Optimization algorithm for the
  generation of ONCV pseudopotentials}},\ }\href
  {https://doi.org/10.1016/j.cpc.2015.05.011} {\bibfield  {journal} {\bibinfo
  {journal} {Computer Physics Communications}\ }\textbf {\bibinfo {volume}
  {196}},\ \bibinfo {pages} {36} (\bibinfo {year} {2015})}\BibitemShut
  {NoStop}%
\bibitem [{\citenamefont {Menahem}\ \emph {et~al.}(2023)\citenamefont
  {Menahem}, \citenamefont {Benshalom}, \citenamefont {Asher}, \citenamefont
  {Aharon}, \citenamefont {Korobko}, \citenamefont {Hellman},\ and\
  \citenamefont {Yaffe}}]{Menahem2023}%
  \BibitemOpen
  \bibfield  {author} {\bibinfo {author} {\bibfnamefont {M.}~\bibnamefont
  {Menahem}}, \bibinfo {author} {\bibfnamefont {N.}~\bibnamefont {Benshalom}},
  \bibinfo {author} {\bibfnamefont {M.}~\bibnamefont {Asher}}, \bibinfo
  {author} {\bibfnamefont {S.}~\bibnamefont {Aharon}}, \bibinfo {author}
  {\bibfnamefont {R.}~\bibnamefont {Korobko}}, \bibinfo {author} {\bibfnamefont
  {O.}~\bibnamefont {Hellman}},\ and\ \bibinfo {author} {\bibfnamefont
  {O.}~\bibnamefont {Yaffe}},\ }\bibfield  {title} {\bibinfo {title} {{Disorder
  origin of Raman scattering in perovskite single crystals}},\ }\href
  {https://doi.org/10.1103/PhysRevMaterials.7.044602} {\bibfield  {journal}
  {\bibinfo  {journal} {Physical Review Materials}\ }\textbf {\bibinfo {volume}
  {7}},\ \bibinfo {pages} {044602} (\bibinfo {year} {2023})}\BibitemShut
  {NoStop}%
\bibitem [{\citenamefont {Asher}\ \emph {et~al.}(2023)\citenamefont {Asher},
  \citenamefont {Bardini}, \citenamefont {Catalano}, \citenamefont {Jouclas},
  \citenamefont {Schweicher}, \citenamefont {Liu}, \citenamefont {Korobko},
  \citenamefont {Cohen}, \citenamefont {Geerts}, \citenamefont {Beljonne},\
  and\ \citenamefont {Yaffe}}]{Asher2023}%
  \BibitemOpen
  \bibfield  {author} {\bibinfo {author} {\bibfnamefont {M.}~\bibnamefont
  {Asher}}, \bibinfo {author} {\bibfnamefont {M.}~\bibnamefont {Bardini}},
  \bibinfo {author} {\bibfnamefont {L.}~\bibnamefont {Catalano}}, \bibinfo
  {author} {\bibfnamefont {R.}~\bibnamefont {Jouclas}}, \bibinfo {author}
  {\bibfnamefont {G.}~\bibnamefont {Schweicher}}, \bibinfo {author}
  {\bibfnamefont {J.}~\bibnamefont {Liu}}, \bibinfo {author} {\bibfnamefont
  {R.}~\bibnamefont {Korobko}}, \bibinfo {author} {\bibfnamefont
  {A.}~\bibnamefont {Cohen}}, \bibinfo {author} {\bibfnamefont
  {Y.}~\bibnamefont {Geerts}}, \bibinfo {author} {\bibfnamefont
  {D.}~\bibnamefont {Beljonne}},\ and\ \bibinfo {author} {\bibfnamefont
  {O.}~\bibnamefont {Yaffe}},\ }\bibfield  {title} {\bibinfo {title}
  {{Mechanistic View on the Order-Disorder Phase Transition in Amphidynamic
  Crystals}},\ }\href {https://doi.org/10.1021/acs.jpclett.2c03316} {\bibfield
  {journal} {\bibinfo  {journal} {The Journal of Physical Chemistry Letters}\
  }\textbf {\bibinfo {volume} {14}},\ \bibinfo {pages} {1570} (\bibinfo {year}
  {2023})}\BibitemShut {NoStop}%
\bibitem [{\citenamefont {Cardona}\ and\ \citenamefont
  {Guntherodt}(1982)}]{Cardona1982}%
  \BibitemOpen
  \bibfield  {author} {\bibinfo {author} {\bibfnamefont {M.}~\bibnamefont
  {Cardona}}\ and\ \bibinfo {author} {\bibfnamefont {G.}~\bibnamefont
  {Guntherodt}},\ }\href {https://doi.org/10.1016/0030-3992(77)90116-5} {\emph
  {\bibinfo {title} {{Light Scattering in Solids II - Basic Concepts and
  Instrumentation}}}}\ (\bibinfo  {publisher} {Springer-Verlag},\ \bibinfo
  {year} {1982})\ pp.\ \bibinfo {pages} {1--252}\BibitemShut {NoStop}%
\bibitem [{\citenamefont {Gross}\ \emph {et~al.}(2017)\citenamefont {Gross},
  \citenamefont {Sun}, \citenamefont {Perera}, \citenamefont {Hui},
  \citenamefont {Wei}, \citenamefont {Zhang}, \citenamefont {Zeng},\ and\
  \citenamefont {Weinstein}}]{Gross2017}%
  \BibitemOpen
  \bibfield  {author} {\bibinfo {author} {\bibfnamefont {N.}~\bibnamefont
  {Gross}}, \bibinfo {author} {\bibfnamefont {Y.-Y.}\ \bibnamefont {Sun}},
  \bibinfo {author} {\bibfnamefont {S.}~\bibnamefont {Perera}}, \bibinfo
  {author} {\bibfnamefont {H.}~\bibnamefont {Hui}}, \bibinfo {author}
  {\bibfnamefont {X.}~\bibnamefont {Wei}}, \bibinfo {author} {\bibfnamefont
  {S.}~\bibnamefont {Zhang}}, \bibinfo {author} {\bibfnamefont
  {H.}~\bibnamefont {Zeng}},\ and\ \bibinfo {author} {\bibfnamefont {B.~A.}\
  \bibnamefont {Weinstein}},\ }\bibfield  {title} {\bibinfo {title} {{Stability
  and Band-Gap Tuning of the Chalcogenide Perovskite \BZS\ in Raman and Optical
  Investigations at High Pressures}},\ }\href
  {https://doi.org/10.1103/PhysRevApplied.8.044014} {\bibfield  {journal}
  {\bibinfo  {journal} {Physical Review Applied}\ }\textbf {\bibinfo {volume}
  {8}},\ \bibinfo {pages} {044014} (\bibinfo {year} {2017})}\BibitemShut
  {NoStop}%
\bibitem [{\citenamefont {Niu}\ \emph {et~al.}(2018)\citenamefont {Niu},
  \citenamefont {Milam-Guerrero}, \citenamefont {Zhou}, \citenamefont {Ye},
  \citenamefont {Zhao}, \citenamefont {Melot},\ and\ \citenamefont
  {Ravichandran}}]{Niu2018}%
  \BibitemOpen
  \bibfield  {author} {\bibinfo {author} {\bibfnamefont {S.}~\bibnamefont
  {Niu}}, \bibinfo {author} {\bibfnamefont {J.}~\bibnamefont {Milam-Guerrero}},
  \bibinfo {author} {\bibfnamefont {Y.}~\bibnamefont {Zhou}}, \bibinfo {author}
  {\bibfnamefont {K.}~\bibnamefont {Ye}}, \bibinfo {author} {\bibfnamefont
  {B.}~\bibnamefont {Zhao}}, \bibinfo {author} {\bibfnamefont {B.~C.}\
  \bibnamefont {Melot}},\ and\ \bibinfo {author} {\bibfnamefont
  {J.}~\bibnamefont {Ravichandran}},\ }\bibfield  {title} {\bibinfo {title}
  {{Thermal stability study of transition metal perovskite sulfides}},\ }\href
  {https://doi.org/10.1557/jmr.2018.419} {\bibfield  {journal} {\bibinfo
  {journal} {Journal of Materials Research}\ }\textbf {\bibinfo {volume}
  {33}},\ \bibinfo {pages} {4135} (\bibinfo {year} {2018})}\BibitemShut
  {NoStop}%
\bibitem [{\citenamefont {Comparotto}\ \emph {et~al.}(2020)\citenamefont
  {Comparotto}, \citenamefont {Davydova}, \citenamefont {Ericson},
  \citenamefont {Riekehr}, \citenamefont {Moro}, \citenamefont {Kubart},\ and\
  \citenamefont {Scragg}}]{Comparotto2020}%
  \BibitemOpen
  \bibfield  {author} {\bibinfo {author} {\bibfnamefont {C.}~\bibnamefont
  {Comparotto}}, \bibinfo {author} {\bibfnamefont {A.}~\bibnamefont
  {Davydova}}, \bibinfo {author} {\bibfnamefont {T.}~\bibnamefont {Ericson}},
  \bibinfo {author} {\bibfnamefont {L.}~\bibnamefont {Riekehr}}, \bibinfo
  {author} {\bibfnamefont {M.~V.}\ \bibnamefont {Moro}}, \bibinfo {author}
  {\bibfnamefont {T.}~\bibnamefont {Kubart}},\ and\ \bibinfo {author}
  {\bibfnamefont {J.}~\bibnamefont {Scragg}},\ }\bibfield  {title} {\bibinfo
  {title} {{Chalcogenide Perovskite BaZrS$_3$ : Thin Film Growth by Sputtering
  and Rapid Thermal Processing}},\ }\href
  {https://doi.org/10.1021/acsaem.9b02428} {\bibfield  {journal} {\bibinfo
  {journal} {ACS Applied Energy Materials}\ }\textbf {\bibinfo {volume} {3}},\
  \bibinfo {pages} {2762} (\bibinfo {year} {2020})}\BibitemShut {NoStop}%
\bibitem [{\citenamefont {Xu}\ \emph {et~al.}(2022)\citenamefont {Xu},
  \citenamefont {Fan}, \citenamefont {Tian}, \citenamefont {Ye}, \citenamefont
  {Zhang}, \citenamefont {Tian}, \citenamefont {Han},\ and\ \citenamefont
  {Shi}}]{Xu2022}%
  \BibitemOpen
  \bibfield  {author} {\bibinfo {author} {\bibfnamefont {J.}~\bibnamefont
  {Xu}}, \bibinfo {author} {\bibfnamefont {Y.}~\bibnamefont {Fan}}, \bibinfo
  {author} {\bibfnamefont {W.}~\bibnamefont {Tian}}, \bibinfo {author}
  {\bibfnamefont {L.}~\bibnamefont {Ye}}, \bibinfo {author} {\bibfnamefont
  {Y.}~\bibnamefont {Zhang}}, \bibinfo {author} {\bibfnamefont
  {Y.}~\bibnamefont {Tian}}, \bibinfo {author} {\bibfnamefont {Y.}~\bibnamefont
  {Han}},\ and\ \bibinfo {author} {\bibfnamefont {Z.}~\bibnamefont {Shi}},\
  }\bibfield  {title} {\bibinfo {title} {{Enhancing the optical absorption of
  chalcogenide perovskite BaZrS$_3$ by optimizing the synthesis and
  post-processing conditions}},\ }\href
  {https://doi.org/10.1016/j.jssc.2021.122872} {\bibfield  {journal} {\bibinfo
  {journal} {Journal of Solid State Chemistry}\ }\textbf {\bibinfo {volume}
  {307}},\ \bibinfo {pages} {122872} (\bibinfo {year} {2022})}\BibitemShut
  {NoStop}%
\bibitem [{\citenamefont {Yu}\ \emph {et~al.}(2021)\citenamefont {Yu},
  \citenamefont {Wei}, \citenamefont {Zheng}, \citenamefont {Hui},
  \citenamefont {Bian}, \citenamefont {Dhole}, \citenamefont {Seo},
  \citenamefont {Sun}, \citenamefont {Jia}, \citenamefont {Zhang},
  \citenamefont {Yang},\ and\ \citenamefont {Zeng}}]{Yu2021}%
  \BibitemOpen
  \bibfield  {author} {\bibinfo {author} {\bibfnamefont {Z.}~\bibnamefont
  {Yu}}, \bibinfo {author} {\bibfnamefont {X.}~\bibnamefont {Wei}}, \bibinfo
  {author} {\bibfnamefont {Y.}~\bibnamefont {Zheng}}, \bibinfo {author}
  {\bibfnamefont {H.}~\bibnamefont {Hui}}, \bibinfo {author} {\bibfnamefont
  {M.}~\bibnamefont {Bian}}, \bibinfo {author} {\bibfnamefont {S.}~\bibnamefont
  {Dhole}}, \bibinfo {author} {\bibfnamefont {J.-H.}\ \bibnamefont {Seo}},
  \bibinfo {author} {\bibfnamefont {Y.-Y.}\ \bibnamefont {Sun}}, \bibinfo
  {author} {\bibfnamefont {Q.}~\bibnamefont {Jia}}, \bibinfo {author}
  {\bibfnamefont {S.}~\bibnamefont {Zhang}}, \bibinfo {author} {\bibfnamefont
  {S.}~\bibnamefont {Yang}},\ and\ \bibinfo {author} {\bibfnamefont
  {H.}~\bibnamefont {Zeng}},\ }\bibfield  {title} {\bibinfo {title}
  {{Chalcogenide perovskite BaZrS$_3$ thin-film electronic and optoelectronic
  devices by low temperature processing}},\ }\href
  {https://doi.org/10.1016/j.nanoen.2021.105959} {\bibfield  {journal}
  {\bibinfo  {journal} {Nano Energy}\ }\textbf {\bibinfo {volume} {85}},\
  \bibinfo {pages} {105959} (\bibinfo {year} {2021})}\BibitemShut {NoStop}%
\bibitem [{\citenamefont {Kwok}(1968)}]{Kwok1968}%
  \BibitemOpen
  \bibfield  {author} {\bibinfo {author} {\bibfnamefont {P.~C.}\ \bibnamefont
  {Kwok}},\ }\bibfield  {title} {\bibinfo {title} {{Green's Function Method in
  Lattice Dynamics}},\ }in\ \href
  {https://doi.org/10.1016/S0081-1947(08)60219-2} {\emph {\bibinfo {booktitle}
  {Solid State Phys.}}}\ (\bibinfo  {publisher} {Academic Press},\ \bibinfo
  {address} {Cambridge, MA, USA},\ \bibinfo {year} {1968})\ pp.\ \bibinfo
  {pages} {213--303}\BibitemShut {NoStop}%
\bibitem [{\citenamefont {Maradudin}\ \emph {et~al.}(1971)\citenamefont
  {Maradudin}, \citenamefont {Montroll}, \citenamefont {Weiss},\ and\
  \citenamefont {Ipatova}}]{Maradudin1971}%
  \BibitemOpen
  \bibfield  {author} {\bibinfo {author} {\bibfnamefont {A.~A.}\ \bibnamefont
  {Maradudin}}, \bibinfo {author} {\bibfnamefont {E.~W.}\ \bibnamefont
  {Montroll}}, \bibinfo {author} {\bibfnamefont {G.~H.}\ \bibnamefont
  {Weiss}},\ and\ \bibinfo {author} {\bibfnamefont {I.~P.}\ \bibnamefont
  {Ipatova}},\ }\bibinfo {title} {{Theory of Lattice Dynamics in the Harmonic
  Approximation}}\ (\bibinfo  {publisher} {Academic Press},\ \bibinfo {address}
  {New York, USA},\ \bibinfo {year} {1971})\ pp.\ \bibinfo {pages} {1--67},\
  \bibinfo {edition} {2nd}\ ed.\BibitemShut {Stop}%
\bibitem [{\citenamefont {Safran}\ \emph {et~al.}(1977)\citenamefont {Safran},
  \citenamefont {Dresselhaus},\ and\ \citenamefont {Lax}}]{Safran1977}%
  \BibitemOpen
  \bibfield  {author} {\bibinfo {author} {\bibfnamefont {S.~A.}\ \bibnamefont
  {Safran}}, \bibinfo {author} {\bibfnamefont {G.}~\bibnamefont
  {Dresselhaus}},\ and\ \bibinfo {author} {\bibfnamefont {B.}~\bibnamefont
  {Lax}},\ }\bibfield  {title} {\bibinfo {title} {{Theory of spin-disorder
  Raman scattering in magnetic semiconductors}},\ }\href
  {https://doi.org/https://doi.org/10.1103/PhysRevB.16.2749} {\bibfield
  {journal} {\bibinfo  {journal} {Phys. Rev. B}\ }\textbf {\bibinfo {volume}
  {16}},\ \bibinfo {pages} {2749} (\bibinfo {year} {1977})}\BibitemShut
  {NoStop}%
\bibitem [{\citenamefont {Filippone}\ \emph {et~al.}(2020)\citenamefont
  {Filippone}, \citenamefont {Zhao}, \citenamefont {Niu}, \citenamefont
  {Koocher}, \citenamefont {Silevitch}, \citenamefont {Fina}, \citenamefont
  {Rondinelli}, \citenamefont {Ravichandran},\ and\ \citenamefont
  {Jaramillo}}]{Filippone2020}%
  \BibitemOpen
  \bibfield  {author} {\bibinfo {author} {\bibfnamefont {S.}~\bibnamefont
  {Filippone}}, \bibinfo {author} {\bibfnamefont {B.}~\bibnamefont {Zhao}},
  \bibinfo {author} {\bibfnamefont {S.}~\bibnamefont {Niu}}, \bibinfo {author}
  {\bibfnamefont {N.~Z.}\ \bibnamefont {Koocher}}, \bibinfo {author}
  {\bibfnamefont {D.}~\bibnamefont {Silevitch}}, \bibinfo {author}
  {\bibfnamefont {I.}~\bibnamefont {Fina}}, \bibinfo {author} {\bibfnamefont
  {J.~M.}\ \bibnamefont {Rondinelli}}, \bibinfo {author} {\bibfnamefont
  {J.}~\bibnamefont {Ravichandran}},\ and\ \bibinfo {author} {\bibfnamefont
  {R.}~\bibnamefont {Jaramillo}},\ }\bibfield  {title} {\bibinfo {title}
  {{Discovery of highly polarizable semiconductors BaZrS$_3$ and
  Ba$_3$Zr$_2$S$_7$}},\ }\bibfield  {journal} {\bibinfo  {journal} {Physical
  Review Materials}\ }\textbf {\bibinfo {volume} {4}},\ \href
  {https://doi.org/10.1103/PhysRevMaterials.4.091601}
  {10.1103/PhysRevMaterials.4.091601} (\bibinfo {year} {2020})\BibitemShut
  {NoStop}%
\end{thebibliography}
